\def\zabs{$z_{\rm abs}$}

\def\kms{km~s$^{-1}$}

\def\ha{H$\alpha$}
\def\hb{H$\beta$}

\def\lya{Ly-$\alpha$}

\def\oii{[O~{\sc ii}]}
\def\oiii{[O~{\sc iii}]}
\def\nii{[N~{\sc ii}]}

\def\hi{H~{\sc i}}

\def\nhi{\mbox{$\sc N(\rm H~{\sc I})$}}

\def\caii{Ca~{\sc ii}}

\def\feii{Fe~{\sc ii}}

\def\mgi{Mg~{\sc i}}
\def\mgii{Mg~{\sc ii}}

\def\mnii{Mn~{\sc ii}}

\def\tiii{Ti~{\sc ii}}

\newcommand{\fSB}{erg~s$^{-1}$~cm$^{-2}$~arcsec$^{-2}$}

\newcommand{\flux}{erg~s$^{-1}$~cm$^{-2}$}

\documentclass[useAMS,usenatbib,epsfig]{mn2e}
\usepackage{rotating}
\usepackage{float}

\title[Multi-phase CGM probed with MUSE and ALMA]{Multi-phase Circum-Galactic Medium probed with MUSE and ALMA\thanks{Based on observations collected at the European Organisation for Astronomical Research in the Southern Hemisphere under ESO programme(s) 67.A-0567, 69.A-0371 and 096.A-0303.} }
\author[C\'eline P\'eroux et al.] {C\'eline P\'eroux$^1$\thanks{e-mail:celine.peroux@gmail.com}, 
Martin A. Zwaan$^2$, Anne Klitsch$^{2,3}$, Ramona Augustin$^{1,2}$, 
\newauthor 
Aleksandra Hamanowicz$^2$, Hadi Rahmani$^{1,4}$, 
Max Pettini$^5$, Varsha Kulkarni$^6$, 
\newauthor
Lorrie A. Straka$^7$, 
Andy D. Biggs$^2$, Donald G. York$^8$, Bruno Milliard$^1$\\
$^1$ Aix Marseille Universit\'e, CNRS, LAM (Laboratoire d'Astrophysique de Marseille) UMR 7326, 13388, Marseille, France.\\
$^2$ European Southern Observatory (ESO), Karl-Schwarzschild-Str.2, D-85748 Garching b. M\"unchen, Germany.\\
$^3$ Centre for Extragalactic Astronomy, Durham University, Department of Physics, South Road, Durham DH1 3LE, UK.\\
$^4$ GEPI, Observatoire de Paris, PSL Universit\'e, CNRS,  5 Place Jules Janssen, 92190 Meudon, France.\\
$^5$ Institute of Astronomy, University of Cambridge, Madingley Road, Cambridge CB3 0HA, UK.\\
$^6$ Dept. of Physics and Astronomy, Univ. of South Carolina, Columbia, SC 29208, USA.\\
$^7$ Sterrewacht Leiden, Leiden University, PO Box 9513, NL-2300 RA Leiden, the Netherlands.\\
$^8$ Dept. of Astronomy and Astrophysics and The Enrico Fermi Institute, University of Chicago, 5640 S. Ellis Ave, Chicago, IL 60637, USA.\\
}

\begin{document}

\date{Accepted 2019 January 16. Received 2018 December 07; in original form 2018 August 1.}

\pagerange{\pageref{firstpage}--\pageref{lastpage}} \pubyear{2016}

\maketitle

\label{firstpage}

\begin{abstract}

Galaxy halos appear to be missing a large fraction of their baryons, most probably hiding in the circumgalactic medium (CGM), a diffuse component within the dark matter halo that extends far from the inner regions of the galaxies. A powerful tool to study the CGM gas is offered by absorption lines in the spectra of background quasars. Here, we present optical (MUSE) and mm (ALMA) observations of the field of the quasar Q1130$-$1449 which includes a log $[N(H I)/cm^{-2}]$=21.71$\pm$0.07 absorber at z=0.313. Ground-based VLT/MUSE 3D spectroscopy shows 11 galaxies at the redshift of the absorber down to a limiting SFR$>$0.01 M$_{\rm \odot} yr^{-1}$ (covering emission lines of \oii, \hb, \oiii, \nii\ and \ha), 7 of which are new discoveries. In particular, we report a new emitter with a smaller impact parameter to the quasar line-of-sight (b=10.6~kpc) than the galaxies detected so far. Three of the objects are also detected in CO(1--0) in our ALMA observations indicating long depletion timescales for the molecular gas and kinematics consistent with the ionised gas. We infer from dedicated numerical cosmological RAMSES zoom-in simulations that the physical properties of these objects qualitatively resemble a small group environment, possibly part of a filamentary structure. Based on metallicity and velocity arguments, we conclude that the neutral gas traced in absorption is only partly related to these emitting galaxies while a larger fraction is likely the signature of gas with surface brightness almost four orders of magnitude fainter that current detection limits. Together, these findings challenge a picture where strong-\nhi\ quasar absorbers are associated with a single bright galaxy and favour a scenario where the \hi\ gas probed in absorption is related to far more complex galaxy structures.
\end{abstract}
\begin{keywords}
galaxies: kinematics and dynamics -- galaxies: abundances -- galaxies: ISM -- quasars: absorption lines -- intergalactic medium
\end{keywords}

\section{Introduction}

Baryons are missing from galaxies in what is known as the galaxy halo missing baryon problem \citep{mcgaugh08}. Indeed, galaxy halos appear to be missing approximately 60\% of their baryons compared to expectations from the cosmological 
mass density, suggesting that these are structures nearly devoid of baryons both in mass and spatial extent. It has become apparent that galaxies also exhibit a diffuse baryonic component within the dark matter halo that extends far from the inner regions to the virial radius and beyond. This halo gas or circumgalactic medium (CGM) has been partly detected in the hot gas emitted around galaxies and detected in the X-rays regime \citep[e.g.][]{anderson10, bregman18, nicastro18} as well as in \lya\ \citep[e.g.][]{wisotzki16, leclercq17, wisotzki18}. On the other hand, a powerful tool to study the cooler gas is offered by absorption lines in the spectra of background quasars as it allows us to study under-dense gas regions with a sensitivity independent of redshift \citep[e.g.][]{morris93, tripp98}. Observations of redshifted UV absorption lines have indeed revealed much lower temperature and density gas in the CGM \citep{steidel10, rudie12, werk13,turner14}.

A remaining challenge is to relate the gas traced in absorption to the galaxy properties. A  particularly powerful technique to bridge the gas component and stellar content of galaxies has come from integral field spectroscopy (IFS). Over the past few years, this technique has been used to study the CGM of high-redshift galaxies using near-infrared Integral Field Units (IFUs) by successfully detecting the galaxies 
responsible for strong \hi\ absorbers at redshifts $z \sim 1$ and $z \sim 2$ in H$\alpha$ and \nii\ emission \citep{bouche07,peroux11a, peroux11b, rudie17}. Together with long slit spectroscopy follow-up, these observations have enabled us to map the kinematics, star formation rate, and metallicity of this emission-line gas, and to estimate the dynamical masses of these galaxies \citep{bouche12, peroux12, bouche13, peroux14}.  As optical IFUs became available, the study of the CGM using absorption techniques has expanded \citep{schroetter15, bielby16, bouche16, fumagalli16, bielby18} to also include lower-redshift objects \citep{peroux17,klitsch18, rahmani18a, peroux18}. 

Here, we present results from new MUSE and ALMA observations of the field of background quasar Q1130$-$1449 with a strong \hi\ absorber at z$_{\rm abs}$=0.313. 
The manuscript is organised as follows: Section 2 presents the observational set-up and data reduction. Section 3 details the analysis and results derived from these data, while Section 4 summarises our findings.
Throughout this paper we adopt an $H_{0}=70$~\kms~Mpc$^{-1}$, $\Omega_{\rm M}=0.3$, and $\Omega_{\rm \Lambda}=0.7$ cosmology. At the redshift of the absorber, 1~arcsec corresponds to 4.6 ~kpc.

\section{Observations of the Q1130$-$1449 Field}

\subsection{Quasar Spectroscopy: Absorption Properties}

The $z_{\rm QSO}=$1.189 (V=16.9\,mag) quasar (J2000 coordinates: 113007.05--144927.38; alternative name: PKS B1127--145) sightline intersects three {\mgii} absorption systems:
$z_{\rm abs}=$0.191 \citep{kacprzak10b}, 0.313 \citep{bergeron91} and 0.328
 \citep{narayanan07}. We here concentrate on the middle redshift absorber, while Hamanowicz et al. (in prep) presents the other two systems.

From a {\it HST}/Faint Object Spectrograph (FOS) UV spectrum, the $z_{\rm abs}=$0.313 absorber was determined
to be a damped Ly$\alpha$ (DLA) system with a large \hi\ column density of neutral gas: log $[N(H I)/cm^{-2}]$=21.71$\pm$0.07 \citep{lane98}. \citet{lane98} also detected 21cm absorption spanning FWHM=42.1$\pm$2.7 ~\kms\ at the redshift of that absorber with the Westerbork Synthesis Radio Telescope (WSRT). This was later confirmed by \citet{kanekar09} who derived a spin temperature T$_{\rm spin}$=820$\pm$145K and covering factor, f=0.9 from Very-long-baseline interferometry (VLBI) observations. \citet{bechtold01} additionally reported strong X-ray absorption in excess of the Galactic value in the Chandra quasar spectrum which they interpreted as associated with that absorber. 
Based on this assumption, they derived the abundance of the oxygen group elements, independently of dust depletion, of 23\% solar. 
\citet{kanekar14} computed the zinc metallicity to be [Zn/H]$=-0.80\pm0.16$ relative to
solar (i.e. the neutral gas metallicity is 16\% solar) from a medium-resolution {\it HST}/STIS spectrum. \cite{guber18} used the VLT/UVES spectrum of the quasar to fit the multi-component profile in Ti, Mn and Ca which are likely depleted into dust grains. In addition, they estimated the \hi\ column density of each component from the observed \caii\ column density pattern, assuming a constant \caii/\hi\ ratio. Interestingly, the thus estimated \hi\ column density for each component, on its own, exceeds the canonical DLA definition (log $[N(H I)/cm^{-2}]>$20.3).

We obtained the VLT/UVES reduced quasar spectra from the ESO advanced data products archives\footnote{http://archive.eso.org/cms.html}. Fig.\ref{f:uves_fit} of the appendix~\ref{app_section:abs} presents our independent fit to the metal absorption profiles, which is consistent with findings from \cite{guber18}, adding lines from \feii, \mgi\ and the saturated \mgii\ doublet. The zero velocity in the figure is set to the systemic redshift of galaxy G0, $z_{G0}=$0.31305 (see later section). Most of the absorption lies blueward of this systemic redshift.

\subsection{Field Imaging: Identifying Absorbing Galaxies}

\subsubsection{New MUSE Observations}

\begin{figure*}
\includegraphics[angle=0,scale=0.78]{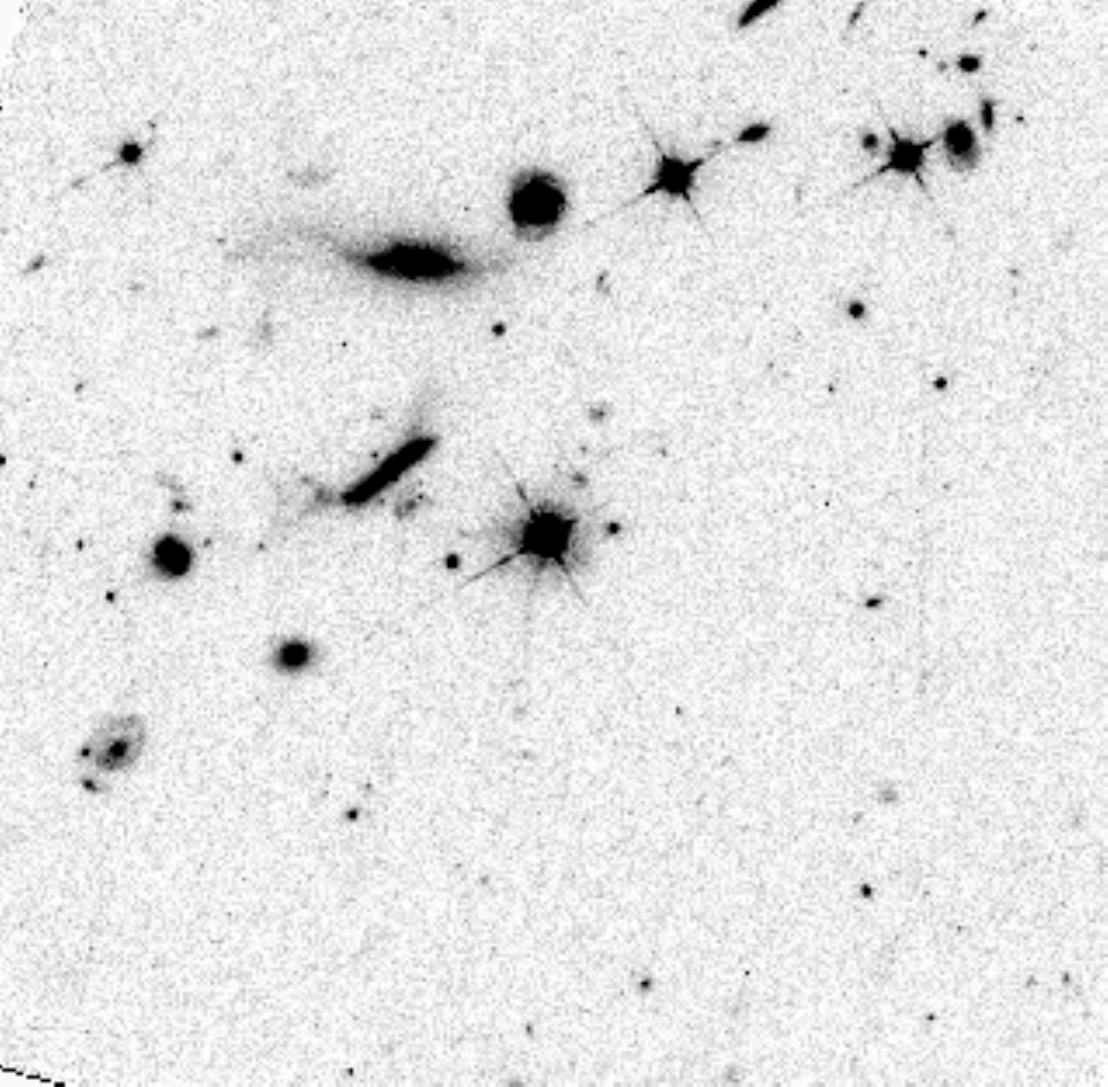}
\includegraphics[angle=0,scale=0.325]{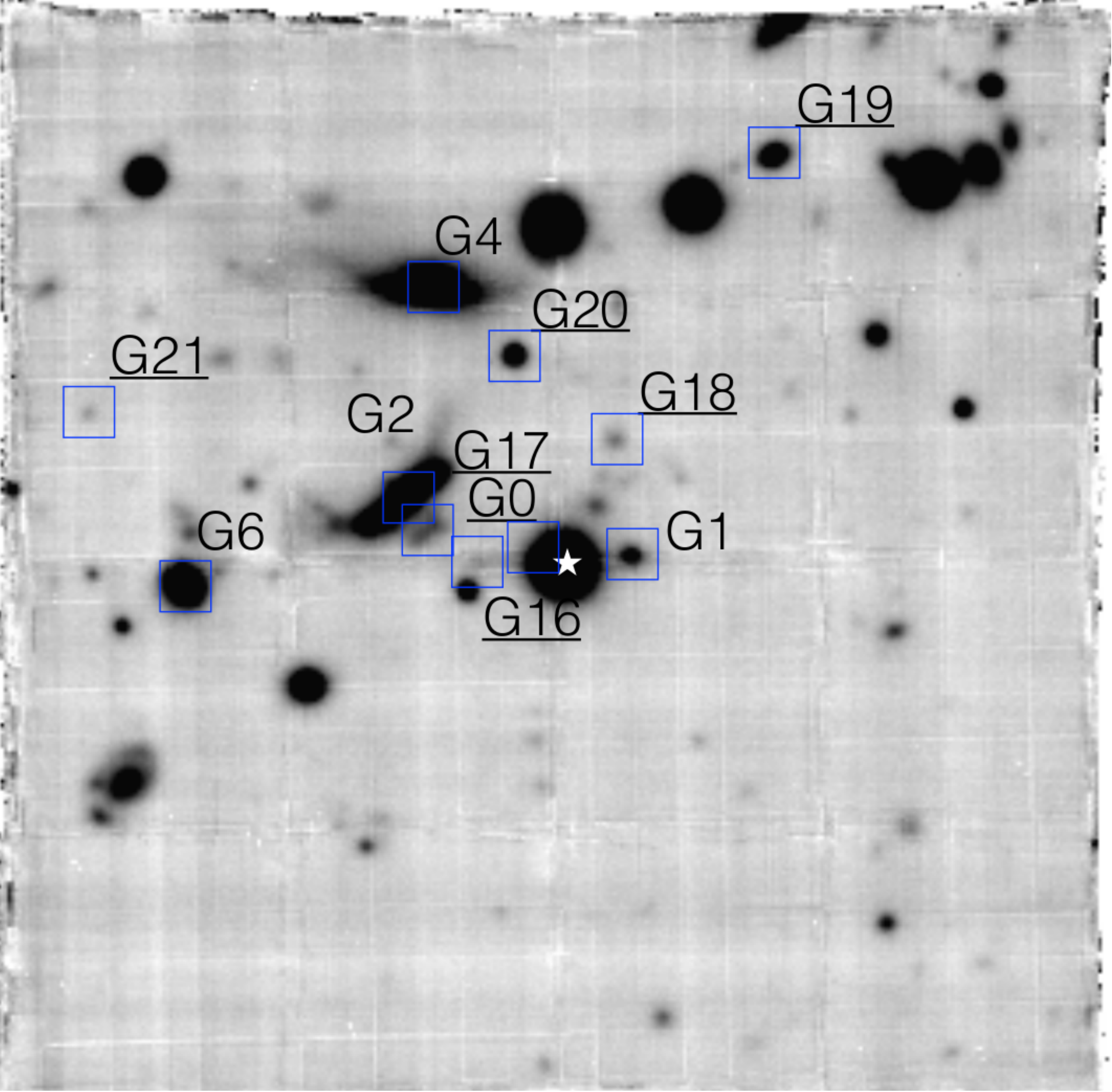}
\caption{{\bf HST/WFPC2 F814W image of the quasar field (left panel) and MUSE white light image (right panel) showing objects with bright continua.} The exposure time for the HST image is 4400 s, while for MUSE it is 14400 s. The quasar is at the center of the image. North is up, East to the left. The fields-of-view are $\sim$1$\times$1'. MUSE observations additionally provide spectroscopic and kinematic information for the majority of the objects in the field. The quasar is shown as a white star and the positions of the galaxies associated with the quasar absorber are marked with light-blue squares. Four of the galaxies previously identified are detected (namely G1, G2, G4 and G6) and their numbering system follows \citet{kacprzak10b}. The other 7 objects (G0 and $>$G16, \underline{underlined}) are new identifications at z$_{\rm abs}$=0.313. Some objects are bright in the continuum but undetected in pseudo narrow-band images because of weak emission lines (i.e. G4, G20 and G21). 
}
\label{f:hst}
\end{figure*}

We report new MUSE medium-deep (5-hours on source) observations of this field. The observations were carried out in service mode (under programme ESO 96.A-0303, PI: C. Peroux) during the nights of 7--8 Dec 2015 (T$_{\rm exp}$=4$\times$960s), 5--6 March 2016 (4$\times$1200s) and 19--20 Feb 2017 (8$\times$1200s). The seeing constraint for these observations was $<0.8~arcsec$ and natural seeing mode was used. The field was rotated by 90 degrees between exposures. These were further divided into two equal sub-exposures, with an additional field rotation of 90 degrees and sub-arcsec dithering offset in 2-step pattern to minimise residuals from the slice pattern. The field of view is 60 ~arcsec $\times$ 60 ~arcsec, corresponding to a 0.2 ~arcsec/pixel scale. We used the "nominal mode" resulting in a spectral coverage of $\sim$4800-9300 \AA. At the redshift of the target ($z_{\rm abs}=$0.313), the data cover emission lines from \oii~$\lambda \lambda 
3727, 3729$ to \ha. The spectral resolution is R=1770 at 4800 \AA\ and R=3590 at 9300 \AA\ resampling the whole spectrum to a spectral sampling of 1.25 \AA/pixel.

The ESO MUSE pipeline \citep{weilbacher15} version v1.6 was used to reduced the data together with additional external routines for 
sky subtraction and extraction of the 1D spectra. We used master bias, flat field images and arc lamp exposures from observations taken closest in time to the science frames to correct each raw cube. We processed the raw science data with the $scibasic$ and $scipost$ recipes, correcting the wavelength calibration to a heliocentric reference. The wavelength solutions were checked using the known wavelengths of the night-sky OH lines and were found to be accurate within 15 ~\kms. 
We centered the individual exposures with the $exp\_align$ recipe using the point sources in the field to ensure accurate relative astrometry. We combined the individual exposures into a single data cube using the $exp\_combine$ recipe. We measured the seeing of the final combined data from the quasar and other bright point sources in the cube. We calculated that the point spread function (PSF) 
has a full width at half maximum of $0.76$~arcsec.
Finally, to estimate the flux uncertainty, we compared the flux calibration with the R magnitude of the quasar and a bright star in the field. We computed the MUSE broad-band fluxes in the R-band filter. We determined the flux error to be $\pm 25$\%. 

The removal of OH emission lines from the night sky was accomplished with an additional purpose-developed code. The $scipost$ recipe was performed with sky-removal method turned off. After selecting sky regions in the field, we created PCA components from the spectra which were further applied to the science datacube to remove sky line residuals \citep{husemann16, peroux17}. The resulting MUSE white-light image is shown on the right panel of Fig.~\ref{f:hst}. We retrieved HST/WFPC2 F814W image of the field from the HST archive, and we show it on the left panel of Fig.~\ref{f:hst}.

\subsubsection{New ALMA Observations}

The field of Q1130$-$1449 was observed with ALMA in Band 3 to cover the CO(1-0) emission lines at the redshift of the $z_{\rm abs}=0.313$ absorber (under programme 2016.1.01250.S, PI: C. Peroux) on 4, 8, and 15 Dec 2016.  The precipitable water vapour (PWV) varied between 1.5 and 5.4~mm and the total on-source observing time was $T_{\rm exp}=3.6$~hrs. A compact antenna configuration (C40-3) resulted in an angular resolution 1.2$\times$1.8$"$. One of the four spectral windows was centred on the redshifted CO(1-0) line frequency of 87.8~GHz and used the high spectral resolution mode, providing 3840 channels of each 0.488~MHz wide. The other three spectral windows were set to low spectral resolution mode and were used for continuum observations of the field. The quasars J1058$+$0133 and J1139$-$1350 were used as amplitude and phase calibrators, respectively.

We started the data reduction with the pipeline-calibrated uv-data sets, as delivered by ALMA. Additional data reduction steps were carried out with the Common Astronomy Software Applications (CASA) software package version 4.7.0. Minor manual flags were added to remove some uv-data points with strongly outlying amplitudes. Next, we applied two cycles of self-calibration on the data. The quasar in the centre of our science field is very bright at mm frequencies ($\sim$800~mJy), which makes the field very suitable for self-calibration of both the phases and amplitudes. 

First, we used the {\it tclean} recipe to Fourier-transform the uv-data into a continuum image. The self-calibration procedure was performed using the tasks {\it gaincal} and {\it applycal} on individual measurement sets to create corrected uv-data, after which the data were re-imaged to create an improved continuum map. We applied one round of phase self-calibration and one round of amplitude and phase self-calibration.The next step was to subtract the bright continuum source from the field using the task {\it uvsub}. We then built a cube with the {\it tclean} task, setting the pixel size to 0.2", so as to oversample the beam sufficiently,  and using a 'robust' weighting scheme with a Briggs parameter of 0.5. One of the four spectral windows was centred on the redshifted CO(1-0) line frequency of 87.8~GHz and used the high spectral resolution mode, using a binning of 8, providing 3840 channels of each 3.906 MHz wide. We applied a spectral binning of seven channels to achieve a velocity resolution of 53~\kms. We removed remaining continuum signatures around the imperfectly subtracted quasar with the {\it uvcontsub} task. Lastly, we produced a cube corrected for the primary beam using the {\it pbcor} task. The final rms noise level of the cube is 0.11~mJy/beam per 53~\kms\ channel, and the angular resolution is $1.4"\times 1.3"$, corresponding to 6~kpc at the redshift of the target. We note that the resulting FWHM of the primary beam of ALMA in band 3 is about $60"$, conveniently matching the MUSE field-of-view.

In addition, we created a small cube around the quasar with the highest spectral resolution ($\sim$7~\kms) to search for absorption lines, but do not detect the CO(1--0) absorption line at the absorber redshift in the spectrum of the background quasar. Given the rms, we compute an optical depth limit of $\tau$=0.0125~\kms\ at 5-$\sigma$ assuming a 10~\kms\ wide absorption line following \cite{zwaan15} calculation. Assuming the excitation temperature equates the $T_{\rm CMB}$ at z=0.313, we derive a stringent limit on the CO column density of N(CO)$<$2$\times$10$^{13}$ cm$^{-2}$ \citep{mangum18}. Using the mean CO/H$_2$ column density ratio of 3$\times$10$^{-6}$ from \cite{burgh07}, this results in a limit of N(H$_2$)$<$7$\times$10$^{18}$ cm$^{-2}$.

\begin{figure*}
\includegraphics[angle=0,scale=0.38]{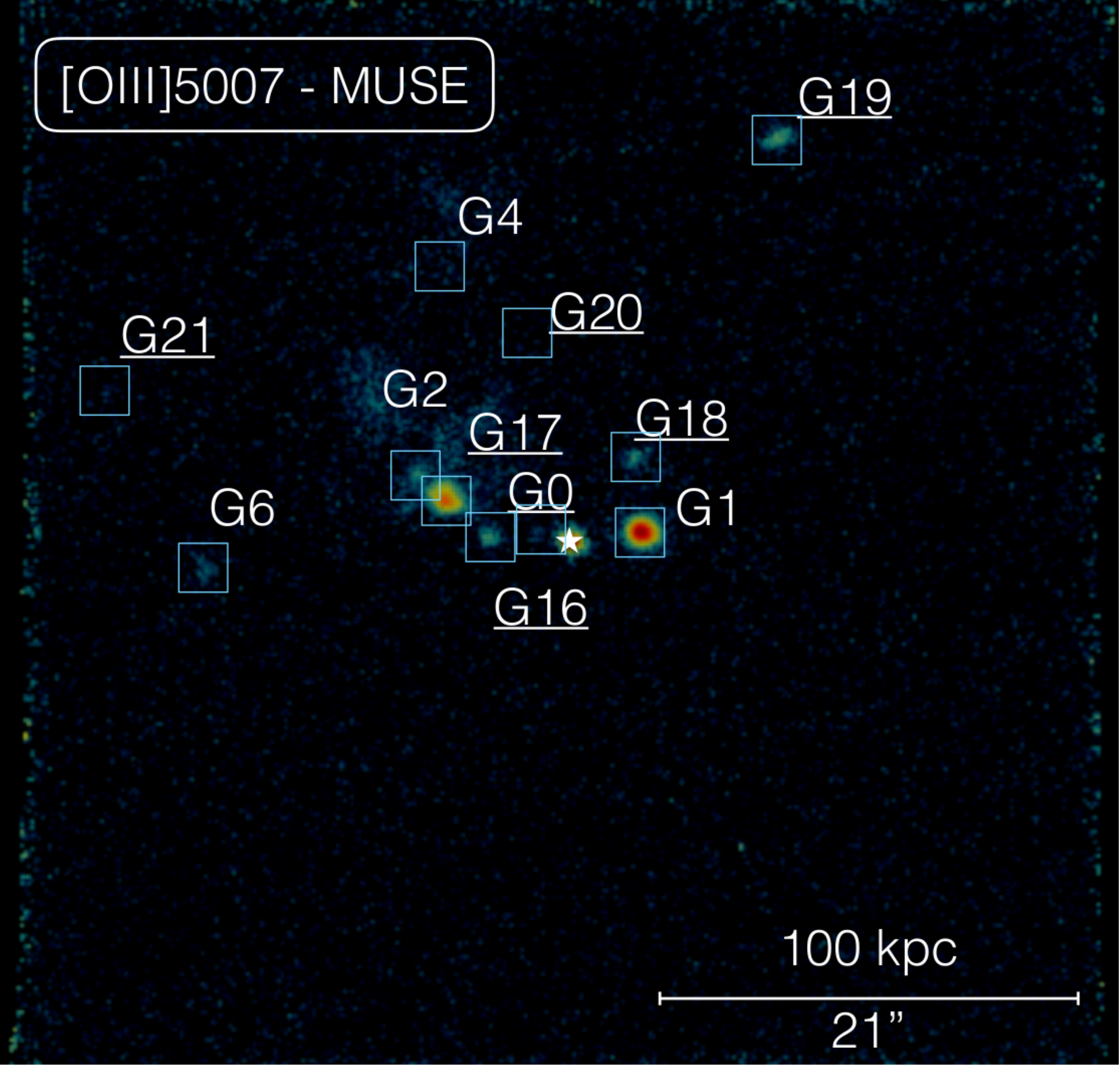}
\includegraphics[angle=0,scale=0.368]{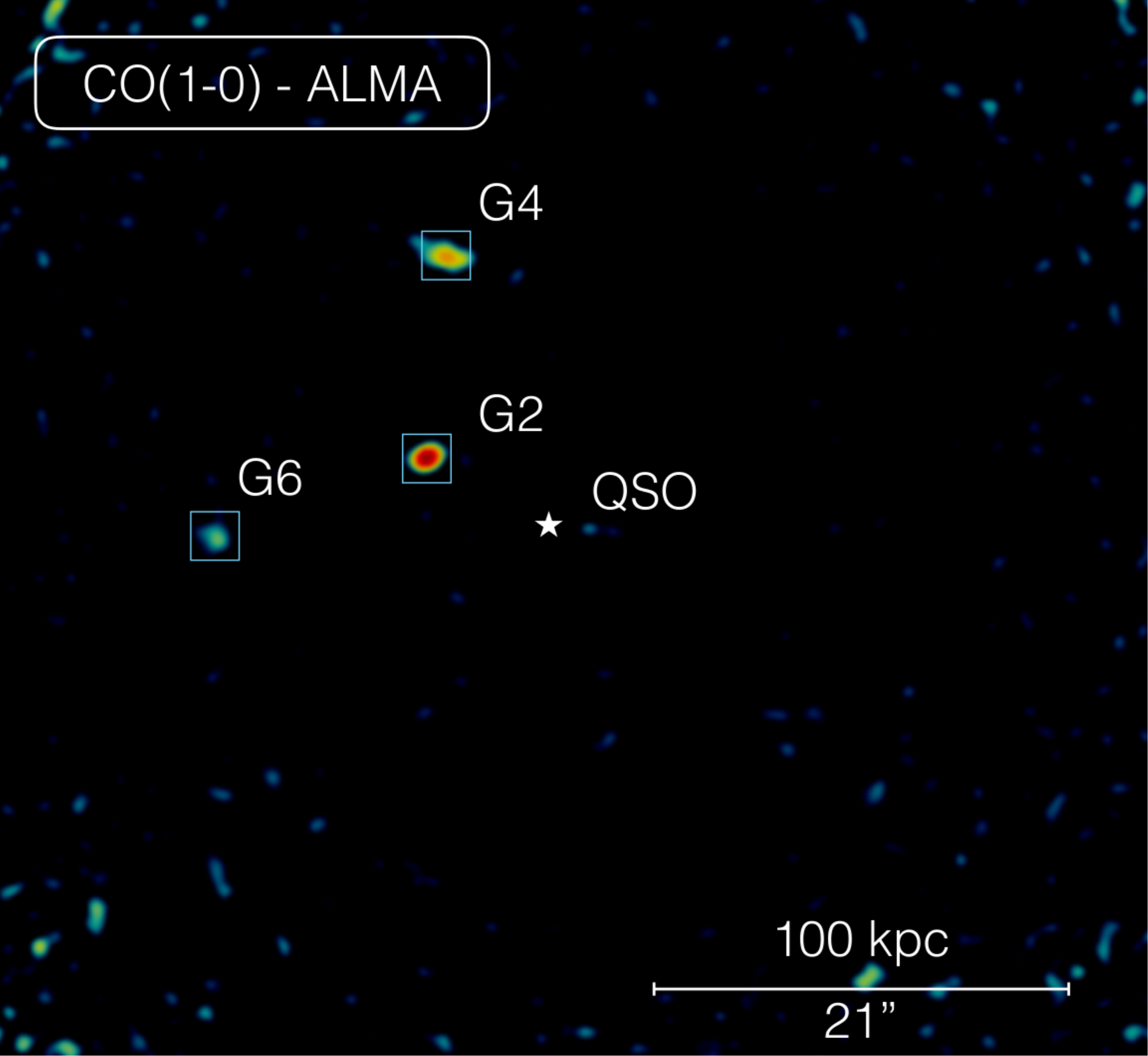}
\caption{{\bf Galaxies at z$_{\rm abs}$=0.313 in the Field of Q1130$-$1449.} {\it Top panel:} Continuum-subtracted MUSE pseudo narrow-band filter around \oiii\ $\lambda 5007$ at the redshift of the absorber. The quasar is shown as a white star and the positions of the galaxies are marked with light-blue squares. Galaxies G4, G20 and G21 have no \oiii\ $\lambda 5007$ emission lines but are detected in continuum. We report also low surface-brightness regions of diffuse gas around G2 and G4 showing all the prominent emission lines such as [O\,II], H$\beta$,  [O\,III],  H$\alpha$ and [N\,II]. {\it Bottom panel:} ALMA Band 3 CO(1-0) 0$^{\rm th}$ moment map of the same field on the same scale. We report three CO (1-0) emission detections matching MUSE galaxy positions (G2, G4 and G6).}
\label{f:OIII_CO}
\end{figure*}

\begin{figure}
\hspace{-1.5cm}
\includegraphics[angle=0, width=12.5cm]{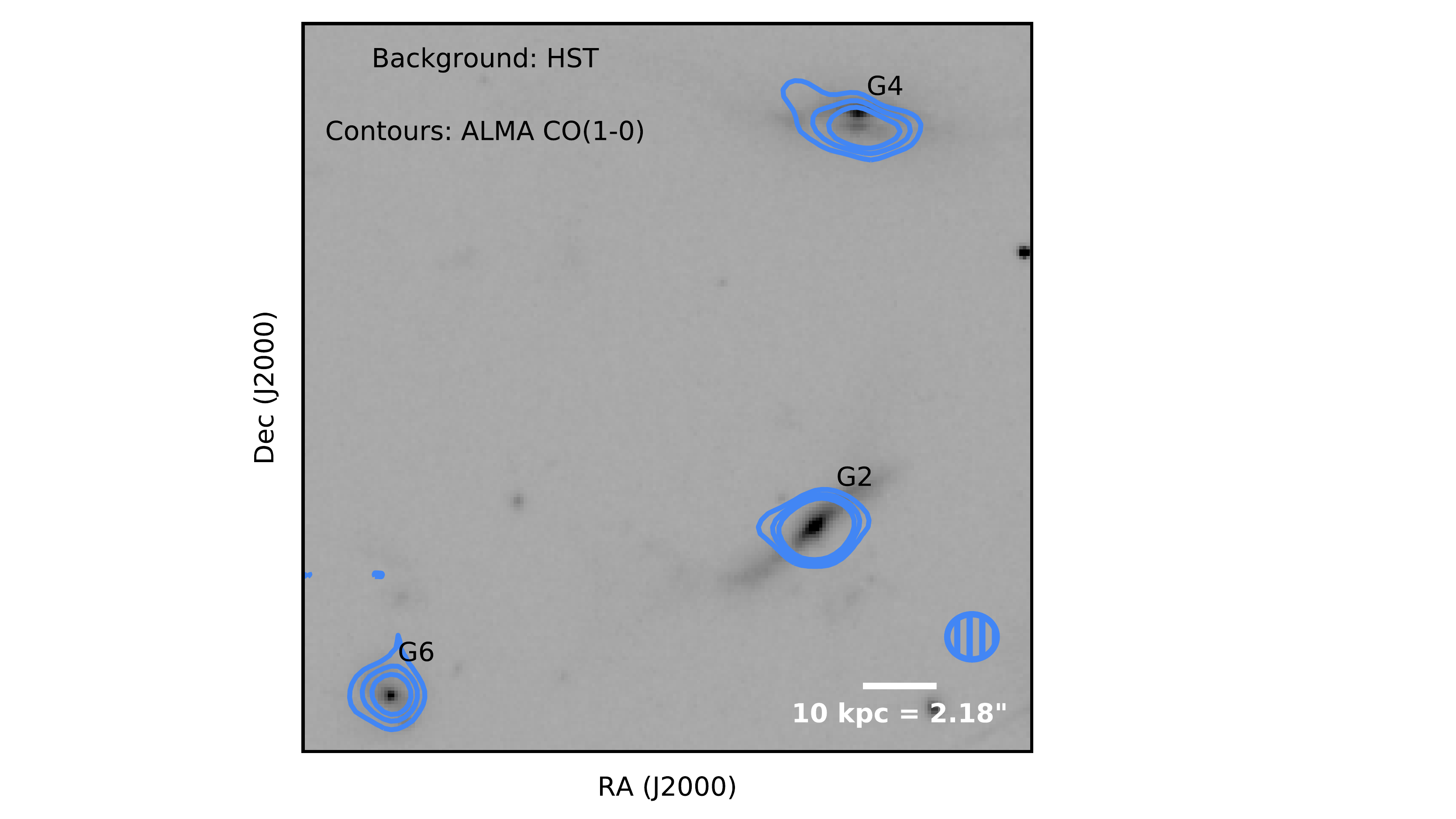}
\caption{{\bf ALMA CO(1--0) contours overlaid on HST image.} The contours show the --3, 3, 5 and 7-$\sigma$ levels, where the dashed negative contours above G6 reflect the noise level in the cube. The hatched ellipse shows the ALMA synthesized beam. 
}
\label{f:ALMA_contours}
\end{figure}

\begin{figure*}
\includegraphics[angle=0,scale=0.244]{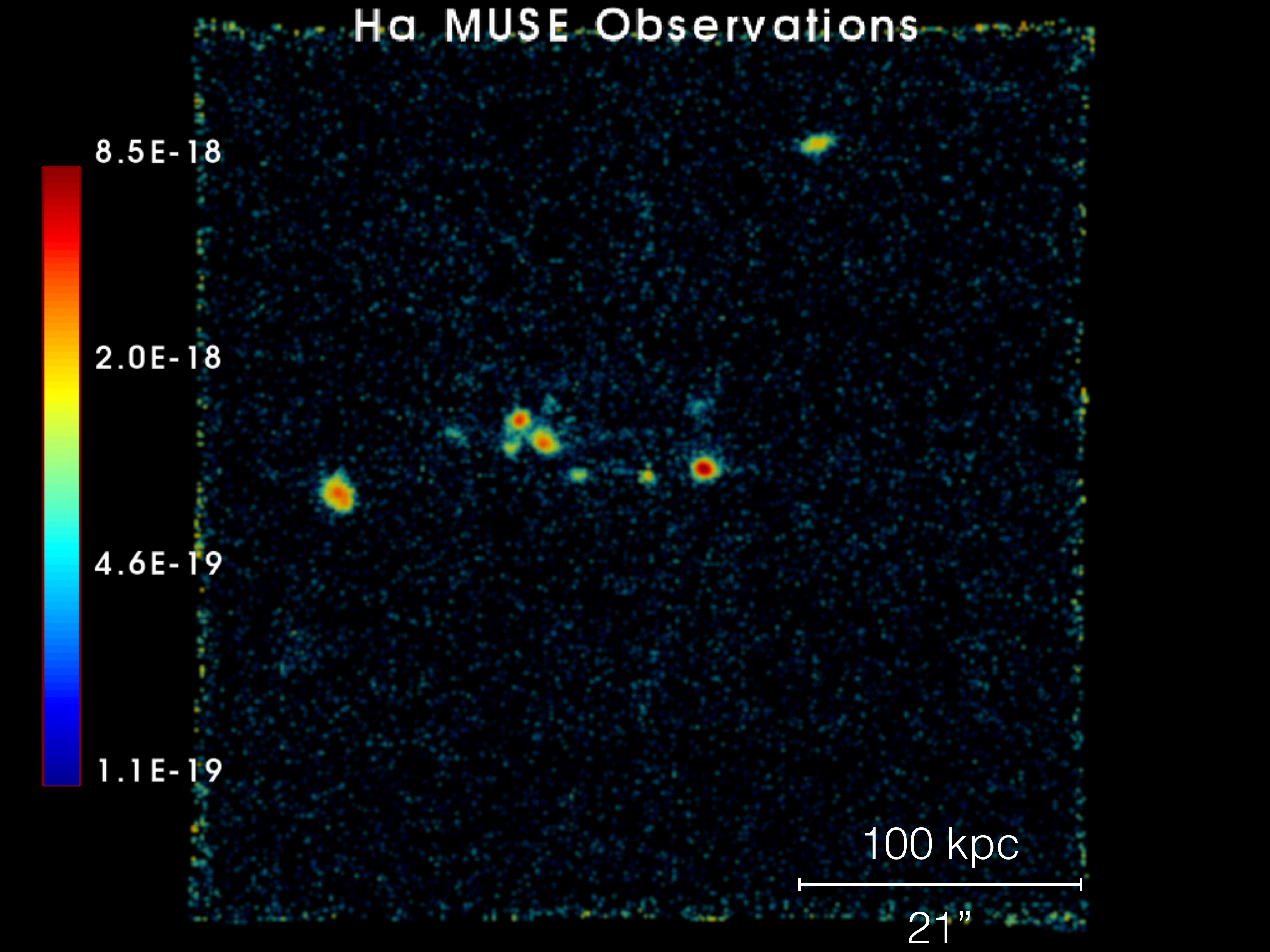}
\includegraphics[angle=0,scale=0.244]{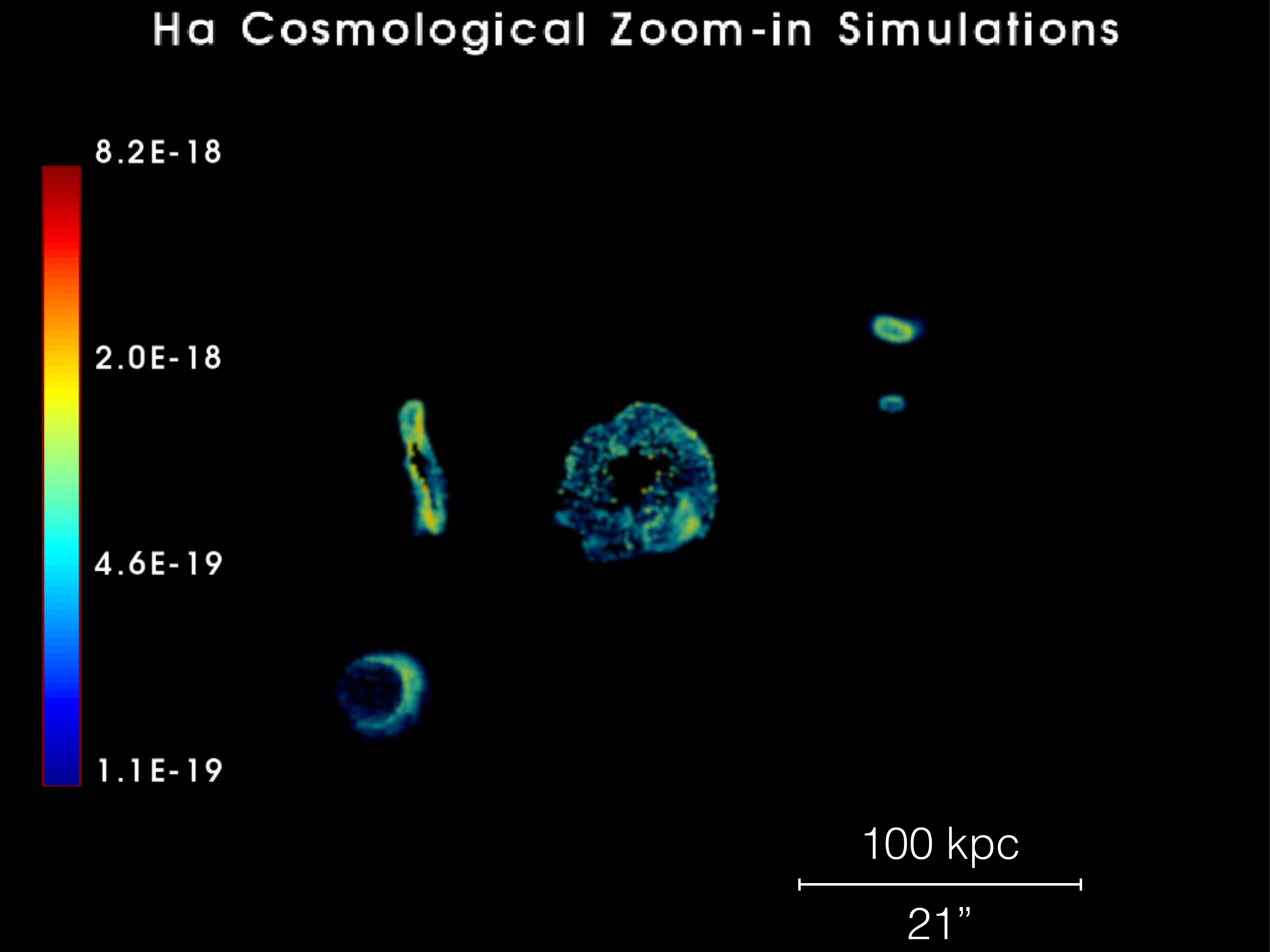}
\includegraphics[angle=0,scale=0.244]{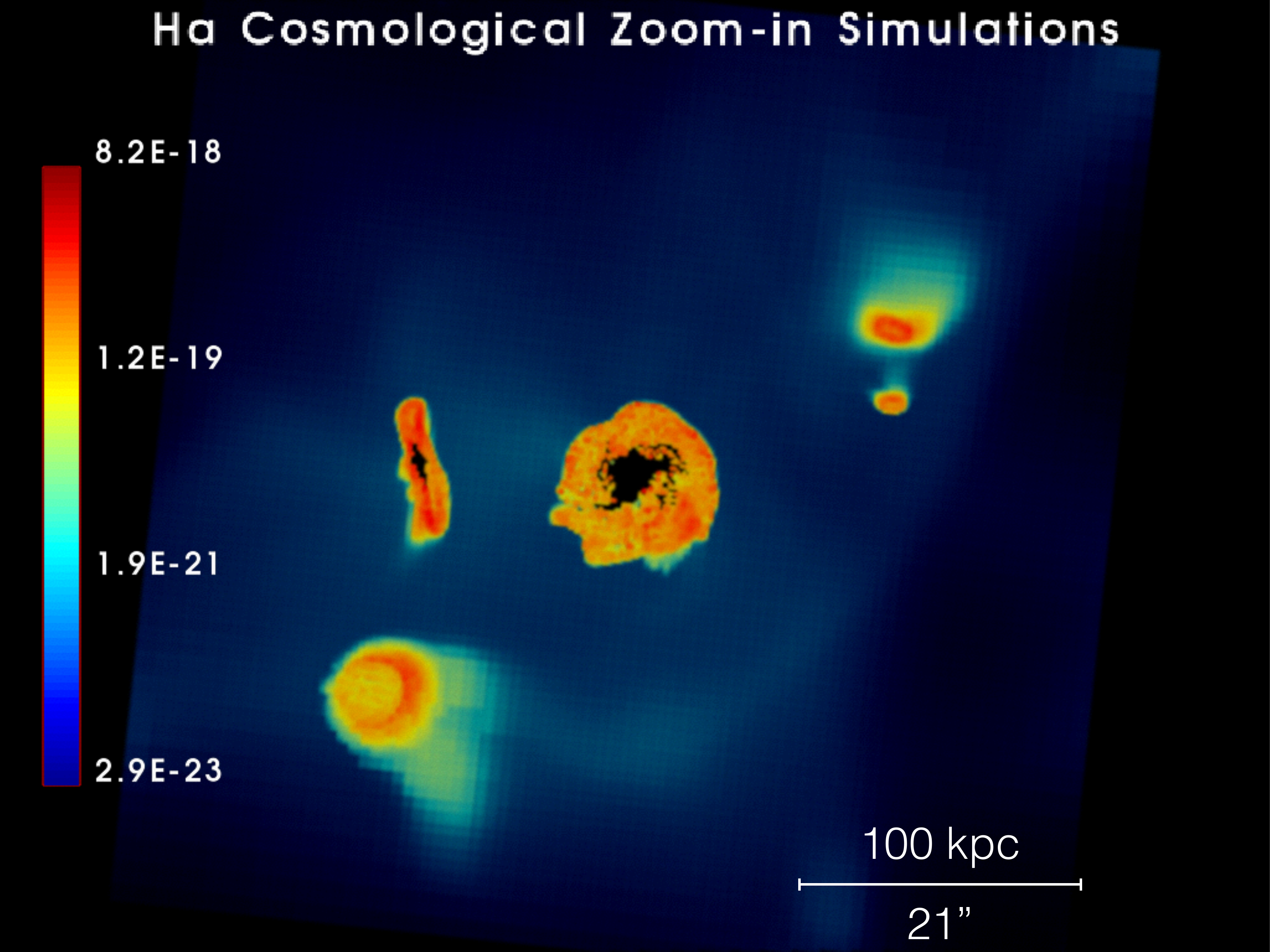}
\vspace{-0.35cm}
\caption{{\bf The environment of z$_{\rm abs}\sim$0.3 galaxies.} {\it Top panel:} MUSE observations of \ha\ emission surface brightness (in \fSB) at z=0.313 in the field of Q1130$-$1449. Every coloured object in this (continuum-subtracted) MUSE pseudo narrow-band image indicates \ha\ emission at the redshift of the absorber. {\it Middle panel:} RAMSES AMR cosmological zoom-in hydrodynamical simulations post-processed with photo-ionisation models (Augustin et al., submitted). This MUSE-size field (1$\times$1', i.e. 300 ~kpc at z=0.313) of \ha\ emission surface brightness (in \fSB) shows a halo with a SFR, stellar and halo masses typical of strong \hi\ quasar absorbers. The flux colour-bar matches the MUSE observations. {\it Bottom panel:} Same figure with the minimum flux cut-off almost 4 orders of magnitude fainter than in the middle panel. The simulations show the presence of multiple galaxies at the same redshift with a broad range of masses as observed in the field of Q1130$-$1449. In addition, the predicted surface brightness from the simulations is comparable to our \ha\ MUSE observations indicating that collisional and photo-ionisation in combination can be the physical processes at the origin of the emission seen in the observations. These results demonstrate that the brightest components of the CGM of low-redshift galaxies can be probed with current observations of the emission, while the diffuse gas in between objects is at present best traced in absorption. These findings add to the paradigm shift where our former view of strong-\nhi\ quasar absorbers being associated with a single bright galaxy changes towards a picture where the \hi\ gas probed in absorption is related to far more complex galaxy structures. }
\label{f:Ha}
\end{figure*}

\section{Analysis and Results}

\subsection{The Environment of z$_{\rm abs}\sim$0.3 \hi-rich Absorbers}

\subsubsection{Galaxies at z$_{\rm abs}$=0.313 in the Field of Q1130$-$1449}

\begin{figure*}
\includegraphics[angle=0, width=14.5cm]{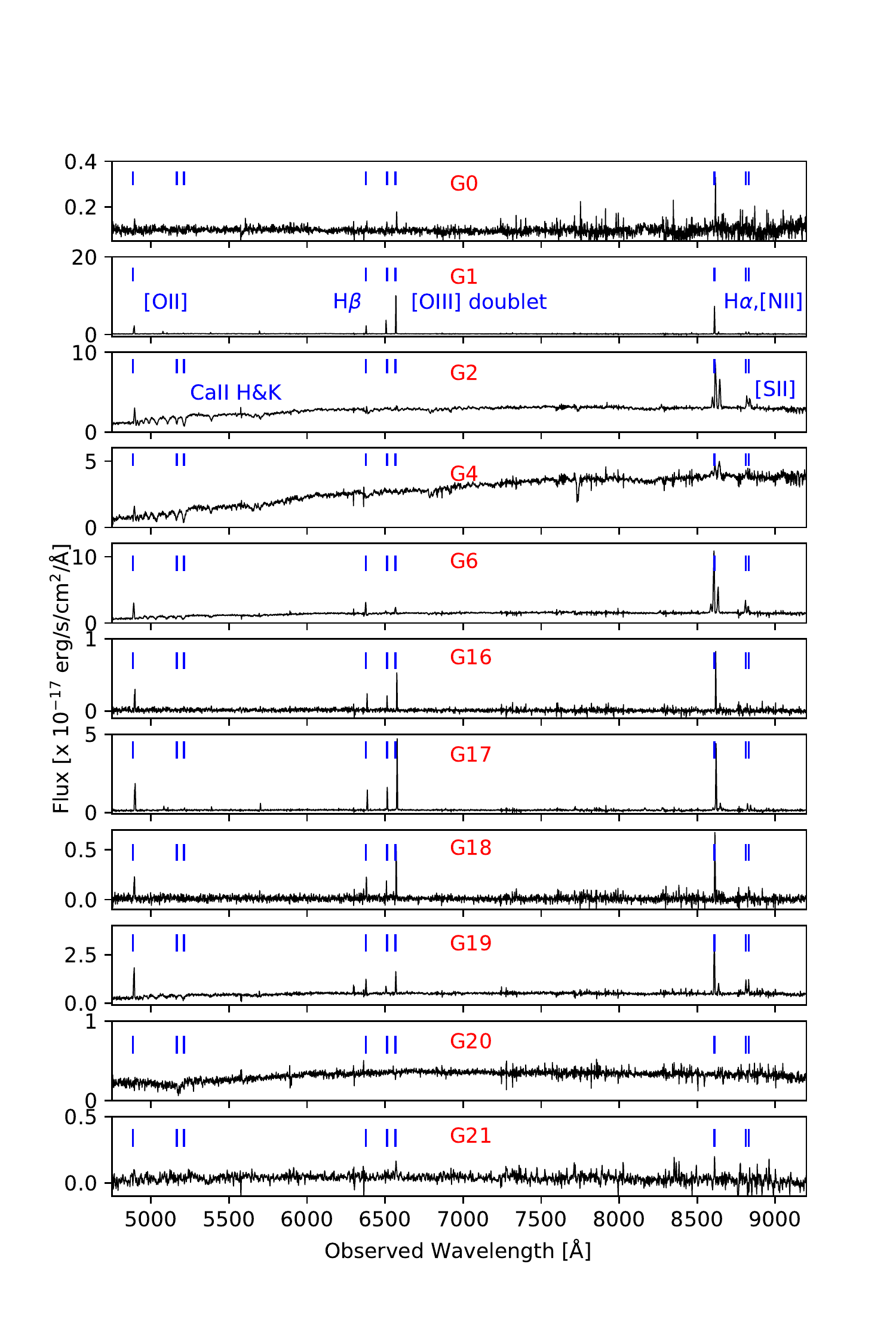}
\vspace{-0.4cm}
\caption{{\bf MUSE spectra of the galaxies at the redshift of the $log [N(H I)/cm^{-2}]$=21.71$\pm$0.07 quasar absorber (z$_{\rm abs}$=0.313).} The panels for G1 and G2 show key emission and absorption lines (at redshifted wavelengths), which lines can also be noted in galaxies plotted below. In some cases, no emission lines were detected, but strong \caii\ absorption lines and a Balmer break matching z$_{\rm abs}$=0.313 were identified (e.g. G20). The spectrum of G21 is smoothed with a  3-pixel boxcar. In other cases, the galaxies have strong emission lines but faint continuum level (e.g. G0, G16 and G17) so that such objects would not appear in broad band imaging. In particular, we discovered a faint galaxy closer to the quasar G0 at impact parameter $b=2.3 ~arcsec$ (10.6 ~kpc). While the object lies under the quasar PSF, several emission lines are clearly identified in the MUSE spectrum. In total, we report 11 objects at the redshift of the absorber in our MUSE data, 7 of which are new discoveries. 
 }
\label{f:MUSE_spectra}
\end{figure*}

The field is part of an observing campaign targeting quasar absorbers whose redshifts allow detection of \ha\ with MUSE, with measured \nhi\ from HST UV spectroscopy and a known associated galaxy. On top of this, the field of Q1130$-$1449 has been the target of a number of observing programs in the past which we summarise here. We use the numbering system introduced by \citet{kacprzak10b} when relevant and add to it for the new detections. Fig.~\ref{f:hst} illustrates this naming scheme. \citet{bergeron91} first reported emission (and absorption) lines at the redshift of the absorber in two objects (G2 and G4). \citet{deharveng95} had successfully detected \hb\ (but no \lya) in G2 from {\it HST}/FOS spectroscopy. \citet{lane98} later
obtained a spectroscopic redshift of another galaxy (G1) which is also consistent with the absorption redshift. \citet{kacprzak10b} spectroscopically identified two additional galaxies (G6 and G14), the latter of which is outside our MUSE field. 

Here, we used the HST images as a prior to extract spectra from the MUSE cube. To this end, we used {\it Sextractor} to identify and deblend faint objects in the broad band WFPC2 image. Using the \textsc{mpdaf}\footnote{http://mpdaf.readthedocs.io/en/latest/index.html} routine \citep{piqueras17}, we extract the pixels at these coordinates above a 0.75-$\sigma$ threshold. We then reviewed the extracted spectra to identify those matching z$_{\rm abs}$=0.313. In some cases, no emission lines were detected, but strong \caii\ absorption lines and a Balmer break matching z$_{\rm abs}$=0.313 were identified (e.g. G20). In addition, we made pseudo narrow-band images of the MUSE cube at the expected position of the emission lines. An example of this is shown in Fig.~\ref{f:OIII_CO} (top panel) for \oiii\ 5007 \AA\ and Fig.~\ref{f:Ha} (top panel) for \ha. This approach reveals additional objects which remain undetected in the HST image because of their faint continuum level (e.g. G0, G16 and G17). 

In particular, we discovered a faint object, G0, closer to the quasar at impact parameter $b=2.3~arcsec$ (10.6 ~kpc). We note that it is not matched by any object detected in Gemini AO-images by \cite{chun06} because of the different wavelength coverage (optical versus H-band) and different sensitivity (emitters versus continuum objects). While the object lies under the quasar PSF, several emission lines are clearly identified in the MUSE spectrum. In the following analysis, we use the systemic redshift of G0 as reference for the analysis of the absorption profile plotted in velocity (see appendix~\ref{app_section:abs}) simply because it has the smallest impact parameter to the quasar of all objects spectroscopically detected in the field. In essence, this system is like the "Galaxies with background QSOs" searches which have found quasars shining through low redshift, foreground galaxies at small impact parameters ($<$10 ~kpc) within the SDSS fibers \citep{york12, straka13, straka15, joshi17}. These, together with recent findings of low impact parameter objects in high spatial resolution HST imaging \citep[i.e.][]{augustin18}, provide observational evidence to the postulate of \cite{york86} that "some quasar absorption-line systems may arise when a quasar sight line intersects a Magellanic-type irregular galaxy (i.e., a gas-rich dwarf)". Fig.~\ref{f:hst} (right panel) summarises the numbering system of the new identifications. The corresponding MUSE spectra are displayed in Fig.~\ref{f:MUSE_spectra}. In total, we report 11 objects at the redshift of the absorber in our MUSE data, 7 of which are new discoveries.

Similarly, we identified the galaxies at the absorber redshift in the ALMA observations by running the {\it Duchamp}\footnote{https://www.atnf.csiro.au/people/Matthew.Whiting/Duchamp/} source finder (version v1.6.2) which performs a three-dimensional search via wavelet reconstruction with a detection threshold of 5-$\sigma$ on the cube before primary beam correction. We find three secure emitters associated with galaxies G2, G4 and G6, respectively. Fig.~\ref{f:ALMA_contours} shows the CO(1--0) contours overlaid on the background HST image. In addition, we performed a visual inspection of the cube at the redshifted frequency of the CO(1--0) line. We detect a fourth emission line, 26.4~arcsec south of the quasar. The MUSE spectrum shows a featureless continuum object consistent with either CO(2--1) at z=1.614 or CO(3--2) at z=2.921 (a higher CO transition would have matching \lya\ line in the MUSE spectrum). Interestingly, one of the continuum spectral window also reveals an emission line at 101.4 GHz which is identified as CO(2--1) at z=1.274 thanks to the detection of the \oii\ doublet in the corresponding MUSE spectrum. These findings illustrate the power of combining ALMA with MUSE observations to securely assess the redshift of single line mm detections.  Fig.~\ref{f:OIII_CO} (bottom panel) shows the 0$^{\rm th}$ moment map of the field. The spectra are extracted within the CASA software and shown in Fig.~\ref{f:ALMA_Kine}. We also built a continuum image of aggregate bandwidth to check for possible detection of the galaxies in continuum. None of the objects are detected in the map with rms=6~$\mu$Jy, corresponding to a 5-$\sigma$ continuum flux limit of 30~$\mu$Jy.

\subsubsection{Cosmological zoom-in Simulations}

In order to reproduce the typical environment of z$\sim$0.3 quasar absorbers, we turn to cosmological hydro-simulations. Our calculations are based on dedicated RAMSES Adaptive Mesh Refinement (AMR) simulations with a total run time of 1.3 million CPU hrs \citep{frank12}. We zoomed on a region around the most massive halo with a box of size of $\sim$14 Mpc/h. The higher-level of refinement allows us to reach a spatial resolution of $\sim$380pc/h (comoving) at z=0. In short, the simulations include non-thermal supernova feedback with "on-the-fly" self-shielding option \citep{teyssier13}. With the goal of reproducing the expected surface brightness in the CGM regions of galaxies, we have post-processed the simulations taking into account gravitational cooling due to collisional ionisation of accreting gas, photo-ionisation by external UV sources and scattering from star forming regions. Details of the simulation are presented in an upcoming paper (Augustin et al. submitted).

Current observations find that the star formation rate (SFR) of strong \hi-absorbers at z$\sim$0.3 is just below 1 M$_{\rm \odot}$/yr \citep{rahmani16}, while their typical stellar masses are measured with values up to 10$^{11}$ M$_{\rm \odot}$ \citep{augustin18} and their halo mass 10$^{12}$ M$_{\rm \odot}$ \citep{peroux13}. To best match these observable constraints, we picked a typical z=0.3 halo in the zoom-in simulations which best mimics the properties of \hi\ absorbers: a SFR=1 M$_{\rm \odot}$/yr), M$_{\rm star}$=10$^{11}$ M$_{\rm \odot}$ and M$_{\rm DM}$=10$^{12}$ M$_{\rm \odot}$ (R$_{\rm vir}$=200 ~kpc/h comoving). Fig.~\ref{f:Ha} presents a MUSE-size FoV H$\alpha$ emission surface brightness map of that simulated halo in units of \fSB. These simulations show the presence of multiple galaxies at the same redshift, typical of a small group. The group members have a broad range of masses so that some of these objects will remain undetected in current state-of-the-art observations. While the simulated objects appear more extended, the predicted surface brightnesses from the simulations are comparable to our \ha\ MUSE observations. The robustness of the surface brightnesses prediction is a remarkable achievement which indicates that collisional and photo-ionisation in combination can be the physical processes at the origin of the emission seen in the observations. In addition, this demonstrates that the brightest components of the CGM of low-redshift galaxies can be probed with current observations. Another important feature of the simulated cube is the presence of faint low surface-brightness gas with large sky cross-section in between galaxies.

\subsubsection{The Typical Environment of low-redshift Absorbers}

\begin{figure}
\includegraphics[angle=0, width=9.cm]{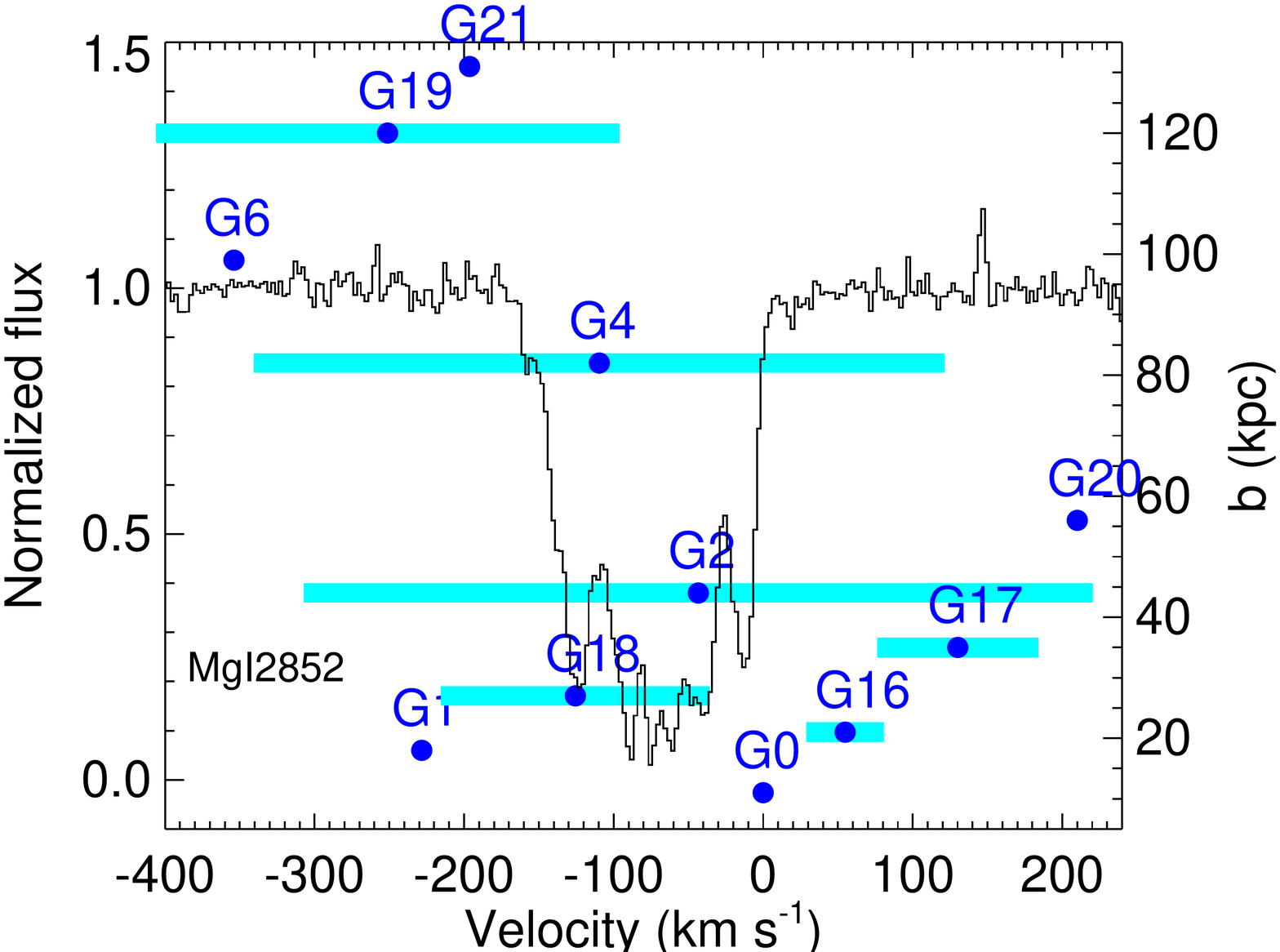}
\caption{{\bf Impact parameters and velocities of galaxies relative to the quasar absorber profile.} Normalised VLT/UVES quasar spectrum of \mgi\ 2852 line at \zabs=0.31305. The systemic redshift of galaxy G0 is used as zero velocity reference. The galaxies are ordered from bottom to top as a function of impact parameter to the quasar line-of-sight (see right y-axis for values in kpc). The light blue horizontal lines indicate the $V_{\rm max}$ extent for galaxies exhibiting ordered rotation. The overlap in velocity space with the closest galaxies is somewhat limited, and much of the absorption bluewards of G0 could be related to the low surface brightness tidal gas predicted in our simulations to be 4 orders of magnitude below the current detection threshold.
 }
\label{f:abs_vel}
\end{figure}

Over the years, there has been mounting evidence that bright galaxies can be found at the redshift of quasar absorbers \citep[see][for a recent review]{krogager17}.
 Tremendous progress has come from IFU observations in particular which have provided a robust means to remove the bright quasar contamination and reach low impact parameters \citep{bouche12, peroux12, schroetter16}. At low redshifts, however, it is becoming clear that the picture is more complex than a simple one-to-one correspondence between a single galaxy and a quasar absorber. Thanks to the MUSE wide FoV, it is now possible to survey a large portion of sky in one observational set-up, with physical sizes (300 ~kpc across) typical of the CGM of galaxies \citep{steidel10}. Clearly, the case presented here does not contain the typical isolated galaxy-absorber match often reported in past searches for absorber counterparts. In part, this is due to deeper data as we are reaching limits of SFR$>$0.01 M$_{\rm \odot} yr^{-1}$, i.e. an order of magnitude deeper than similar z$\sim$1 studies. In addition, we report the detection of passive galaxies whose redshift estimate comes solely from Ca H\&K absorption features and Balmer breaks detected in their continuum. While galaxies at these low redshifts will be more clustered, finding simlar objects at higher redshifts is challenging as the continuum flux gets fainter. 

While we note the possible alignments of the objects in what could be a filament with a north-west to south-east orientation \citep{moller01, fumagalli16, peroux17}, it is also possible that these objects are part of a galaxy group. Assuming G2 as the central member galaxy, we note that G6, G4, G19 and G21 (lying to the north) are at velocity positions blueshifted with respect to G2 while G0, G16 and G17 (lying to the south) are redshifted with respect to G2. This alignment shows an overall direction of rotation in the halo of the group. The velocity dispersion of the system is $\sim$170~\kms\ which is in the range of small galaxy groups. Based on such dispersion and assuming a virialised spherical system, we obtain a virial radius of 420 ~kpc so that the MUSE FoV probes 1/3$^{rd}$ of the virial radius of the group. The virial mass for such system will be 2.9$\times$10$^{12} M_{\odot}$. A number of studies have already reported several counterparts to low-redshift strong-\hi\ absorbers \citep{peroux17, bielby17, rahmani18a, klitsch18, rahmani18b, klitsch19, borthakur19} possibly tracing small groups. 

In fact, these findings are in line with predictions from most recent numerical simulations of galaxy formation. At redshift z=3, cosmological simulations have  demonstrated the connection of quasar absorbers with large central halos \citep{bird14, rahmati14}. The column densities observed also fall in the range expected from accretion occurring in the form of cold flows \citep{keres05,nelson13}. Similarly, galactic winds could produce some of this gas, where the most energetic systems will have the higher sky coverage. Recently, several groups have presented zoom-in hydrodynamical simulations which focus more resolution into the CGM regions to explore smaller physical scales \citep{vandeVoort19, peeples19, suresh19, hummels19}. The cosmological zoom-in simulations presented in the section 3.1.2 are specifically evolved to redshift zero with the goal to predict the emissivity of CGM gas around low-redshift galaxies thus enabling a more direct comparison with the observations presented here. They indicate that strong-\hi\ absorbers could be associated with faint low surface-brightness intra-group gas. In the simulations, this gas is typically remnant tidal debris from previous interactions between the main galaxy and other smaller satellite galaxies. Indeed, the selection of absorbing systems on the basis of their sky coverage will favour interacting systems that have spread their gas around. The MUSE white light image (Fig.~\ref{f:hst}) shows clear indications of extended gas around G2 and G4 in particular.

This lay-out bears resemblance to cold gas streams similar to the Magellanic Stream \citep{richter14}. Low-redshift analogues of such groups include the M81 (NGC 3031)/M82 (NGC 3034)/NGC 3077 complex where optical observations tracing starlight indicate well-separated objects. Twenty-one cm maps however show \hi\ gas distributed in between these objects dominated by filamentary structures, clearly demonstrating the violent disruption of this system by tidal interaction \citep{yun93, yun94}. Similarly, \cite{deblok14a} report the detection of a cloud of neutral gas with admittedly lower column density outside the main \hi\ disk of NGC2403 which they argue could be either accreting from the intergalactic medium, or the result of a minor interaction with a neighbouring dwarf galaxy. Interestingly, they note that the velocities of the \hi\ in NGC 2366 in the same group completely overlap with those of the \hi\ in NGC 2403, which would make their absorption signature indistinguishable in the spectrum of a background object. Twenty five years ago, \cite{morris94} had already calculated that the space density of \lya\ absorbers and local small groups matched, thus proposing that quasar absorbers can be produced by tidal debris in groups of galaxies.

These results pose a new challenge to the interpretation of the neutral gas probed in absorption. Indeed, most studies of the absorbing galaxies have concentrated on the brightest objects which are likely to be the main group member. Others have reasonably argued that the absorption is likely to arise from the galaxy with the smallest impact parameter to the quasar line-of-sight. Indeed, we note that lower-mass galaxies such as G0 revealed here by the new MUSE observations of Q1130$-$1449 are expected to be one of the main contributors to galactic winds, given that the gas can more easily escape the galaxies' smaller potential wells. Fig.~\ref{f:abs_vel} shows the galaxies' positions with respect to the absorption profile in velocity space. The galaxies are ordered from bottom to top as a function of impact parameter to the quasar line-of-sight (see right y-axis for values in kpc). The light blue horizontal lines indicate the $V_{\rm max}$ extent for galaxies indicating rotation. The overlap in velocity space with the closest galaxies is somewhat limited, and much of the absorption bluewards of G0 could be related to the low surface brightness tidal gas predicted in our simulations to be four orders of magnitude below the current detection threshold. These results strenghten previous reports of low-redshift strong-\hi\ absorbers associated with groups \citep{peroux17, bielby17, rahmani18a, klitsch18, rahmani18b, klitsch19, borthakur19}. Altogether, these findings add to the paradigm shift where our former view of strong-\nhi\ quasar absorbers being associated to a single bright galaxy changes towards a picture where the \hi\ gas probed in absorption is related to far more complex galaxy structures. Since both winds and interactions are predicted to be enhanced in the distant Universe, these factors will likely become more important at higher redshifts but harder to discern.

\subsection{Physical Properties of the Galaxy Group} 

\begin{table*}
\begin{center}
\caption{{\bf Dust-corrected nebular emission line fluxes.} The fluxes are expressed in units of \flux. The quoted errors are 1$\sigma$ uncertainties. Galaxies G1 to G6 (shown in italics) were previously identified, while all the remaining are new identifications. 
}
\begin{tabular}{lcccccc}
\hline\hline
Galaxy		 &F(\oii)  &F(\hb)   &F(\oiii)   &F(\oiii)   &F(\ha) &F(\nii)	 \\
\hline
G0		& 1.6$\pm$0.4   & 0.7$\pm$0.2  & 1.0$\pm$0.25 & 1.6$\pm$0.4    & 4.5$\pm$1.2	 & 1.9$\pm$0.5	\\	
{\it G1}	&63.7$\pm$15.9  &40.6$\pm$10.1 &73.5$\pm$18.4 &221.4$\pm$55.3  &240.9$\pm$60.2   &17.3$\pm$4.3	\\
{\it G2}$^a$	&11.7$\pm$2.9   & 3.9$\pm$1.0  &$<$0.9        & 2.2$\pm$0.5    &76.4$\pm$19.1	 &53.1$\pm$13.3	\\
{\it G4}$^a$	&$<$23.5        &$<$4.4        &$<$5.0        &$<$5.0          &31.0$\pm$7.7	 &--	        \\
{\it G6}$^a$	&27.8$\pm$31.9  &19.3$\pm$4.8  & 3.5$\pm$0.9  &12.1$\pm$3.0    &200.6$\pm$50.1   &83.2$\pm$20.8  \\
G16		& 9.6$\pm$2.4   & 3.9$\pm$1.0  & 3.3$\pm$0.8  &10.4$\pm$2.6    &23.5$\pm$5.9	 &1.5$\pm$0.4	 \\
G17		&70.9$\pm$17.7  &35.3$\pm$8.8  &42.8$\pm$10.7 &135.7$\pm$38.4  &200.4$\pm$50.1   &22.1$\pm$5.5	 \\
G18		& 6.8$\pm$1.7   & 3.5$\pm$0.9  & 2.4$\pm$0.6  & 9.7$\pm$2.4    &21.7$\pm$5.4	 &3.8$\pm$0.9	 \\
G19		&25.6$\pm$6.4   & 9.2$\pm$2.3  & 6.0$\pm$1.5  &16.7$\pm$4.2    &72.0$\pm$18.0	 &12.5$\pm$3.1	  \\
G20$^b$		&$<$0.4         &$<$0.2        &$<$0.2        &$<$0.2          &$<$0.3	         &$<$0.3	\\
G21$^b$	        &$<$0.4         &$<$0.3        &$<$0.3        & 0.7$\pm$0.2    & 0.7$\pm$0.2	 & $<$0.7	\\
\hline\hline 				       			 	 
\label{t:GalFlux}
\end{tabular}			       			 	 
\end{center}			       			 	 
\begin{minipage}{180mm}
Note: $^a$ Galaxies showing NaID absorption features in their spectra.
$^b$ The flux limits are not corrected for dust.
\end{minipage}
\end{table*}

\begin{table*}
\begin{center}
\caption{{\bf Ionised gas properties of the galaxies detected at the redshift of the z$_{\rm abs}$=0.313 absorber in the MUSE observations.} We use the numbering system introduced by \citet{kacprzak10b} when relevant and extend it for the seven new detections. $\delta$ is the angular distance from the quasar in arcsec and $b$ is the impact parameter in kpc.  SFR estimates are dust-corrected. G20 and G21 have no, or weak, emission lines. Galaxies G1 to G6 (shown in italics) were previously identified. }
\begin{tabular}{lcccclccc}
\hline\hline
Galaxy 		 &RA &Dec  &$\delta$ &$b$ &$z_{\rm gal}$  &E(B-V) &SFR  	&12+log(O/H)\\
	&	  &  &[arcsec] &[kpc] &&[mag]&[$M_{\odot}$/yr]&\\
\hline
G0		&11 30 07.16	&$-$14 49 27.13	&2.3		&10.6 &0.31305 &0.143  &0.1$\pm$0.1 &8.58$\pm$0.14\\
{\it G1}	&11 30 06.74	&$-$14 49 27.33	&3.8		&17.6 &0.31205 &0.098  &3.4$\pm$0.8 &8.10$\pm$0.08\\
{\it G2}	&11 30 07.63	&$-$14 49 23.93	&9.5		&43.9 &0.31286 &0.595  &1.1$\pm$0.3 &8.73$\pm$0.05\\
{\it G4}	&11 30 07.54	&$-$14 49 11.93	&17.7	&81.8 &0.31257 &$>$0.171 &$>$0.4&$<$8.65$^b$\\   
{\it G6}	&11 30 08.48	&$-$14 49 29.13	&21.3	&98.5 &0.31150 &0.330  &2.9$\pm$0.7  &8.75$\pm$0.16\\   
G16		&11 30 07.34	&$-$14 49 27.73	&4.5		&20.8 &0.31329 &0.099  &0.3$\pm$0.1  &8.12$\pm$0.14\\   
G17		&11 30 07.52	&$-$14 49 25.13	&7.6		&35.1 &0.31362 &0.079  &2.8$\pm$0.7  &8.09$\pm$0.17\\      
G18		&11 30 06.77	&$-$14 49 22.93	&5.9		&27.3 &0.31250 &0.111  &0.3$\pm$0.1  &8.49$\pm$0.18\\    
G19		&11 30 06.21	&$-$14 49 04.53	&26.0	&120.2&0.31195 &0.214 &1.0$\pm$0.2  &8.50$\pm$0.16\\     
G20		&11 30 07.20	&$-$14 49 15.93	&12.1	&55.9 &0.31397$^a$       &--     &$<$0.01&--\\   
G21		&11 30 08.88	&$-$14 49 19.13	&28.4	&131.3 &0.31219 &--      &0.01$\pm$0.01 &8.56$\pm$0.17$^b$\\   
\hline\hline 				       			 	 
\label{t:GalDetect}
\end{tabular}			       			 	 
\begin{minipage}{180mm}
Note: $^a$ Redshift based on \caii\ absorption features in the galaxy spectrum. $^b$ Metallicity based on N2 indicator only.\\
\end{minipage}
\end{center}			       			 	 
\end{table*}

The MUSE observations provide a wealth of information on the stellar properties of the galaxies in the group. Accurate redshifts are determined from multiple lines detected in the 1D MUSE spectra calibrated in air. We measured the emission fluxes from a Gaussian fit to the nebular lines. The 3$\sigma$ upper limits for non-detection are computed for an unresolved source spread over 6 spatial pixels and spectral FWHM = 2.4 pixels = 3 \AA. We followed the method described in \cite{peroux14} based on \ha\ and \hb\ to estimate moderate E(B-V) and further correct the observed fluxes for dust attenuation. The resulting fluxes are tabulated in Table~\ref{t:GalFlux}. The SFR with 1$\sigma$ uncertainties were estimated from these \ha\ fluxes following the relation of \citet{kennicutt98}. The emission line galaxies have values of SFR ranging from 0.1 to a few solar masses per year (see Table~\ref{t:GalDetect}).

Following the methodology of \cite{peroux17}, the HII region metallicities were measured from the strong-line indices (N2, O3N2 and R$_{\rm 23}$ most relevant branch) following prescription by \cite{kobulnicky99} and the mean value of each of these was computed. Our estimates agreed within the errors with earlier measurements of G2 and G6 metallicities by \cite{kacprzak10b}. The metallicity of the galaxies in the group ranges from [X/H]=$-$0.57$\pm$0.17 to 0.09$\pm$0.16. The total metallicity of the neutral gas probed in absorption, [Zn/H] = $-$0.80$\pm$0.16, is therefore consistent within the error bars with the lowest of the metallicities found in the galaxies in the MUSE image.

While metallicity gradients in galaxies could partly reconcile these measurements \citep{carton18}, we note that the low dust-corrected abundances measured in absorption resemble the properties of the extragalactic diffuse gas produced in our numerical simulations. Furthermore, the nebular emission line ratios place all but one galaxy into the star-forming region of the BPT diagram \citep{baldwin81}. Galaxy G1 shows evidence for ionisation by an AGN with \oiii/\hb=5.45 and \nii/\ha=0.07. The galaxy emission properties are listed in Table~\ref{t:GalDetect}.

\begin{table*}
\begin{center}
\caption[]{{\bf CO emitters at z$_{\rm abs}$=0.313 in the ALMA observations.} 
Flux densities, velocity widths, CO luminosities, molecular masses expressed in units of $\alpha_{\rm CO}$ and depletion timescales are provided for each detected galaxy.  
}
\begin{tabular}{cccccccccc}
\hline\hline
Galaxy & RA &Dec &$\delta$ &$b$ &S$_{\rm CO}$ &FWHM & L$_{\rm CO}$ &M$_{\rm mol}$$\times$ ($\alpha_{\rm CO}$/1 K km/s pc$^2$)  &$\tau_{\rm depl}$\\
& & &[''] &[kpc] &[Jy km/s]&[km/s] &[K km/s pc$^2$] &[$\times$ 10$^9$ M$_{\odot}$] &[Gyr]\\
\hline
G2 &11 30 07.66  &$-$14 49 23.41  &9.7    &44.8   &0.63$\pm$0.01  &250$\pm$50     &(3.1$\pm$0.1)$\times$10$^{9}$    &3.16$\pm$0.07 &13$\pm$5\\
G4 &11 30 07.62  &$-$14 49 11.44  &18.0  &83.2   &0.42$\pm$0.03  &535$\pm$50     &(2.1$\pm$0.1)$\times$10$^{9}$    &1.99$\pm$0.15 &$<$23\\
G6 &11 30 08.53  &$-$14 49 28.54  &21.5  &99.4   &0.20$\pm$0.01  &205$\pm$50     &(1.0$\pm$0.1)$\times$10$^{9}$    &1.00$\pm$0.05 &1.5$\pm$0.5\\
\hline\hline 				       			 	 
\label{t:ALMA}
\end{tabular}			       			 	 
\end{center}			       			 	 
\end{table*}			       	
 
We used the ALMA observations to constrain the molecular gas content of the galaxy group. We measured the flux density in the lines by integrating
the CO(1--0) emission lines in the extracted spectra (Fig.~\ref{f:ALMA_Kine}). We then computed the corresponding luminosities using the following formula: L$_{\rm CO}$ = (3.25 $\times$ 10$^7$ $\times$ dL [Mpc]$^2$ $\times$ F$_{\rm int}$) / (f$_{\rm obs}$ [GHz]$^2$ $\times$ (1+z)$^3$), where dL is the luminosity distance, F$_{\rm int}$ the flux density and f$_{\rm obs}$=87.8GHz is the observed frequency \citep{solomon92}. Because we observe the CO(1-0) transition directly, our data naturally overcome uncertainties related to unknown CO line ratios. Estimates of the corresponding H$_2$ molecular masses however rely on the rather uncertain CO-H$_2$ conversion factor \citep{bolatto13}. Previous works have used Galactic CO-H$_2$ conversion factor of $\alpha_{\rm CO}$=4.6 M$_{\odot}$ (K km/s pc$^2$)$^{-1}$. Recent results by \cite{klitsch19} find that this is not always appropriate for absorbing-galaxies so that here we opt to express the molecular masses as a function of $\alpha_{\rm CO}$. The resulting molecular masses range from
$M_{\rm mol}$ = (1.00$\pm$0.05) to (3.16$\pm$0.07) $\times$ ($\alpha_{\rm CO}$/1 K km/s pc$^2$) $\times$ 10$^9$ M$_{\odot}$ and are listed Table~\ref{t:ALMA}.

The ALMA observations indicate substantial cold gas reservoirs in three of the galaxies in the group. These molecular masses are large in comparison with the typical stellar mass of absorbing galaxies \citep{augustin18}. In fact, a number of recent works have reported CO associated with quasar absorbers from ALMA observations with a high rate of incidence and derived equally high molecular masses \citep{neeleman16, moller18, neeleman18, kanekar18, klitsch18, dodorico18, klitsch19}. However, it is unclear why \hi-rich absorbing galaxies should have large reservoir of H$_2$ gas. At first sight, this might appear at odds with searches for H$_2$ rotational and vibrational transitions at UV restframe wavelengths which have only reported both low incident rates (10\%) and low molecular gas content ($f_{\rm mol} \le 0.01$) with respect to the neutral gas \citep{ledoux03, noterdaeme08}. While the most abundant molecule after H$_2$, CO is known to be less abundant with a CO/H$_2$ column density ratio ranging from 10$^{-7}$ to 10$^{-5}$ and a mean value of 3$\times$10$^{-6}$ \citep{burgh07}. However, it must be borne in mind that the neutral atomic phase of the gas traces significantly lower gas densities than the molecular gas traced by CO \citep{snow06}. Indeed the observations of nearby galaxies indicate that \hi\ and H$_2$ are not co-spatial \citep{schruba11}, with the molecular gas concentrated in the central part of galaxies so that the detection of H$_2$ in absorption at an impact parameter distance at the edges of galaxies is less likely \citep{zwaan06, obreschkow09}.

 In the field of Q1130$-$1449, we detected three objects in CO at the absorber redshift, two of which have large impact parameters (b$>$50 ~kpc). The objects close to the quasar (i.e. G0 and G1) do not show CO(1--0) emission, although G1 is the most actively star-forming member of the group. We calculated the molecular depletion times for the three objects detected with ALMA following: $\tau_{\rm depl}$=M$_{H2}$/SFR=1/SFE, where SFE is the star formation efficiency. To compute the depletion time scale, we assumed a Galactic CO-H$_2$ conversion factor of $\alpha_{\rm CO}$=4.6 M$_{\odot}$ (K km/s pc$^2$)$^{-1}$. We derive $\tau_{\rm depl}$ ranging from 1.7$\pm$0.5 to 12$\pm$5 Gyr (see Table~\ref{t:ALMA}). The gas depletion times are long  compared to the timescales of the processes driving the evolution of the interstellar medium state \citep{leroy13}. \cite{semenov17} argue that star-forming gas converts only a small fraction ($\sim$1\%) of its mass into stars while most of it ($\sim$90\%) is dispersed by dynamical and feedback processes. Selection based on strong \hi-absorbers thus possibly preferentially finds galaxies with large molecular gas reservoirs (at given SFR) which inefficiently convert their gas into stars \citep{kanekar18}.

\subsection{Morphological and Kinematical Properties of Group Members}

\begin{table*}
\caption{{\bf Nebular emission lines and cold molecular gas (\bf{bold}) morpho-kinematic properties.} The results of the 3-dimensional MCMC forward modelling of the morphological and kinematical parameters listed are all inclination corrected. Inclination and PA (NoE) are derived from a {\it Sextractor} fit to the broad band HST/WFPC2 image when the algorithm did not converge. When the object is detected in both MUSE and ALMA, i.e. G2, G4 and G6, the kinematic models (bold) are sometimes found to be in good agreement indicating a relation between the stellar content and the molecular gas in some of these galaxies}. The maximum velocity of the disks are used to derive the dynamical and halo masses. Galaxies G1 to G6 (shown in italics) were previously identified.
\label{t:Kine}
\centering
\begin{tabular}{lcccccc}
\hline\hline
		&$r_{1/2}$ 	&$i$  		&PA 		& $V_{\rm max}$ &log $M_{\rm dyn}$ &log $M_{\rm halo}$ \\
		&[kpc]		&[deg] 			&[deg] 				   &[km~s$^{-1}$]  		   &[M$_{\odot}$] &[M$_{\odot}$] \\
\hline
G0   	     	&under quasar PSF &67		&122			&--&--&--   \\
{\it G1}   	&dispersion dominated &51      &91			&--&--&--   \\
{\it G2}   	&14$\pm$2		&77$\pm$2 &131$\pm$2	&264$\pm$14 	&11.3$\pm$0.2 &12.9$\pm$0.1  \\
{\bf G2}     &2$\pm$1 	        &76$\pm$3 &117$\pm$2		&134$\pm$14	&--                    &--   \\
{\it G4}   	&9$\pm$2 	    	&54$\pm$2 &86$\pm$2	&231$\pm$12 	&11.1$\pm$0.2 &12.7$\pm$0.1  \\
{\bf G4}     &6$\pm$1 	         &82$\pm$4 &84$\pm$2		&290$\pm$19 	&-- 	                 &-- \\
{\it G6}   	&face-on 			&18 		&33			&--&--&--   \\
{\bf G6}   	&face-on 			&--		&--			&--&--&--   \\
G16   	&3$\pm$2 	    	&33$\pm$25 &66$\pm$33	&$<$26 	&$<$8.7&$<$9.8  \\
G17   	&3$\pm$1 	    	&52$\pm$1 &54$\pm$1	&54$\pm$13 	&9.4$\pm$0.3 &10.8$\pm$0.1  \\
G18   	&5$\pm$1 	    	&32$\pm$10 &103$\pm$11	&90$\pm$34 	 &9.4$\pm$0.9 &11.5$\pm$0.4  \\
G19   	&3$\pm$1	    	        	&63$\pm$4 &110$\pm$4	&155$\pm$68 	 &10.3$\pm$0.5 &12.2$\pm$0.3  \\
G20   	&continuum-detected &10		 &130		&--&--&--   \\
G21   	&faint emission lines &28 		&147			&--&--&--   \\
\hline\hline
\end{tabular}\par
\end{table*}

MUSE and ALMA observations of the field are also complementary in that they both provide information on the kinematics of the stars and molecular gas respectively. We performed a 3-dimensional morpho-kinematic analysis of the objects detected in MUSE and ALMA, using the method described in \citet{peroux17}. In brief, we used the GalPak$^{\rm 3D}$ algorithm \citep{bouche15} which compares directly the data cube with a parametric model mapped in $x,y,\lambda$ coordinates. The algorithm uses a Markov Chain Monte Carlo (MCMC) approach with a proposed distribution of the parameters in order to efficiently probe the parameter space. The code has the advantages that it fits the galaxy in 3-d space and provides a robust description of the morpho-kinematics of the data. We assessed the quality of the fit from a $\chi^2$ minimisation as well as the inspection of the residual maps. We mostly used the brightest \oiii\ 5007 \AA\ line to perform the analysis. Our results are tabulated in Table~\ref{t:Kine} and illustrated in Fig.~\ref{f:MUSE_Kine} and \ref{f:ALMA_Kine}  in appendix~\ref{app_section:kine} for the MUSE and ALMA detections respectively. The CO velocity fields are well resolved in our ALMA observations. As expected, the stellar sizes as measured by r$_{1/2}$ are larger than the molecular gas sizes. Our analysis further provides a determination of the inclination and position angle (PA) of the galaxies. In cases where GalPak$^{\rm 3D}$ did not converge (namely for G0, G1, G6, G20 and G21), we use {\it Sextractor} to derive these parameters from the broad band HST/WFPC2 image. 

The analysis of galaxy G0 is hampered by the proximity to the bright quasar (2.3~arcsec). G1 shows no indication of a velocity shear notwithstanding the bright emission lines in the object and a favourable inclination parameter (i=51 deg) as measured in the HST/WFPC2 broad band image. Indeed, the nebular emission line ratios in the object show evidence for ionisation by an AGN with \ha\ emission produced in the central engine \citep{baldwin81}. G2 is a large galaxy (half-light radius  r$_{\rm 1/2}$=14$\pm$2~kpc) dominated by rotation in both \oiii\ and CO as indicated by the dispersion velocity for G2 peaking in the center of the system beyond beam smearing. A smaller nearby galaxy, G17, presumably interacts with this system. The MUSE white light image (Fig.~\ref{f:hst}) also shows extended gas around the galaxy.
For galaxy G4, we used the \ha\ emission line rather than the fainter \oiii\ 5007 \AA. The ALMA observations of G4 indicate two, spectrally and spatially well-separated components (Fig.~\ref{f:ALMA_contours}) clearly seen in the spectrum (Fig.~\ref{f:ALMA_Kine}) with wide FWHM=530$\pm$50~\kms. This is reflected in the kinematics analysis where we derive a large $V_{\rm max}$=290$\pm$4~\kms\ for this system. We tentatively interpret these characteristics as a signature of merging and note extended diffuse gas is also present in the MUSE white light image (Fig. B1). Galaxy G6 is almost face-on (i=18 deg) hindering the kinematical analysis of this system. An inspection of the velocity field of G16 reveals that it has yet to converge to its $V_{\rm max}$ value, which should be taken as an upper limit. The faint galaxy G17 on the other hand has a well-converged velocity field as illustrated in Fig.~\ref{f:MUSE_Kine}. Our detection of galaxy G19 indicates that our observations have not fully reached $V_{\rm max}$ either, although rotation can be securely reported. Galaxy G20 is a continuum-detected object while G21 has faint emission lines thus precluding further kinematic analysis. Overall, the model converges for most galaxies. While not a unique solution, these findings indicate that the emission lines are well described by disk rotation. However, such an analysis might be rather insensitive to interactions of large objects with lower mass galaxies (say, as in the case of G2 and G17).

We derived the dynamical mass of each galaxy from the enclosed mass: $M_{\rm dyn} = V_{max}^2~r_{1/2}~/~G$ \citep{peroux11b}. Assuming a spherical virialised collapse model \citep{mo02}, we further computed the halo mass related to each object: $M_{\rm halo}= 0.1 H_o^{-1} G^{-1} \Omega_m^{-0.5} (1+z)^{-1.5} V_{max}^3$. The highest of these values are representative within the errors of our estimate of the halo mass of the whole group (log $M_{\rm halo}$=12.5 M$_{\rm \odot}$). For one of these objects (G1), \cite{christensen14} performed a Spectral Energy Distribution (SED) fit to the optical and near-infrared magnitudes and derived a low stellar mass of log M$_{\rm *}$=8.29$\pm$0.09 M$_{\rm \odot}$. 

When an object is detected from its nebular emission lines and in CO(1--0), i.e. G2, G4 and G6, both the ionised gas traced by nebular lines and the CO molecular line widths (FWHM=200--530~\kms) are significantly broader than the neutral gas traced by absorption at some distance from the galaxies ($\sim$150~\kms) and especially the 21cm line absorption (FWHM=42.1$\pm$2.7~\kms). The kinematics of the stars and molecular gas in terms of orientation (PA) deviate by 7-$\sigma$ (G2) or rather similar (G4). The maximum velocities differ significantly: 264$\pm$3~\kms\ (H$\alpha$) vs 134$\pm$2~\kms\ (CO) for G2 and 231$\pm$4~\kms\ (H$\alpha$) vs 290$\pm$4~\kms\ (CO) for G4. \cite{levy18} made a detailed comparison of the molecular and ionised gas (traced by H$\alpha$) kinematics in a sample of local galaxies. They find that $\sim$75\% of their sample galaxies have smaller ionised gas rotation velocities than the molecular gas in the outer part of the rotation curve. They report no case where the molecular gas rotation velocity is measurably lower than that of the ionised gas unlike what we observe in the case of galaxy G2.

\section{Conclusion}

In this paper, we have presented deep MUSE and ALMA observations of the field of the QSO Q1130$-$1449 which shows a log $[N(H I)/cm^{-2}]$=21.71$\pm$0.07 damped Ly$\alpha$ system at \zabs = 0.313. Our main findings can be summarised as follow:

\begin{itemize}

\item Our MUSE observations cover 11 galaxies at the redshift of the absorber, 7 of which are new discoveries. In particular, we report a new object with the smallest impact parameter to the quasar line-of-sight (b=10.6~kpc). Three of the objects are also detected in CO(1--0) in our ALMA observations.

\item Using dedicated numerical cosmological simulations, we infer that the physical properties of these galaxies qualitatively resemble a small group environment, possibly part of a filamentary structure.

\item Based on metallicity and velocity arguments, we conclude that the neutral gas traced by strong \hi\ absorbers is only partly related to these emitting galaxies while a larger fraction is likely the signature of low surface brightness emitting gas four orders of magnitude fainter than the current detection limits outside the Local Group. 

\item We report large molecular gas reservoirs with long depletion timescales in the three galaxies detected with ALMA. These results together with other reports of large molecular masses in strong \hi-absorption systems indicate that selection based on absorption preferentially picks galaxies which inefficiently convert their gas into stars. 

\item Detailed kinematics analysis of both the ionised and molecular component of these galaxies shows signatures of past interactions and possible merging between various members of the group. While the stellar component is spatially more extended, the resolved molecular lines are broader in velocity space. 

\item Together with other similar reports, our findings challenge a picture where strong-\nhi\ quasar absorbers are associated with a single bright galaxy and favour a scenario where the \hi\ gas probed in absorption is related to far more complex galaxy structures.

\end{itemize}

\section*{Acknowledgements}
We thank Thomas Ott, Nicolas Bouch\'e and Jean-Charles Lambert for developing and distributing the QFitsView, GalPak$^{\rm 3D}$ and glnemo software, respectively. We thank Chian-Chou Chen (TC) for help with the CO column density calculation. The authors wish to thank the anonymous
referee for insightful comments which have led to significant
improvement in the quality of the manuscript.

CP thanks the Alexander von Humboldt Foundation for the granting of a Bessel Research Award held at MPA. CP is also grateful to the ESO and the DFG cluster of excellence "Origin and Structure of the Universe" for support.
AK acknowledges STFC grant ST/P000541/1 and Durham University. RA thanks CNRS and CNES (Centre National d'Etudes Spatiales) for support for her PhD. VPK acknowledges partial support from NASA grant NNX14AG74G and NASA/Space Telescope Science Institute support for Hubble Space Telescope programs GO-13733 and 13801. LAS acknowledges support from ERC Grant agreement 278594-GasAroundGalaxies. This work has been carried out thanks to the support of the OCEVU Labex (ANR-11-LABX-0060) and the A*MIDEX project (ANR-11-IDEX-0001-02) funded by the "Investissements d'Avenir" French government program managed by the ANR.

We are grateful to the Paranal and Garching staff at ESO for performing the observations in service mode and the instrument team for making a reliable instrument. 
This paper makes use of the following ALMA data:
ADS/JAO.ALMA\#2016.1.01250.S. ALMA is a partnership of ESO (representing its member states), 
NSF (USA) and NINS (Japan), together with NRC (Canada), NSC and ASIAA (Taiwan), and KASI 
(Republic of Korea), in cooperation with the Republic of Chile. 
The Joint ALMA Observatory is operated by ESO, AUI/NRAO and NAOJ.

\bibliographystyle{mn2e}
\bibliography{bibliography.bib}  


\newpage
\appendix

\section{Gas Absorption Profiles}
\label{app_section:abs}

\begin{figure*}
\includegraphics[angle=0,scale=0.8]{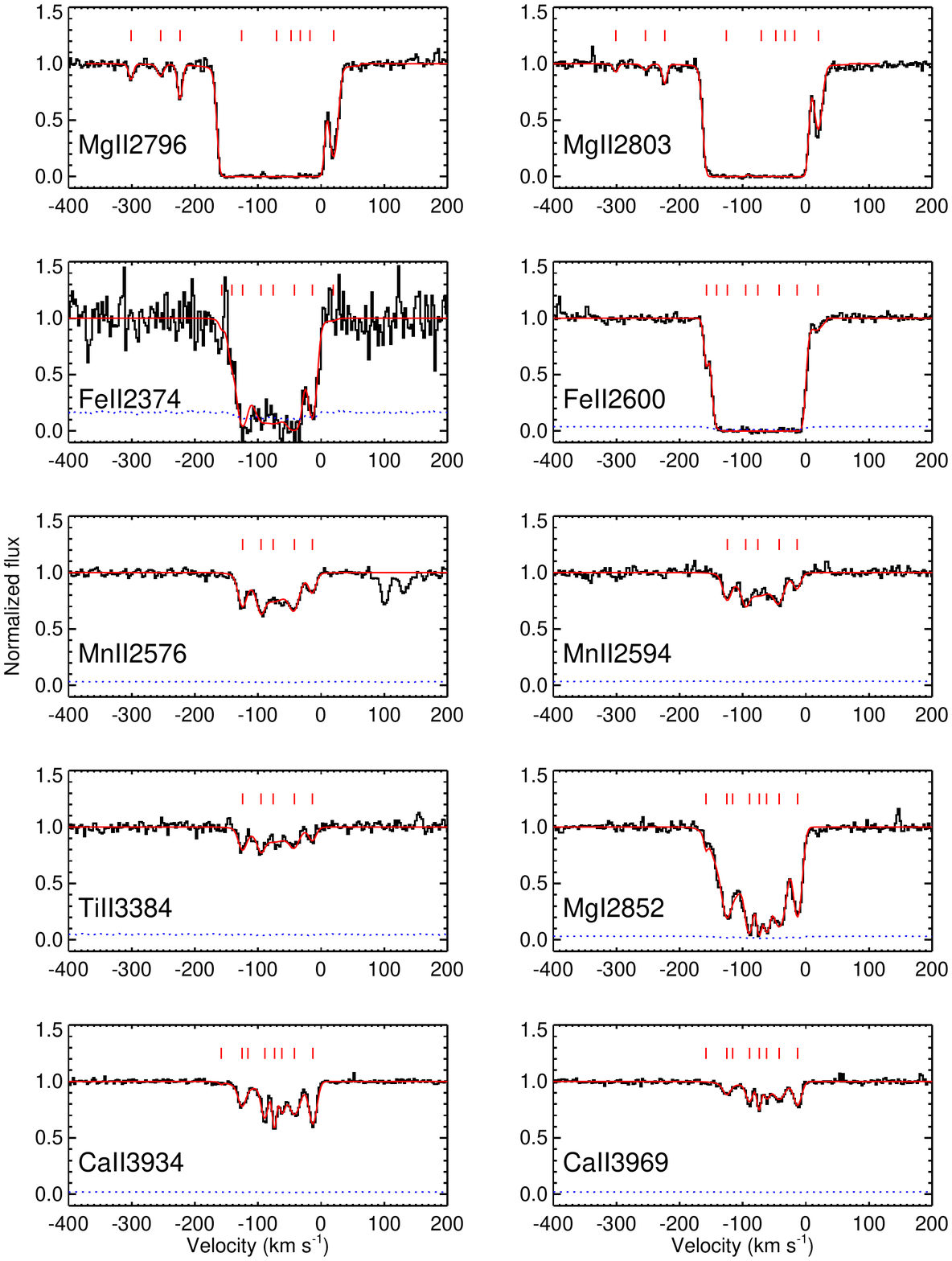}
\caption{{\bf VLT/UVES normalised spectrum of quasar Q1130$-$1449.} The zero velocity is set to the systemic redshift of galaxy G0 $z_{gal}=$0.31305. Most of the absorption lies bluewards of this systemic redshift. Voigt profile fits to the absorption profile are shown in red and the fitted components are shown as tick marks above the spectrum. The error array is shown as a dotted blue line. \mgii\ and \feii\ lines are saturated, but a good fit could be obtained from weaker elements like \mnii, \tiii\ and \caii. 
}
\label{f:uves_fit}
\end{figure*}

We obtained the VLT/UVES reduced quasar spectra from the ESO advanced data products archives\footnote{http://archive.eso.org/cms.html}. 
We applied a heliocentric correction, combined the individual exposures weighting by SNR and normalised the combined quasar continuum. 
Fig.\ref{f:uves_fit} shows portions of the VLT/UVES normalised spectrum of the quasar showing the metal absorption profiles where the zero velocity is set to the systemic redshift of galaxy G0, $z_{gal}=$0.31305. The profiles span a wide velocity range of $\sim250$ ~\kms\ with the weakest components at $-350<v<-200$ ~\kms\ only seen in \mgii\ 2796,2803. 
We modelled the absorption lines with Voigt profiles using \textsc{VPFIT}\footnote{http://www.ast.cam.ac.uk/~rfc/vpfit.html} v10.0. The resulting metallicity derived from all \feii, \mnii\ and \tiii\ are consistent with each other leading to [X/H]=$-$1.94$\pm$0.08 or 1\% solar. This is significantly lower than the measurement based on Zn from {\it HST}/STIS, [Zn/H]$=-0.80\pm0.16$ (16\% solar). This would conventionally be explained by postulating a modest amount of dust along this line-of-sight, which would not affect Zn but only refractory elements such as Fe, Mn, Ti and Ca. We note however recent work by \citet{skuladottir18} who warn that stars in the Sculptor dwarf spheroidal galaxy indicate that Zn and Fe do not trace all the same nucleosynthetic production channels, so that a direct comparison might not be appropriate. It is also important to note that the [Zn/H] measurement is based on {\it HST}/STIS which have a significantly lower spectral resolution than the VLT/UVES spectrum studied here.

\section{Morpho-kinematics Analysis of Galaxies in the Field}
\label{app_section:kine}

The section displays the flux and velocity maps of individual galaxies observed at $z_{gal}$=0.313 with MUSE and ALMA.

\begin{figure*}
\begin{center}
\includegraphics[width=5.8cm, angle=0]{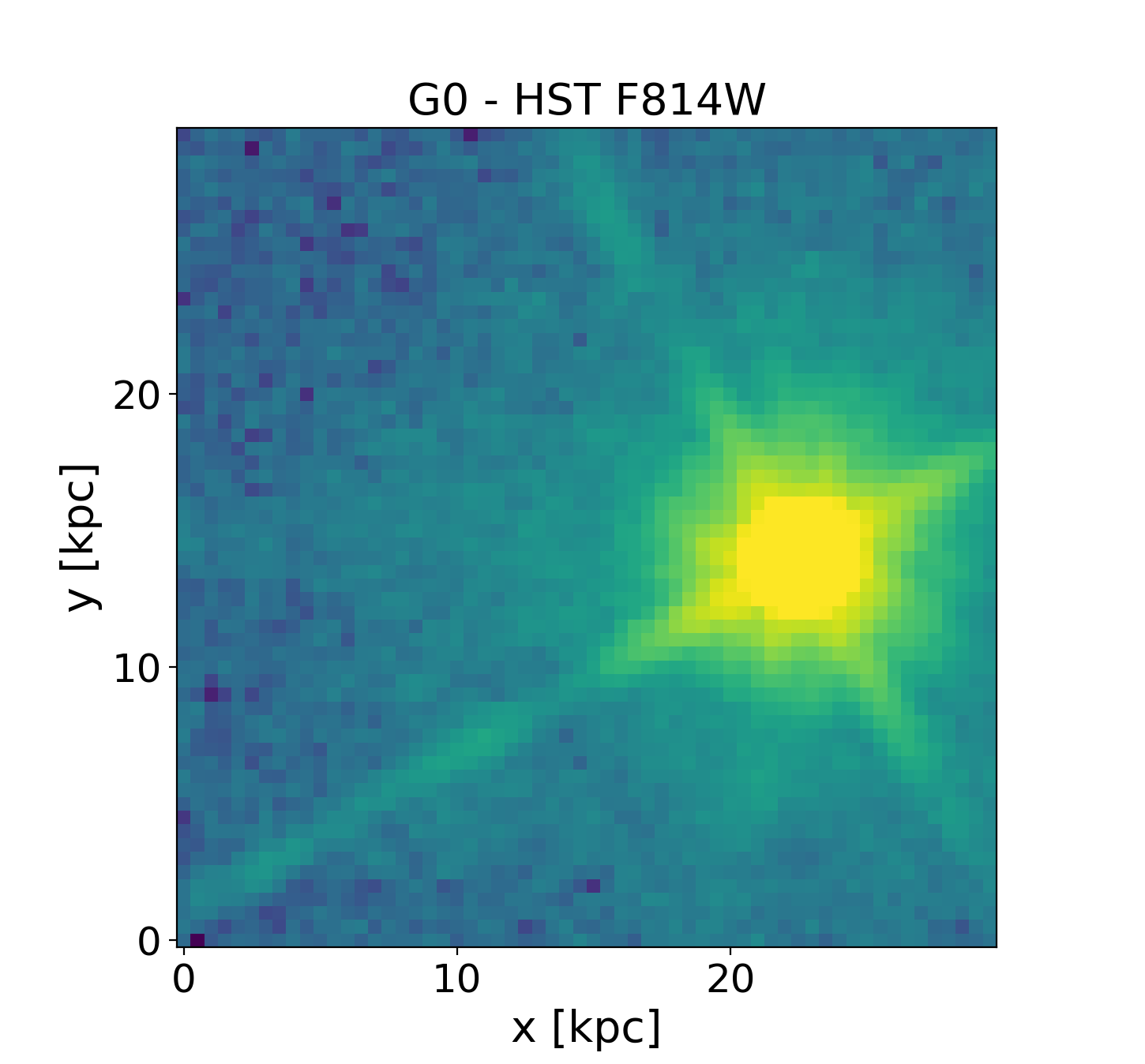}
\includegraphics[width=5.8cm, angle=0]{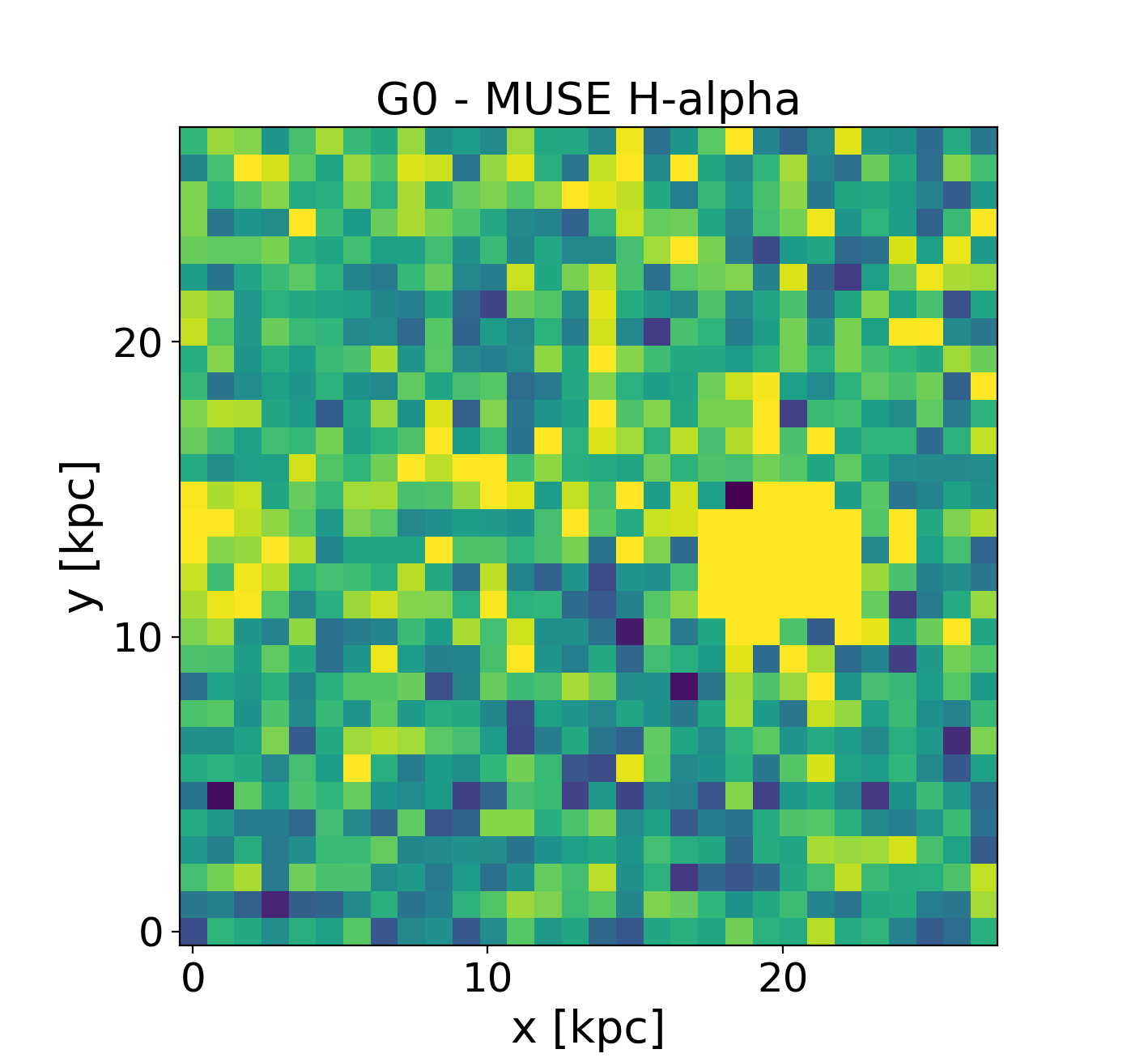}
\hspace{5.8cm}

\includegraphics[width=5.8cm, angle=0]{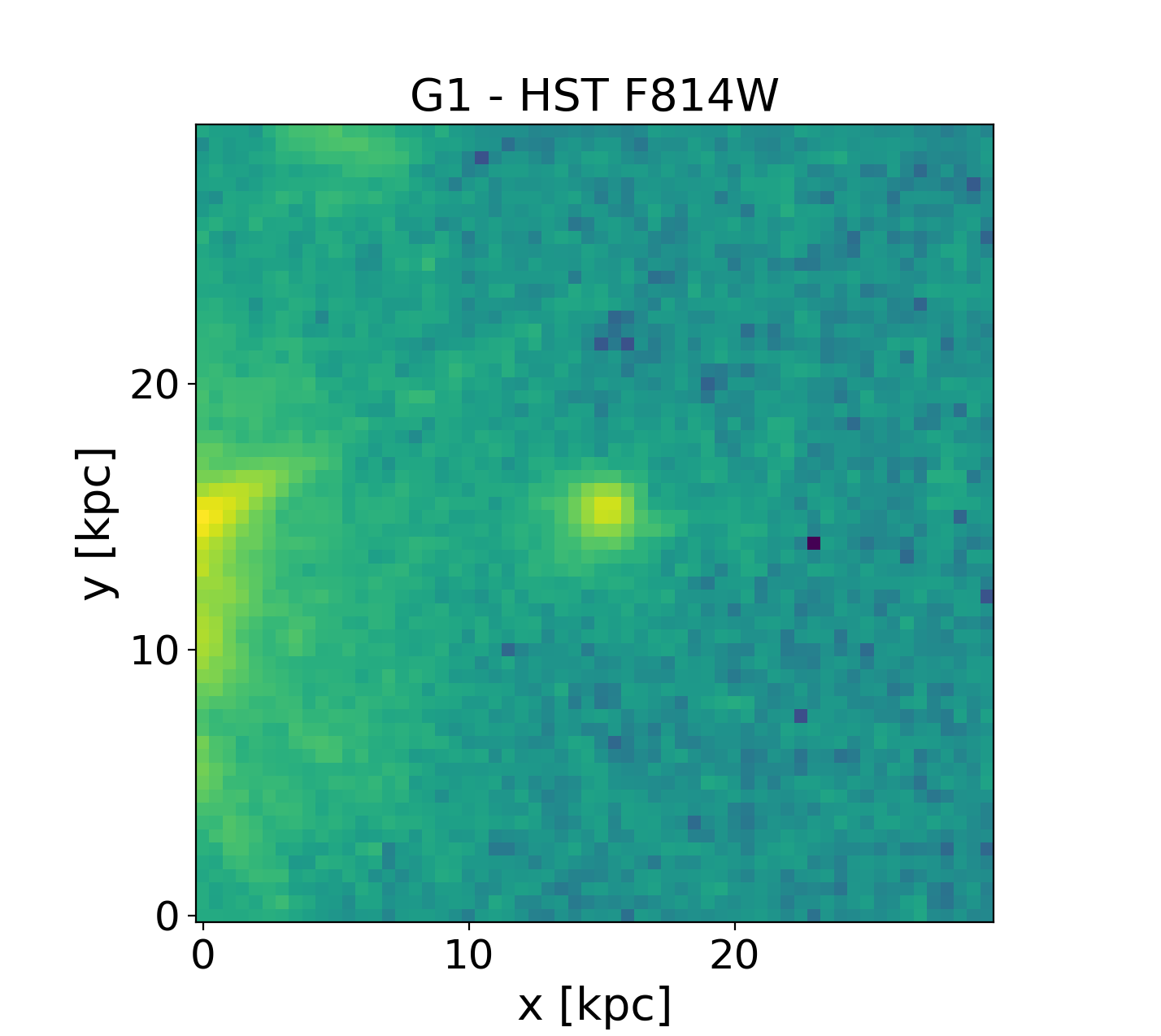}
\includegraphics[width=5.8cm, angle=0]{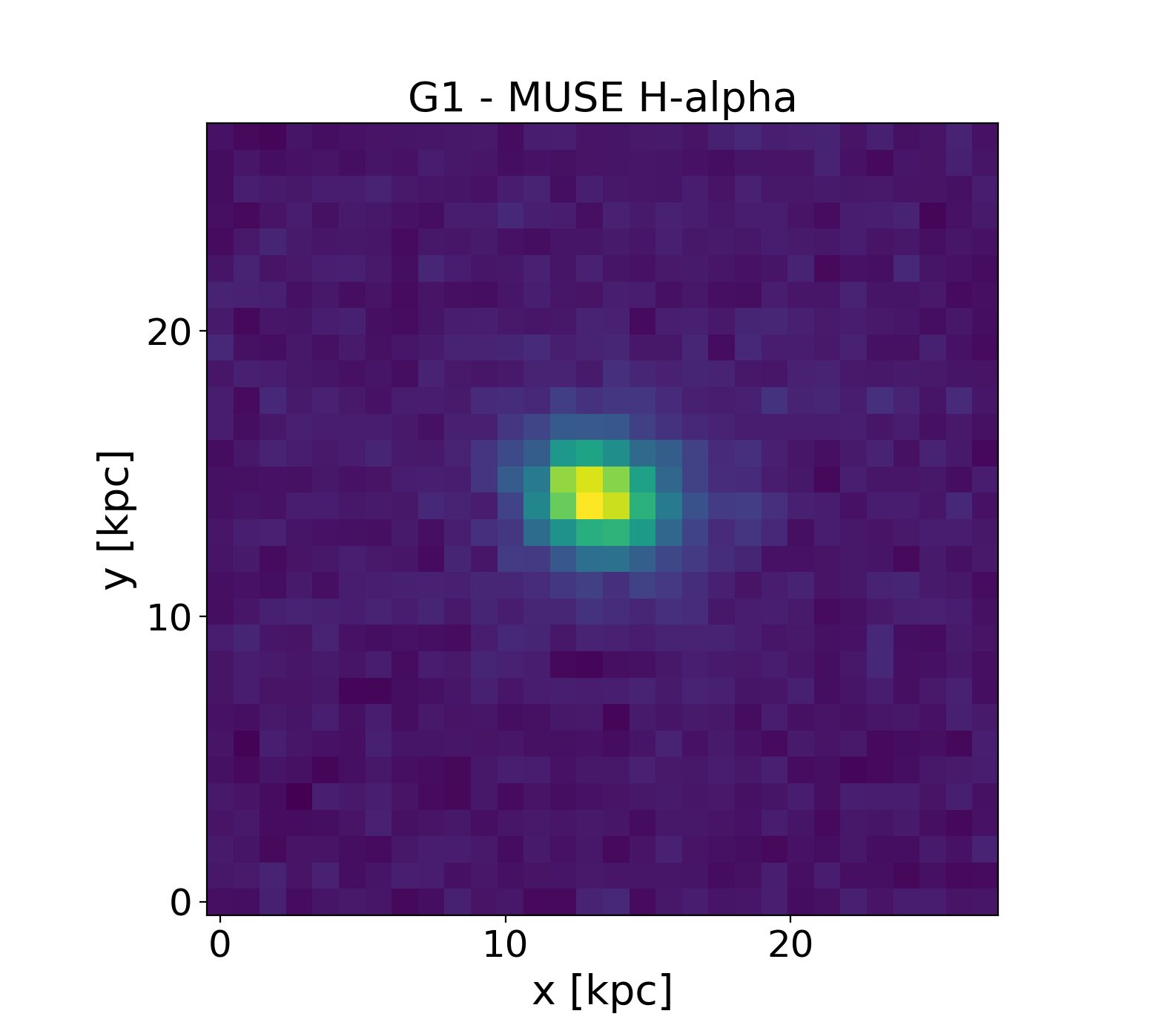}
\hspace{5.8cm}

\includegraphics[width=5.8cm, angle=0]{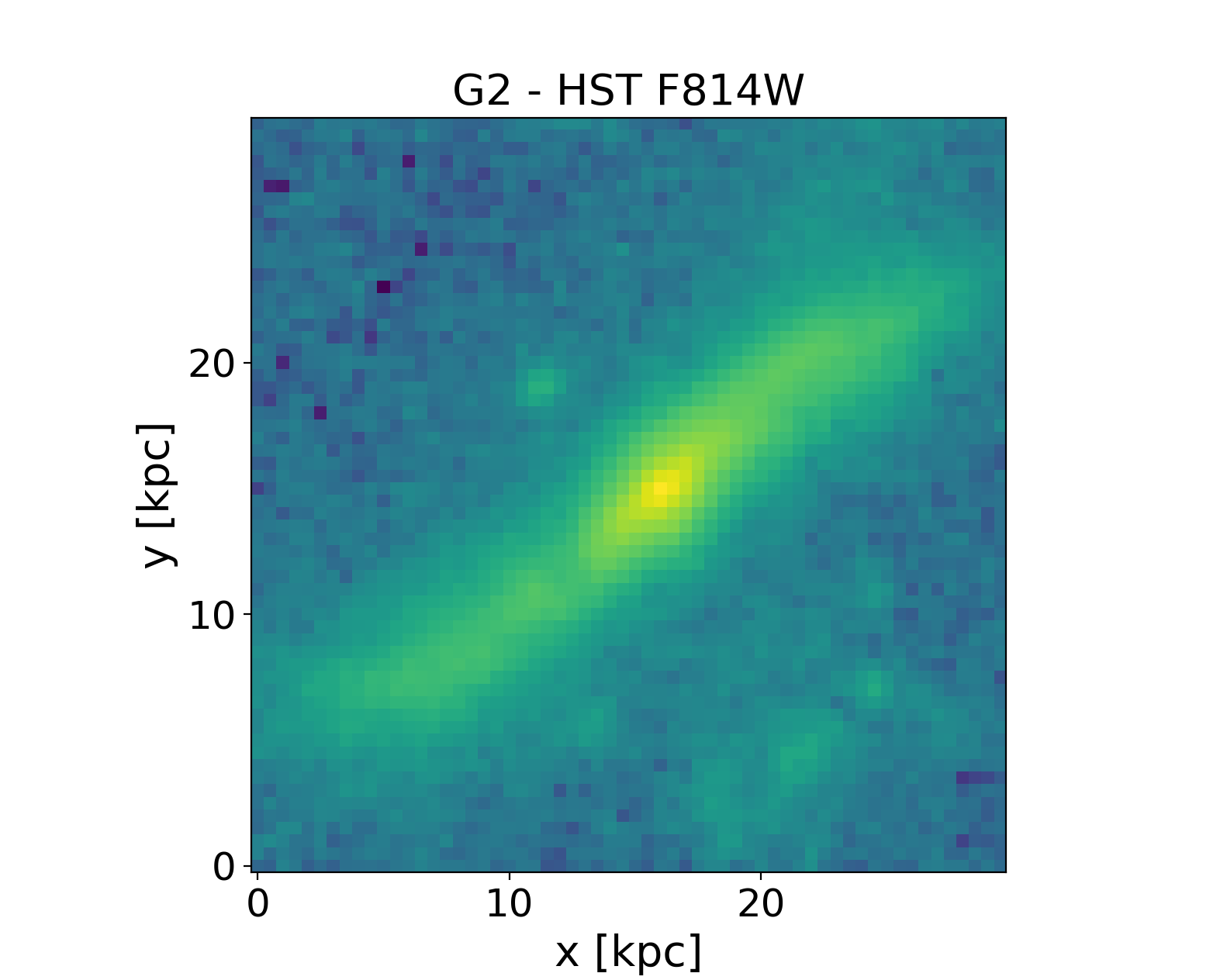}
\includegraphics[width=5.8cm, angle=0]{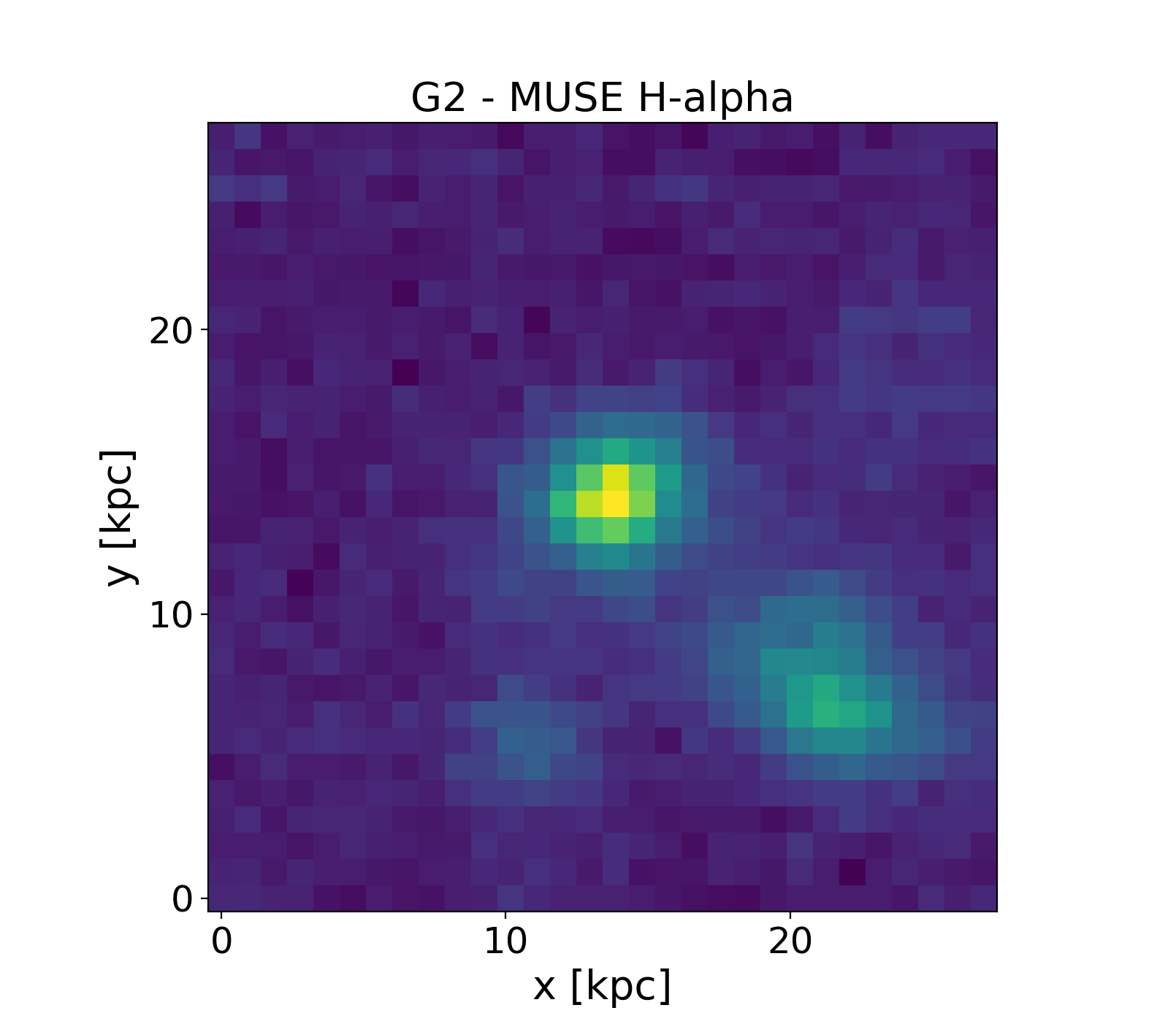}
\includegraphics[width=5.3cm, angle=0]{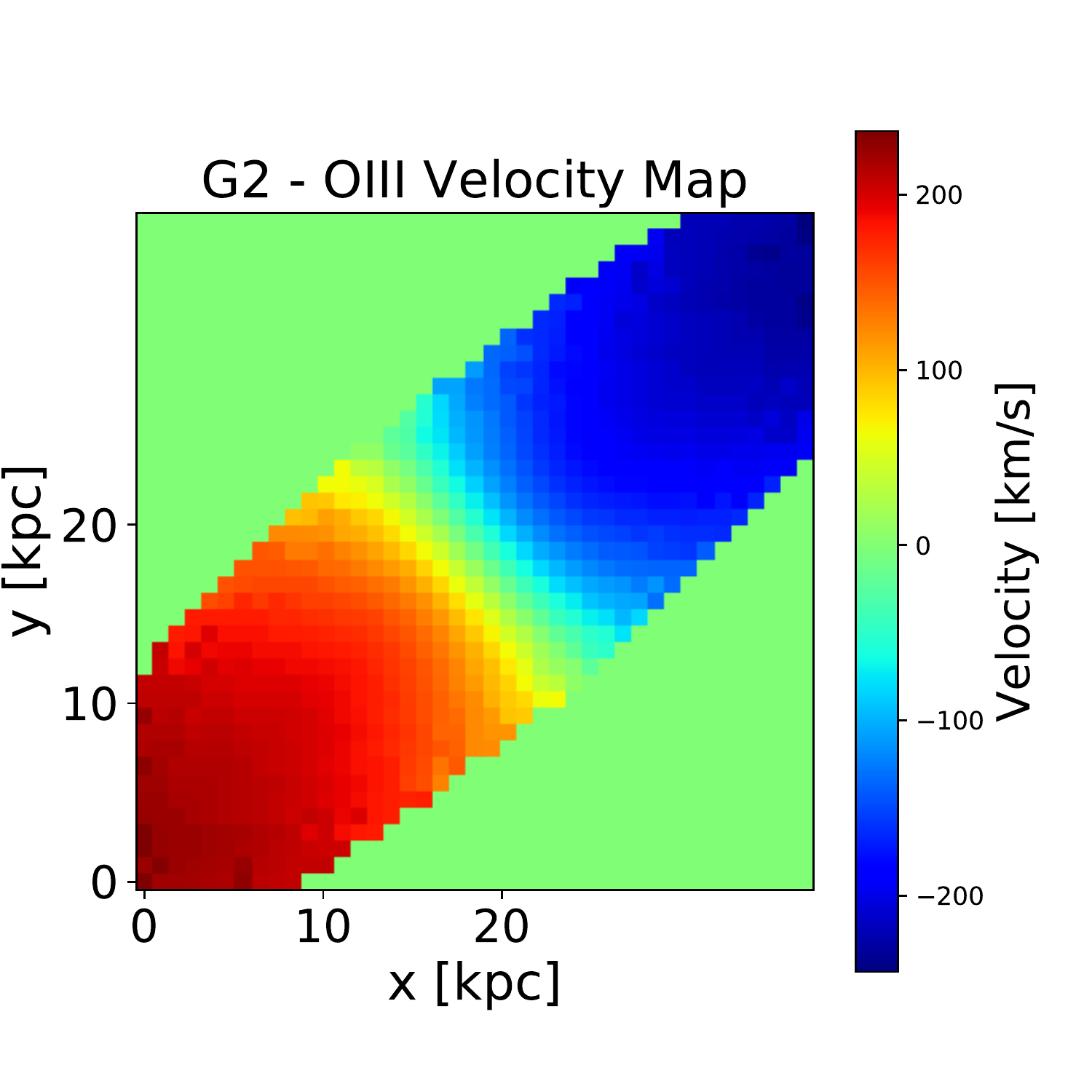}

\includegraphics[width=5.8cm, angle=0]{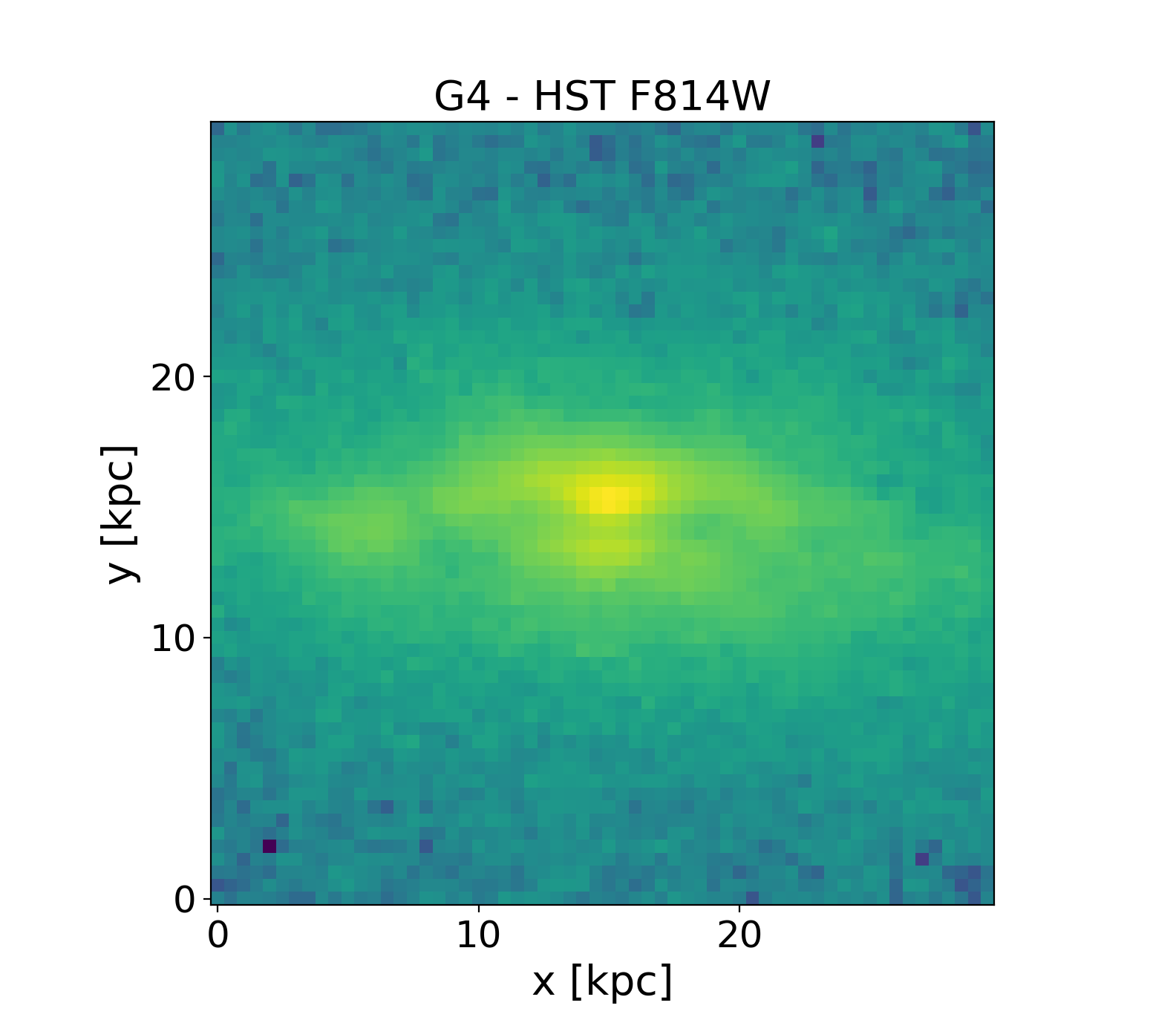}
\includegraphics[width=5.2cm, angle=0]{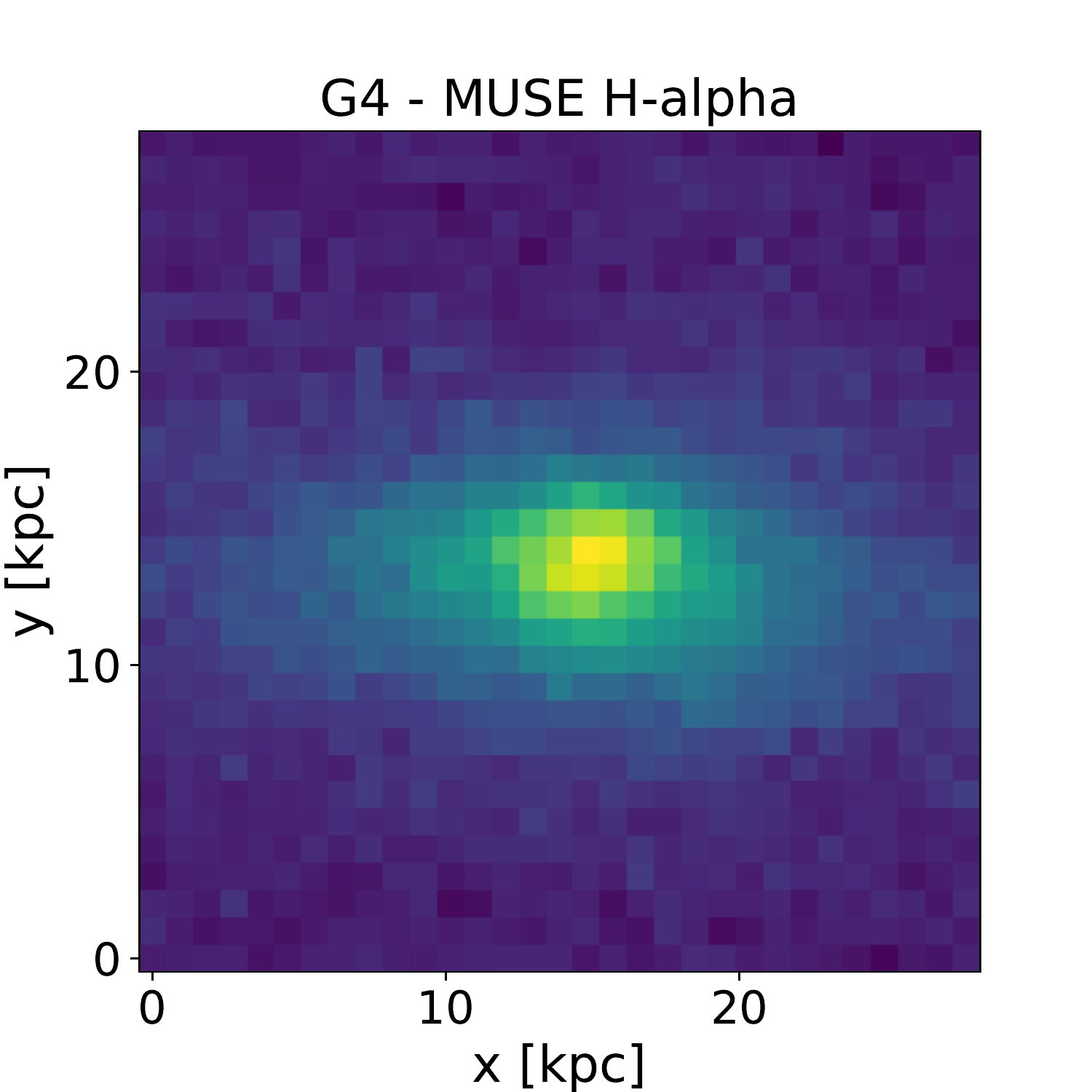}
\includegraphics[width=5.3cm, angle=0]{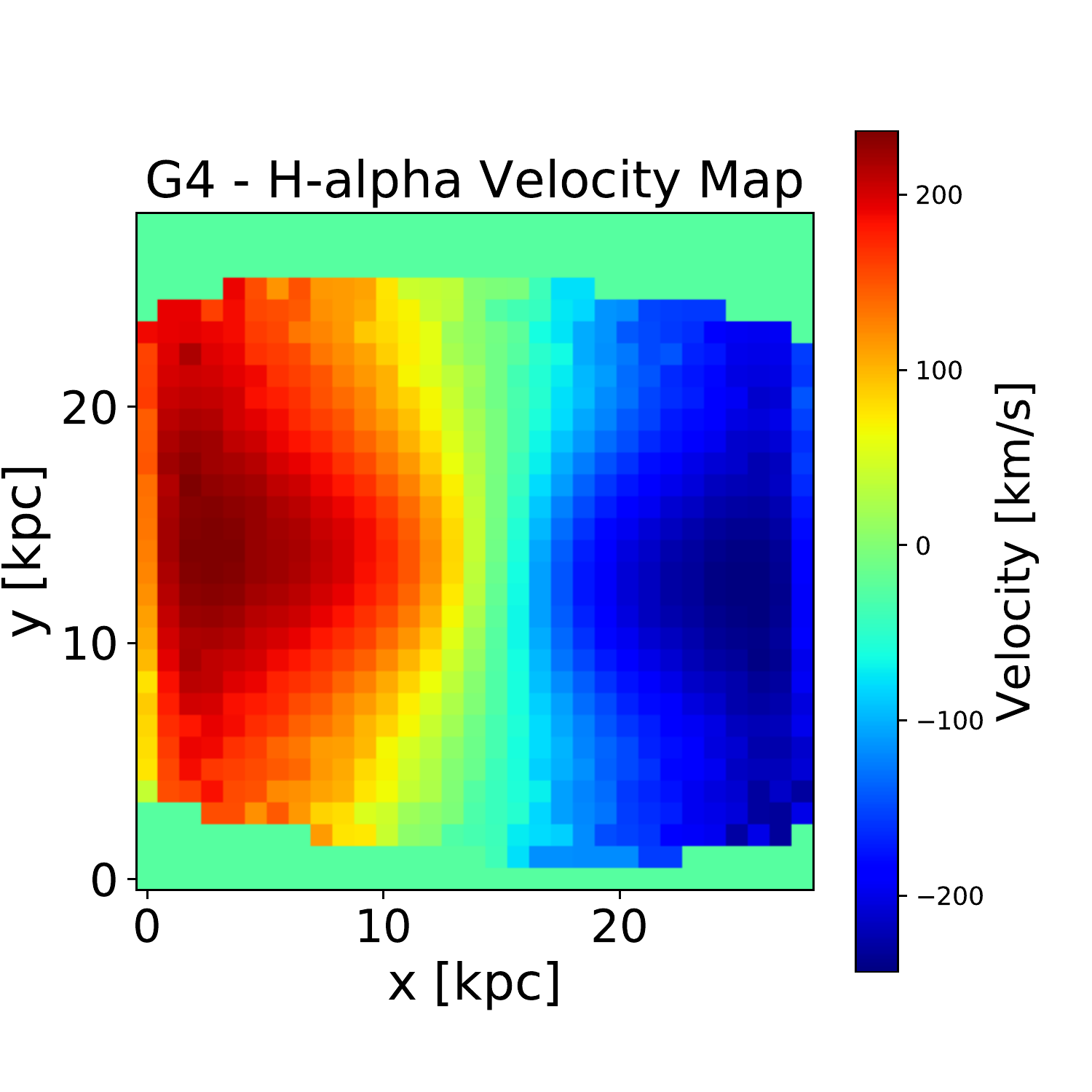}

\caption{{\bf Gas content of individual galaxies detected in MUSE at z$_{\rm abs}$=0.313.} Postage stamps of the stellar continuum observed with {\it HST}/WFPC2 (left), nebular emission of \oiii\ or \ha\ from MUSE narrow-band images (center) and modeled velocity maps convolved with the instrumental PSF (right). North is up, east to the left. The signal in galaxy G0 panel is dominated by the bright nearby quasar (slightly off-center to the West). While the object lies under the quasar PSF, several emission lines are clearly identified in the MUSE spectrum after quasar spectral-PFS subtraction (see Fig.~\ref{f:MUSE_spectra}). Some object are bright in continuum but have weak or undetected emission lines (e.g. G20) so that they are apparent in the HST data but weak in MUSE narrow-band. Conversely, some objects show strong emission lines with faint continuum (e.g. G0 and G16).}

\end{center}
\end{figure*}

\begin{figure*}
\begin{center}
\includegraphics[width=5.8cm, angle=0]{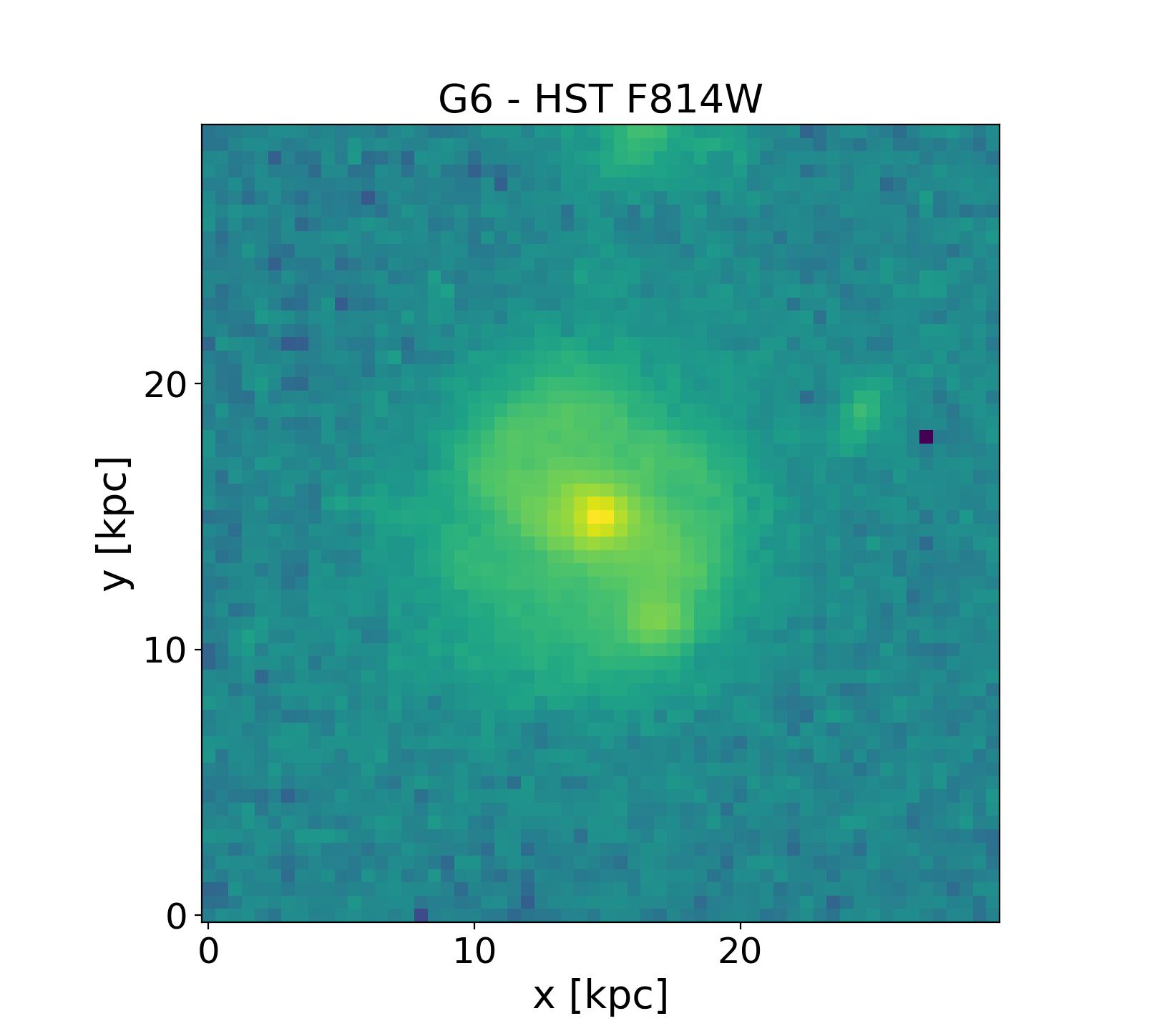}
\includegraphics[width=5.5cm, angle=0]{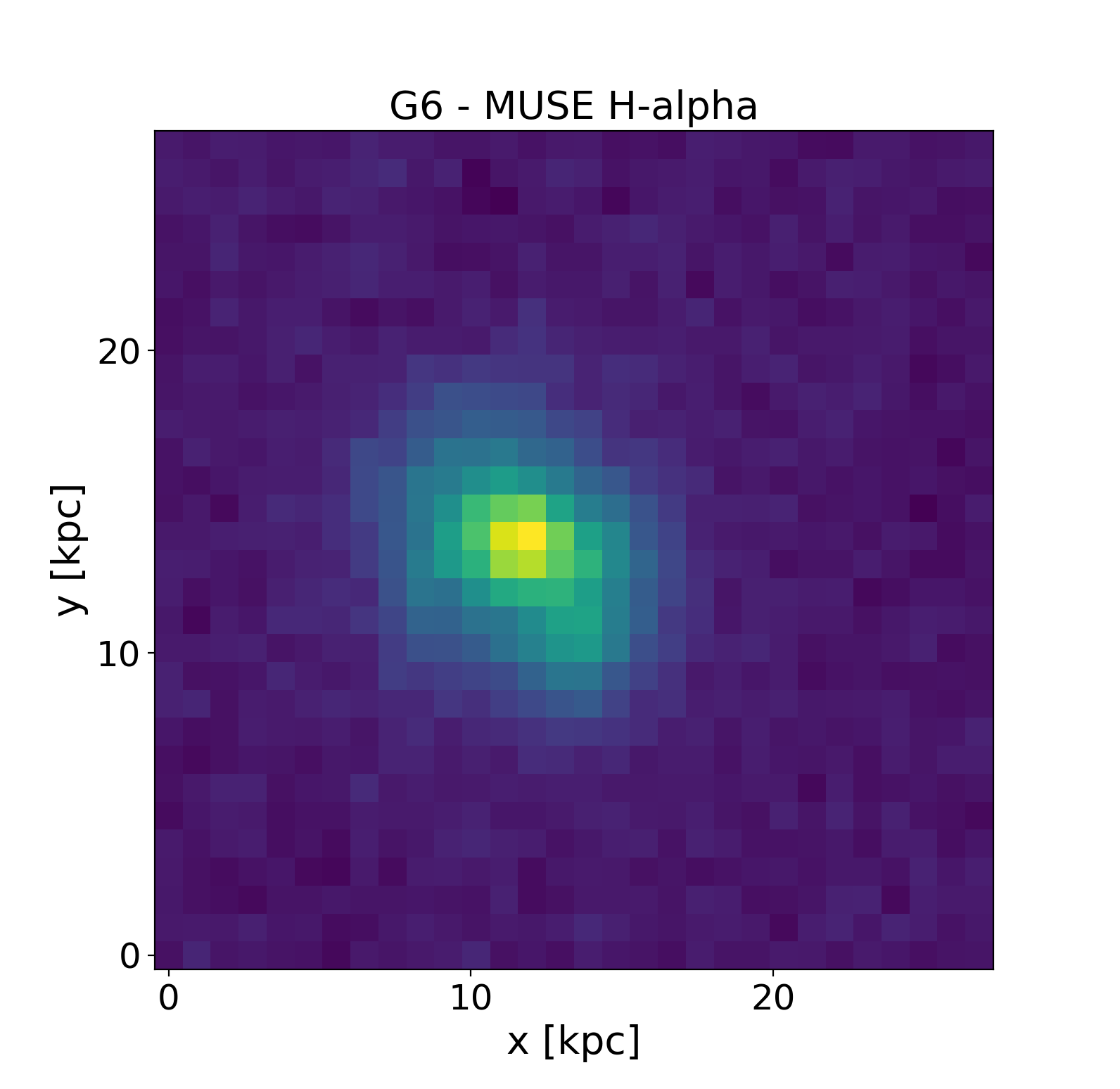}
\hspace{5.8cm}

\includegraphics[width=5.7cm, angle=0]{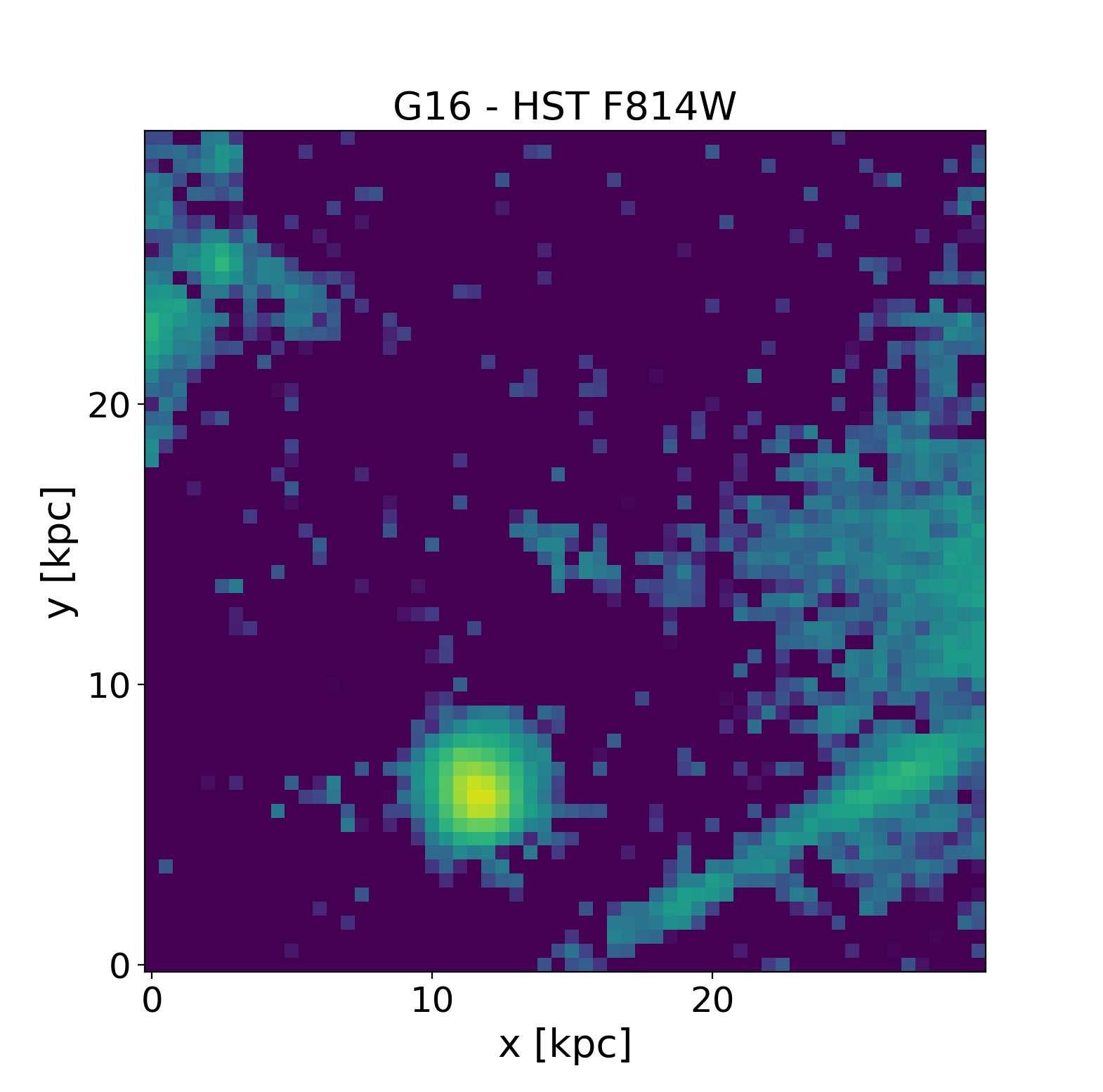}
\includegraphics[width=5.3cm, angle=0]{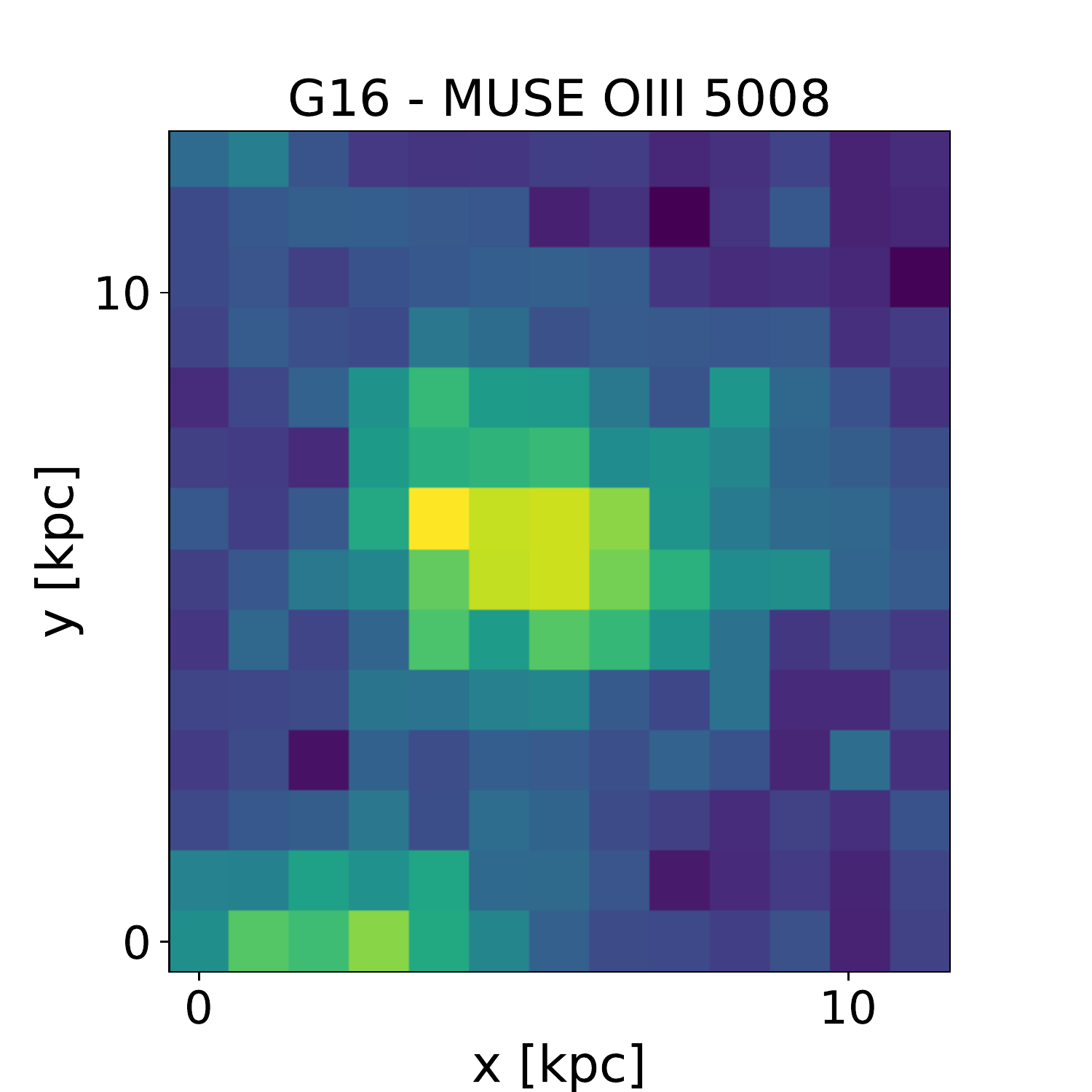}
\includegraphics[width=5.3cm, angle=0]{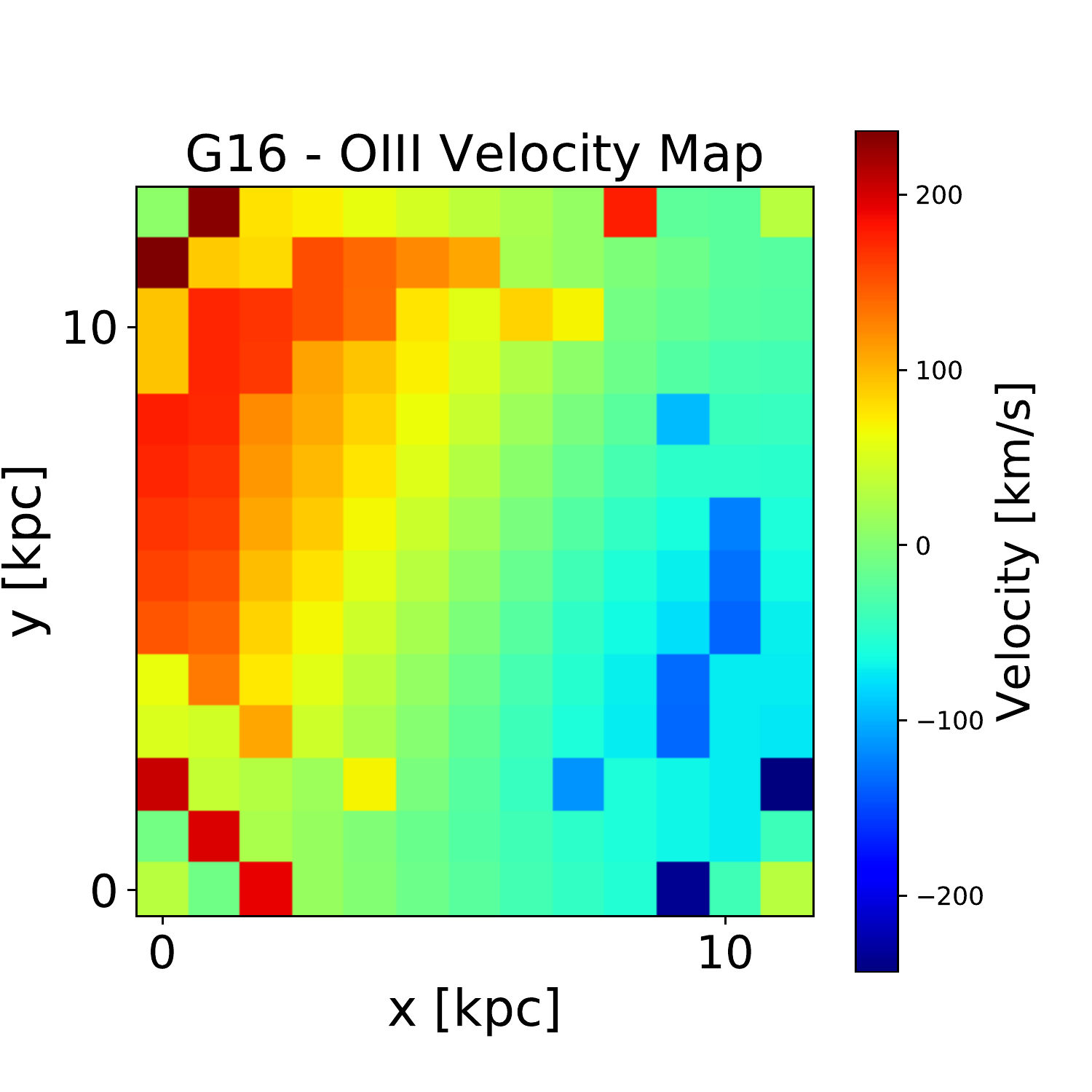}

\includegraphics[width=5.8cm, angle=0]{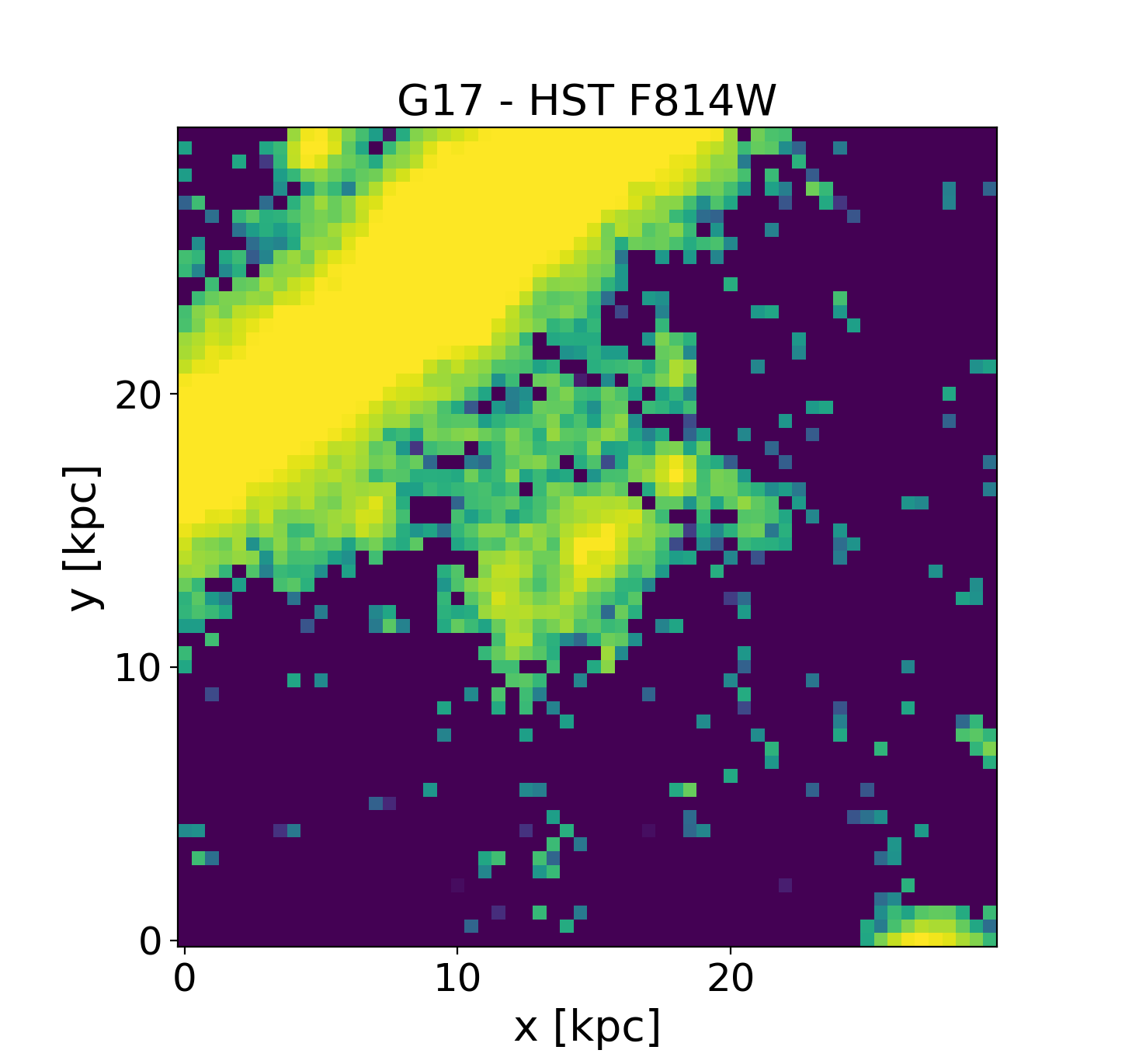}
\includegraphics[width=5.8cm, angle=0]{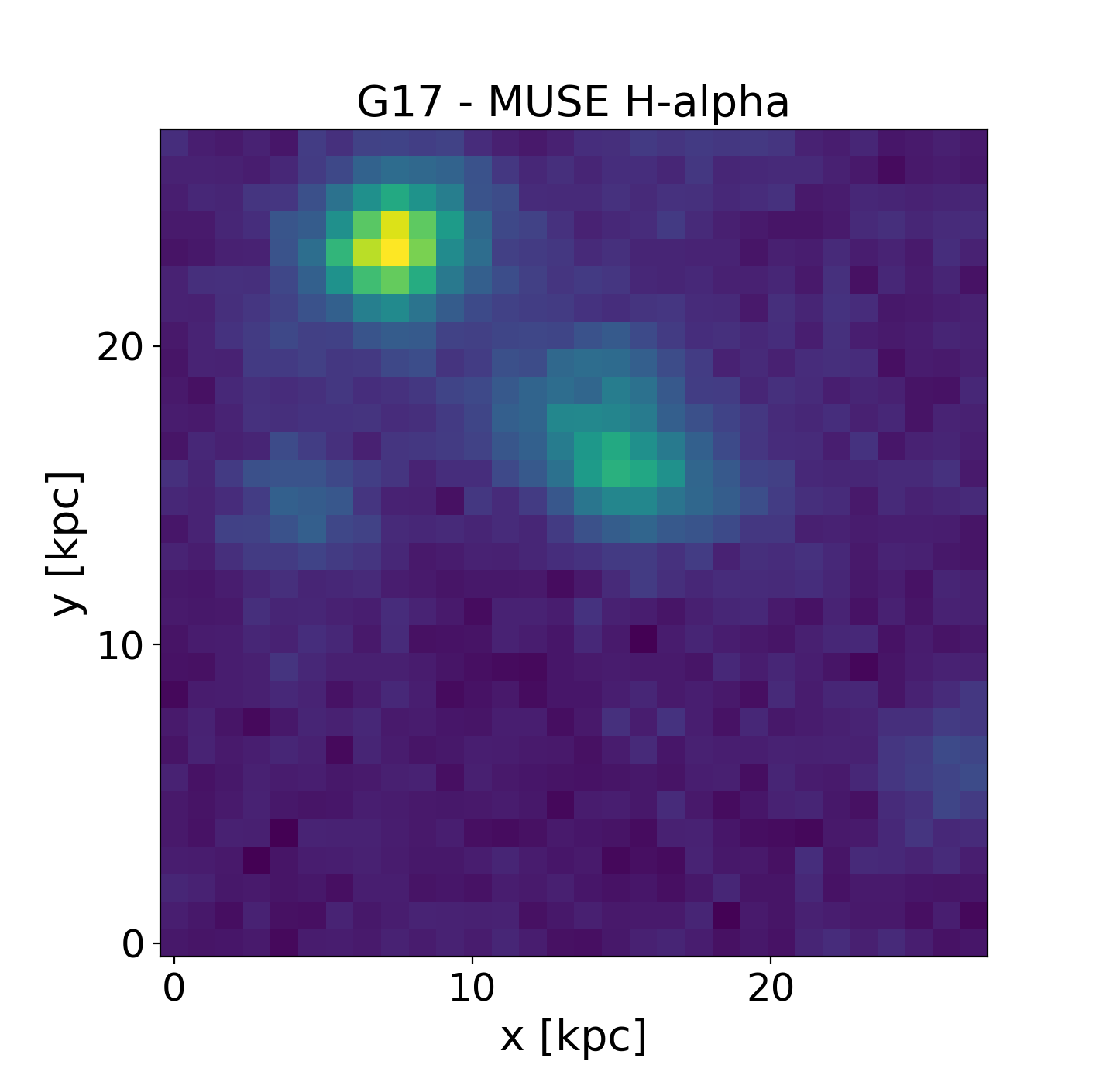}
\includegraphics[width=5.3cm, angle=0]{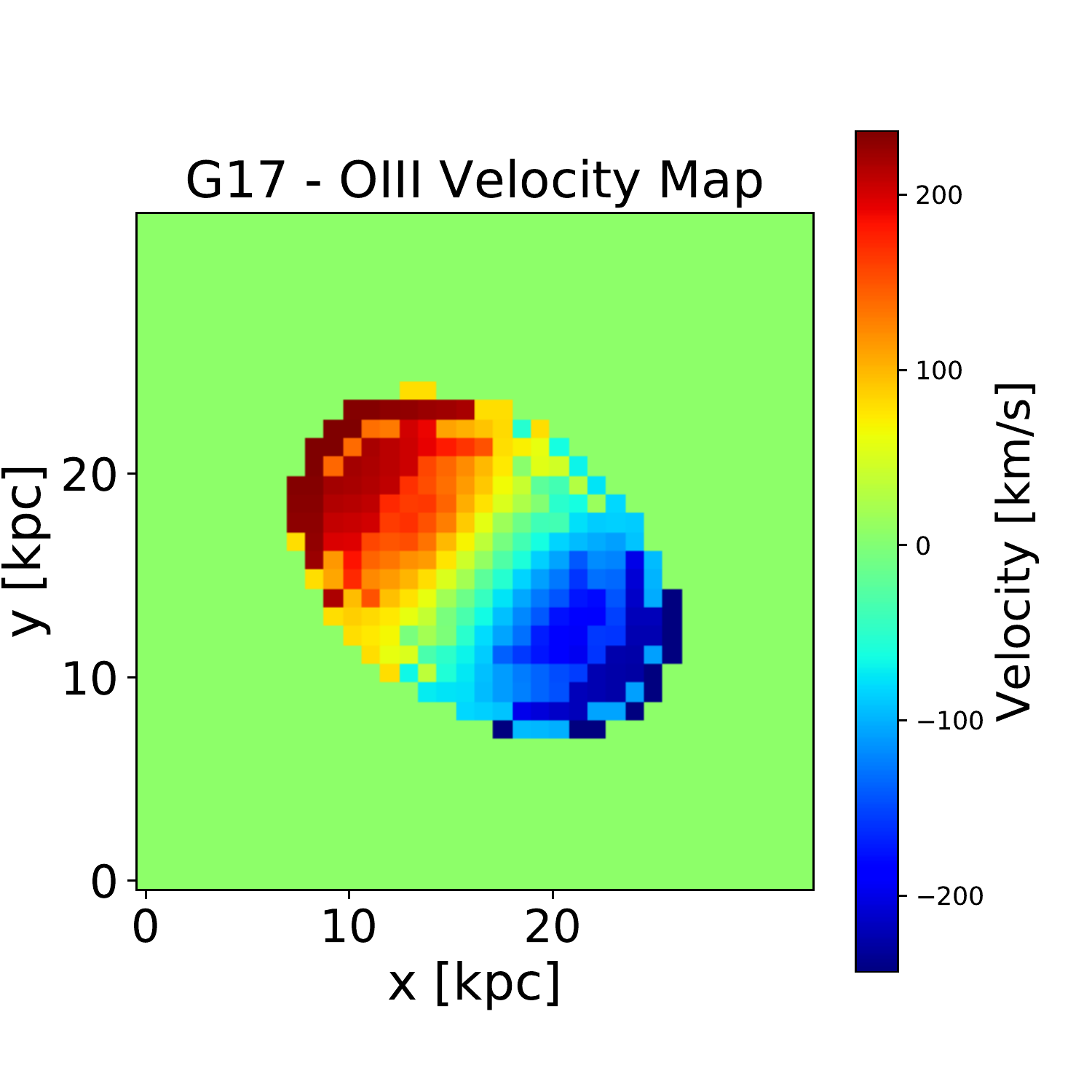}

\includegraphics[width=5.7cm, angle=0]{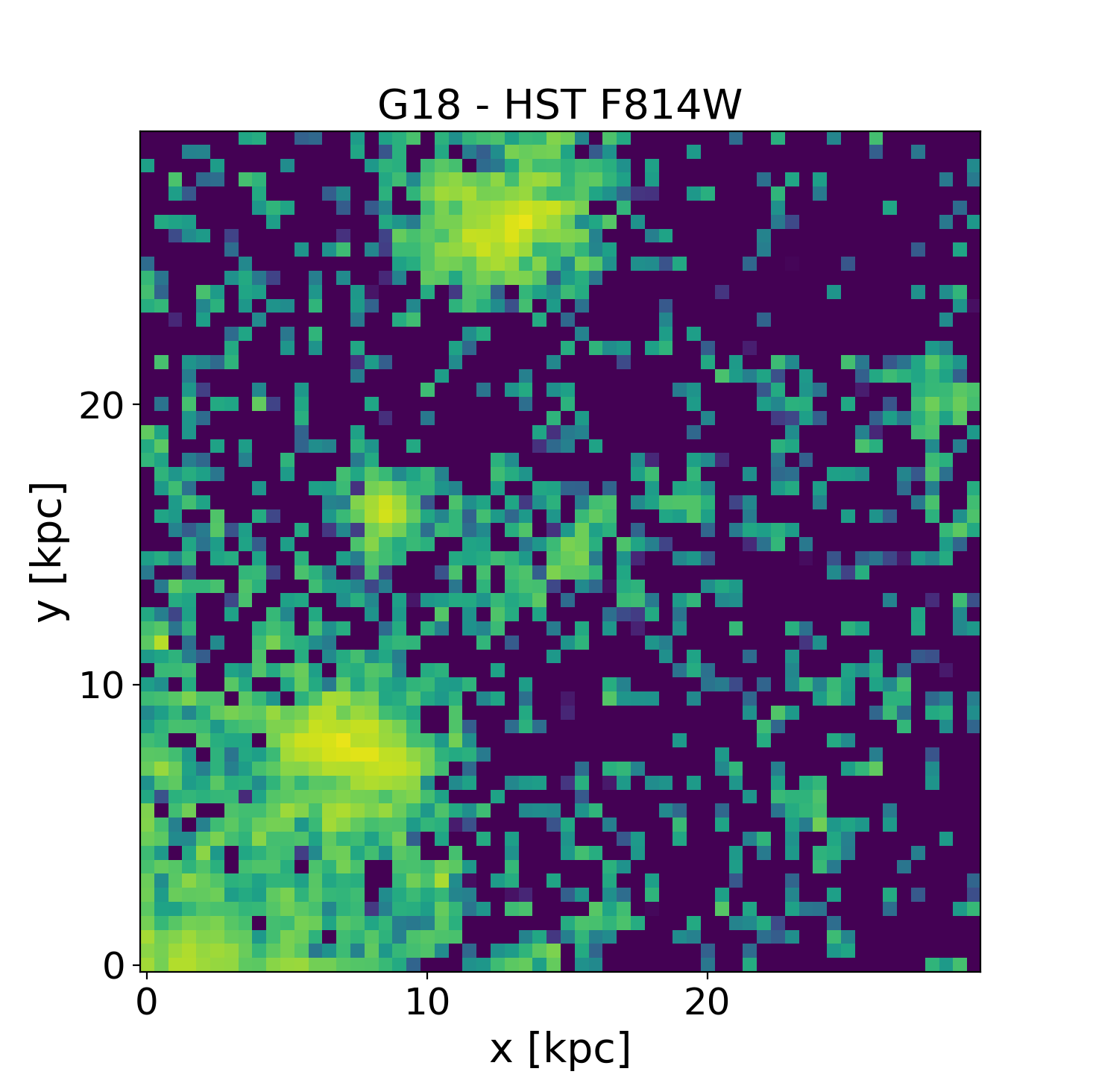}
\includegraphics[width=5.3cm, angle=0]{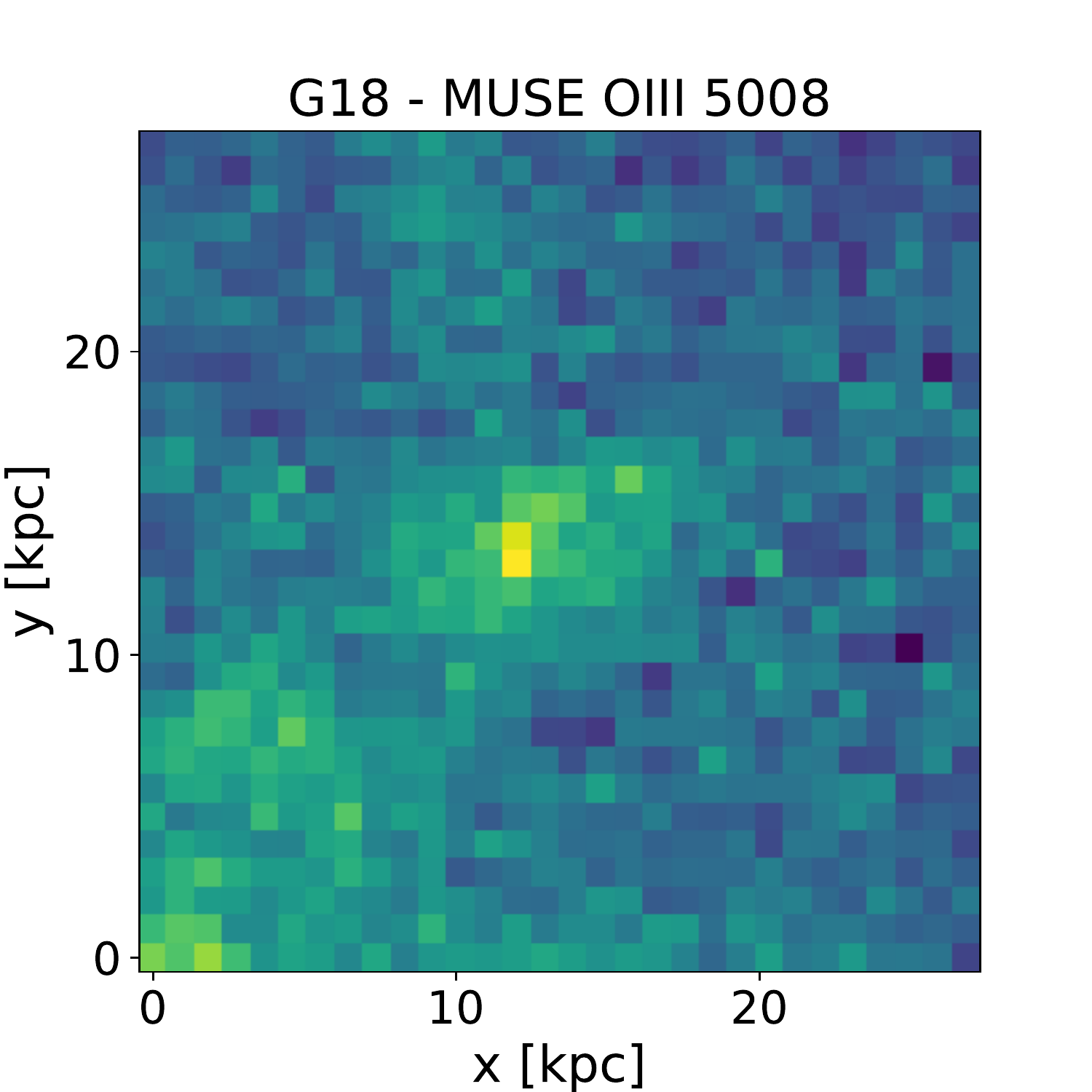}
\includegraphics[width=5.3cm, angle=0]{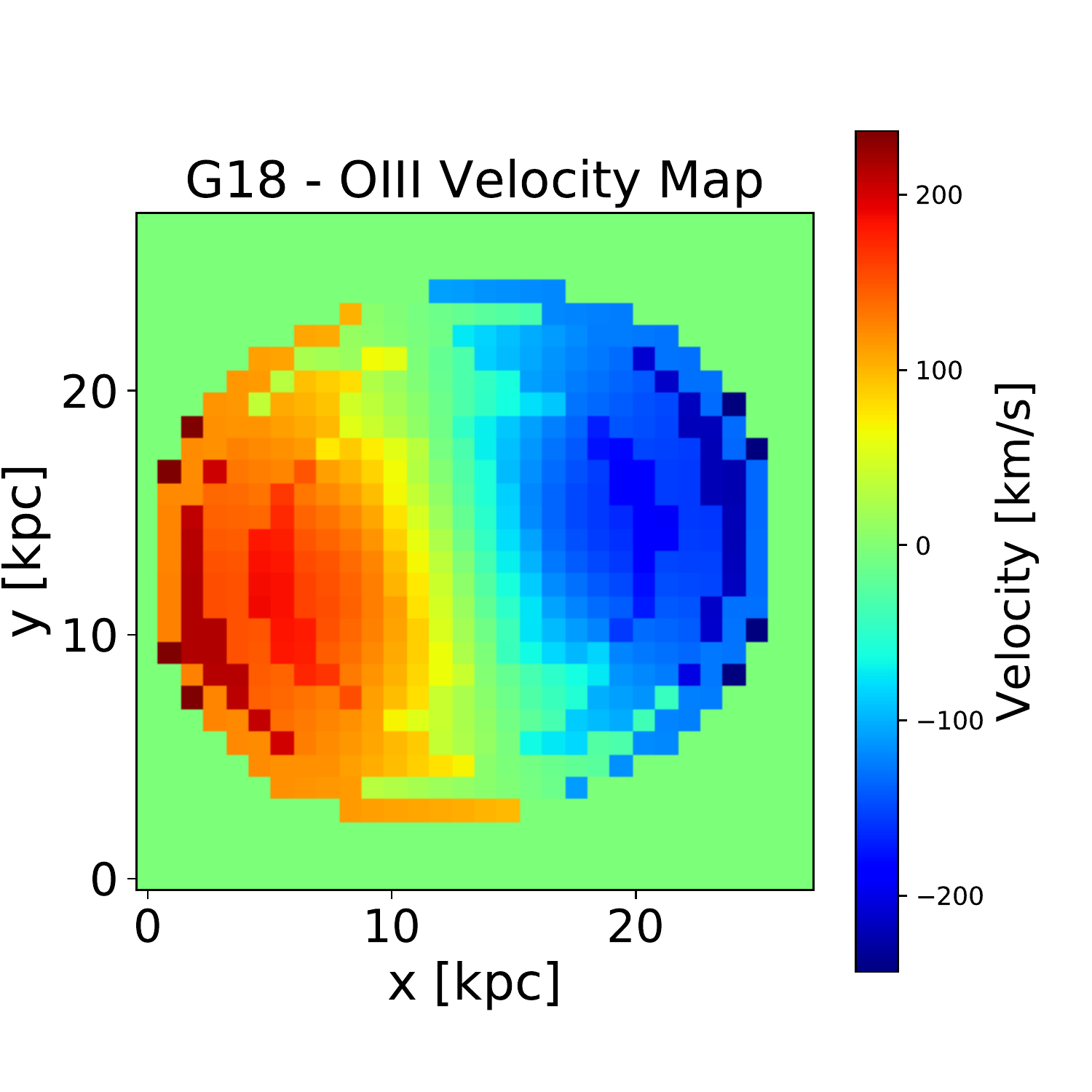}

\addtocounter{figure}{-1}
\caption{{\it continued}}

\end{center}
\end{figure*}

\begin{figure*}
\begin{center}
\includegraphics[width=5.8cm, angle=0]{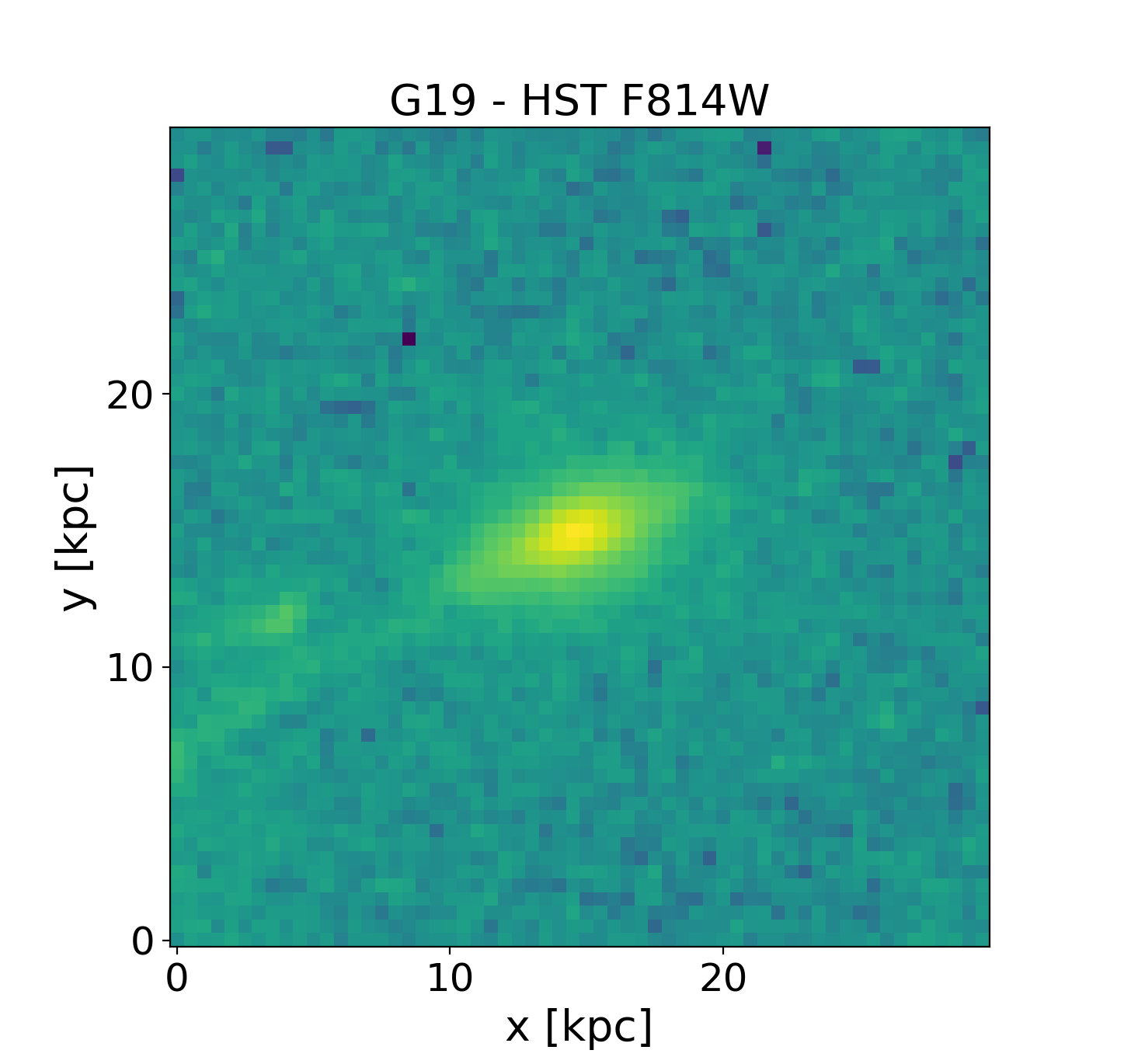}
\includegraphics[width=5.2cm, angle=0]{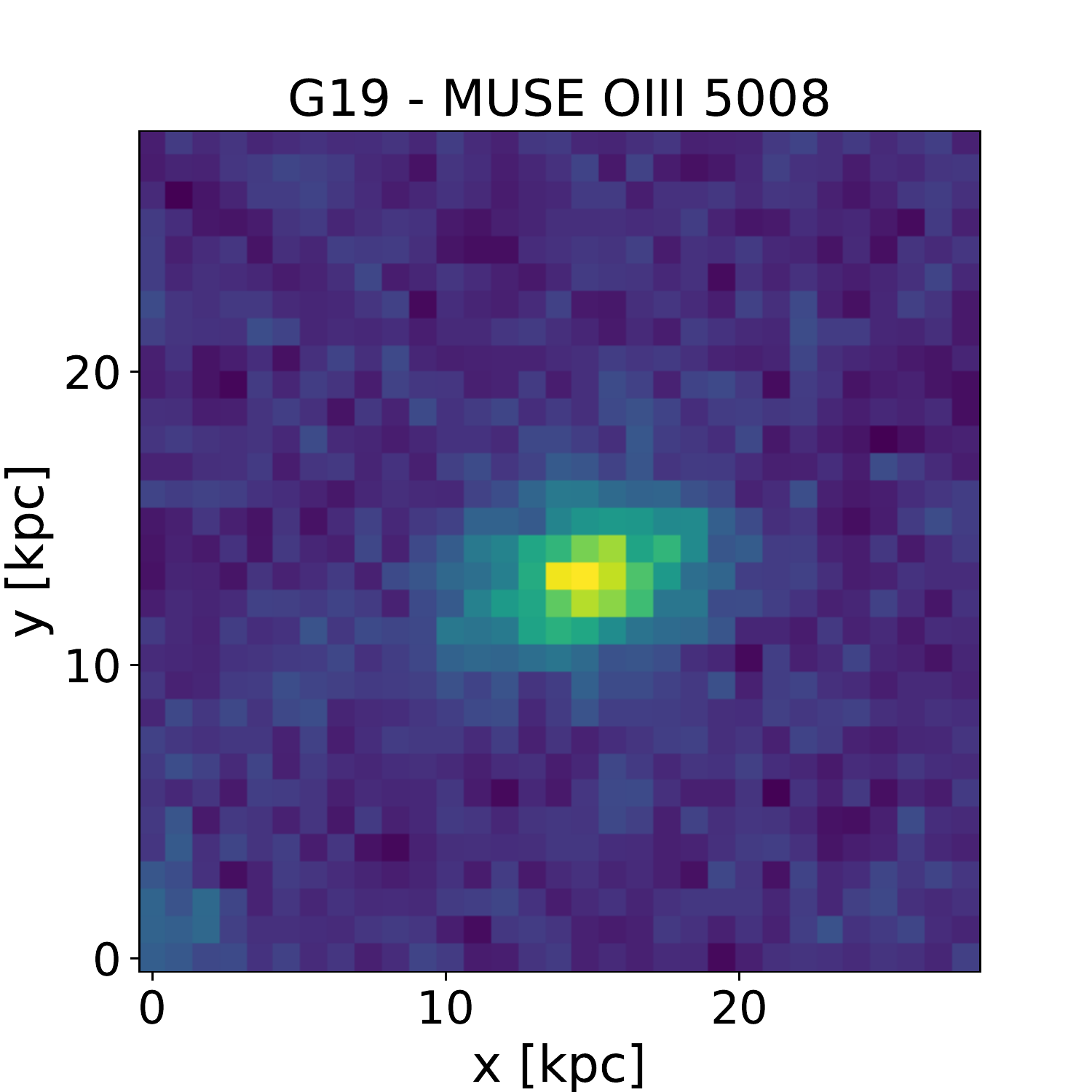}
\includegraphics[width=5.3cm, angle=0]{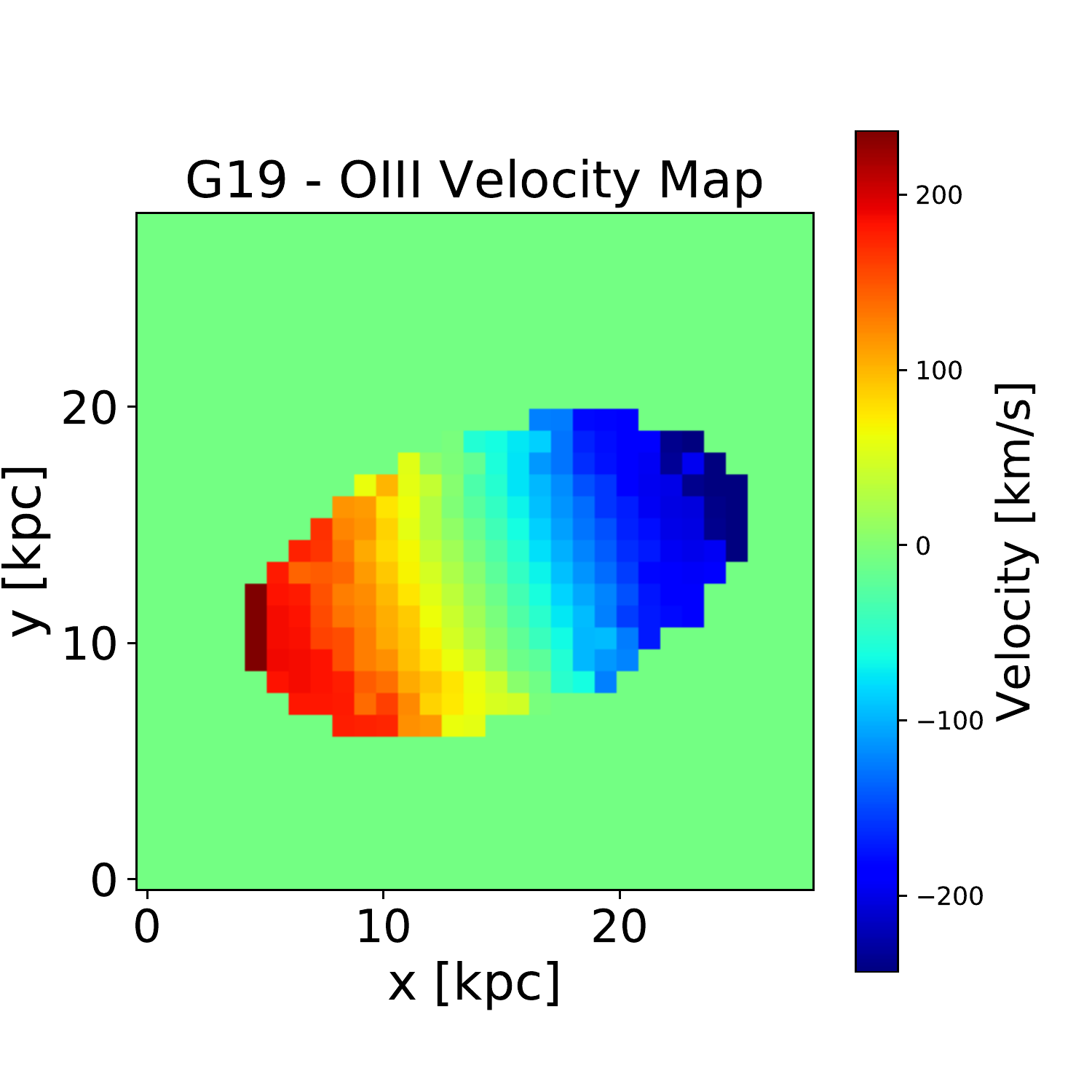}

\includegraphics[width=5.8cm, angle=0]{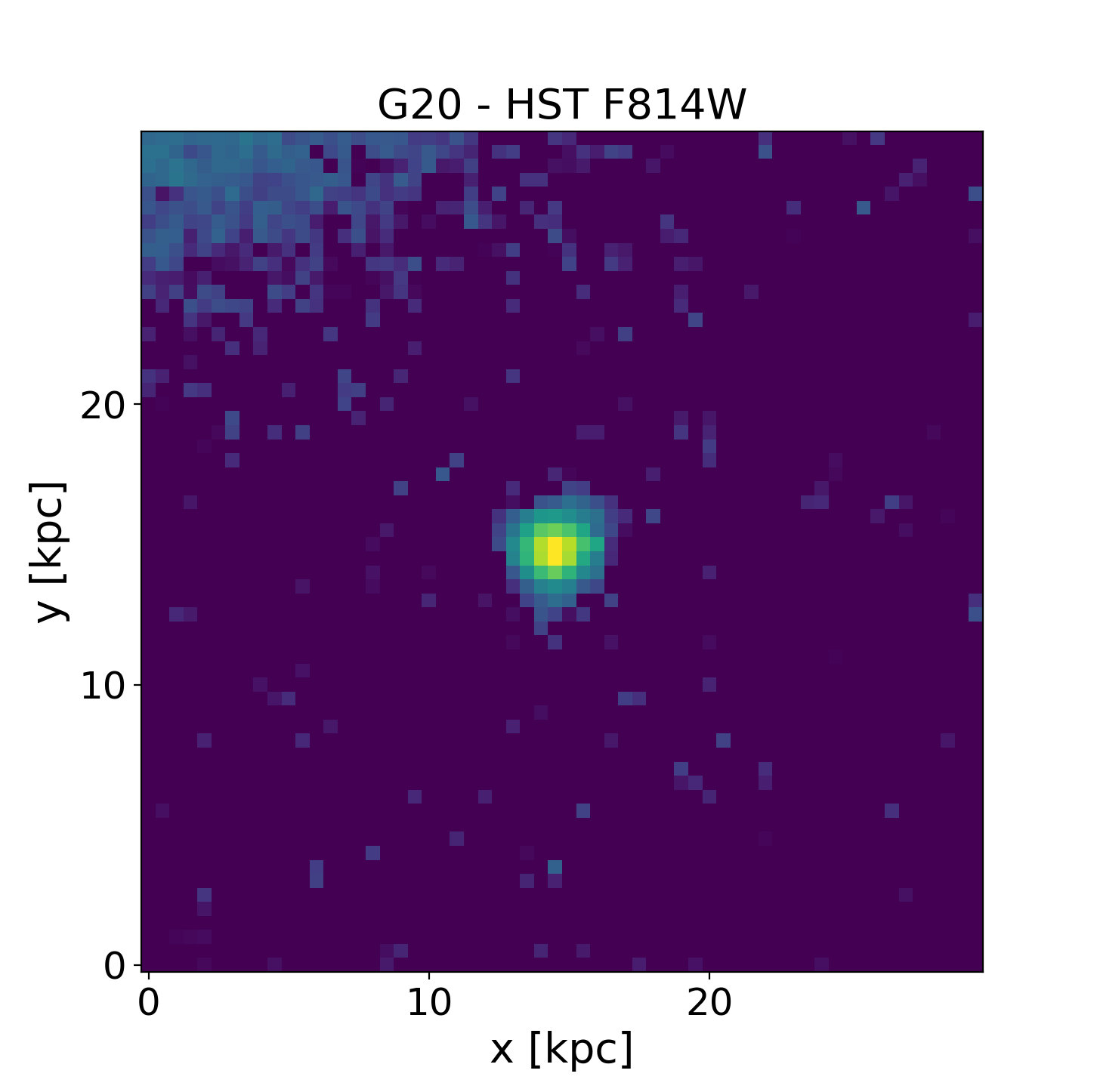}
\includegraphics[width=5.8cm, angle=0]{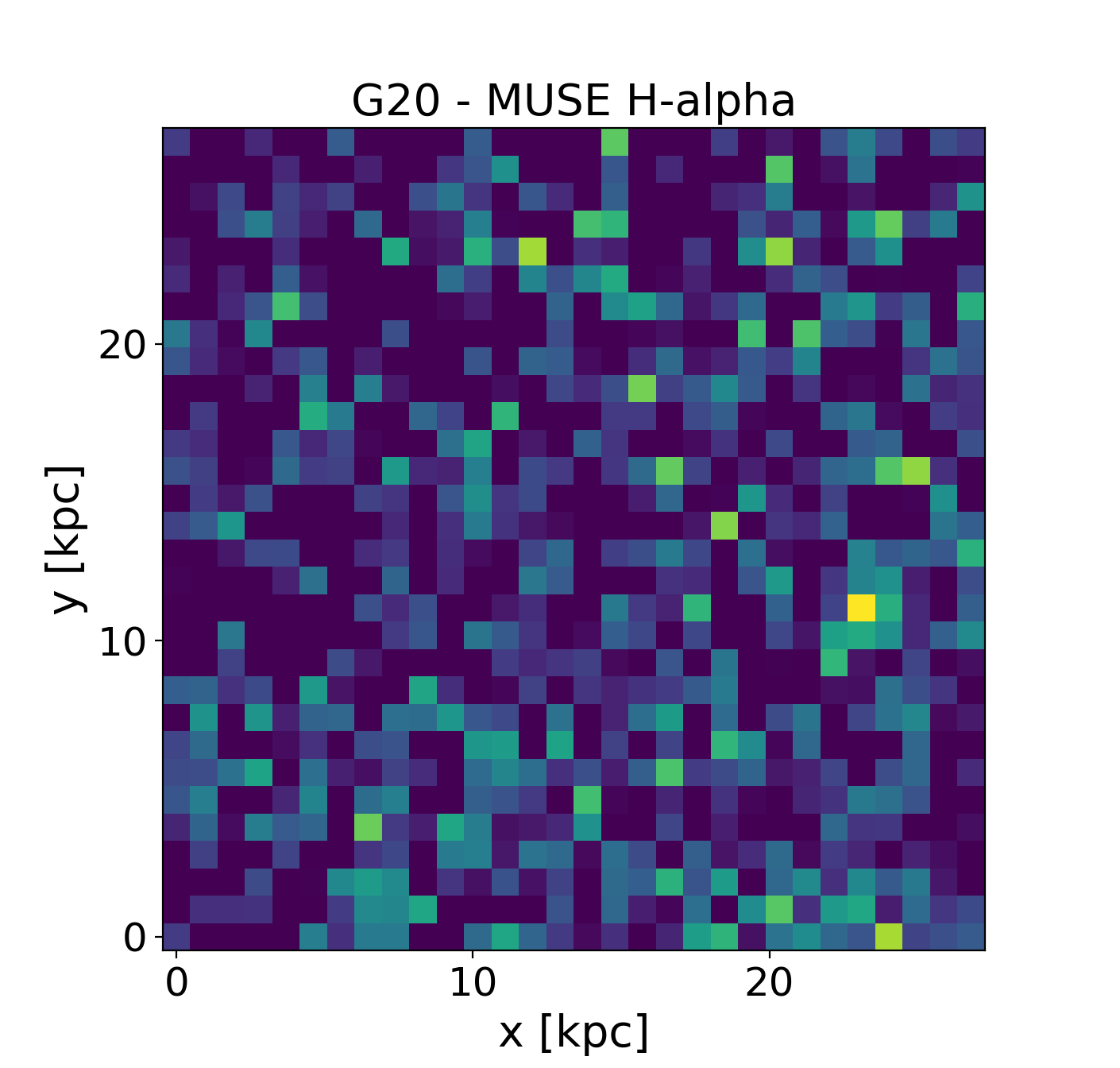}
\hspace{5.8cm}

\includegraphics[width=5.8cm, angle=0]{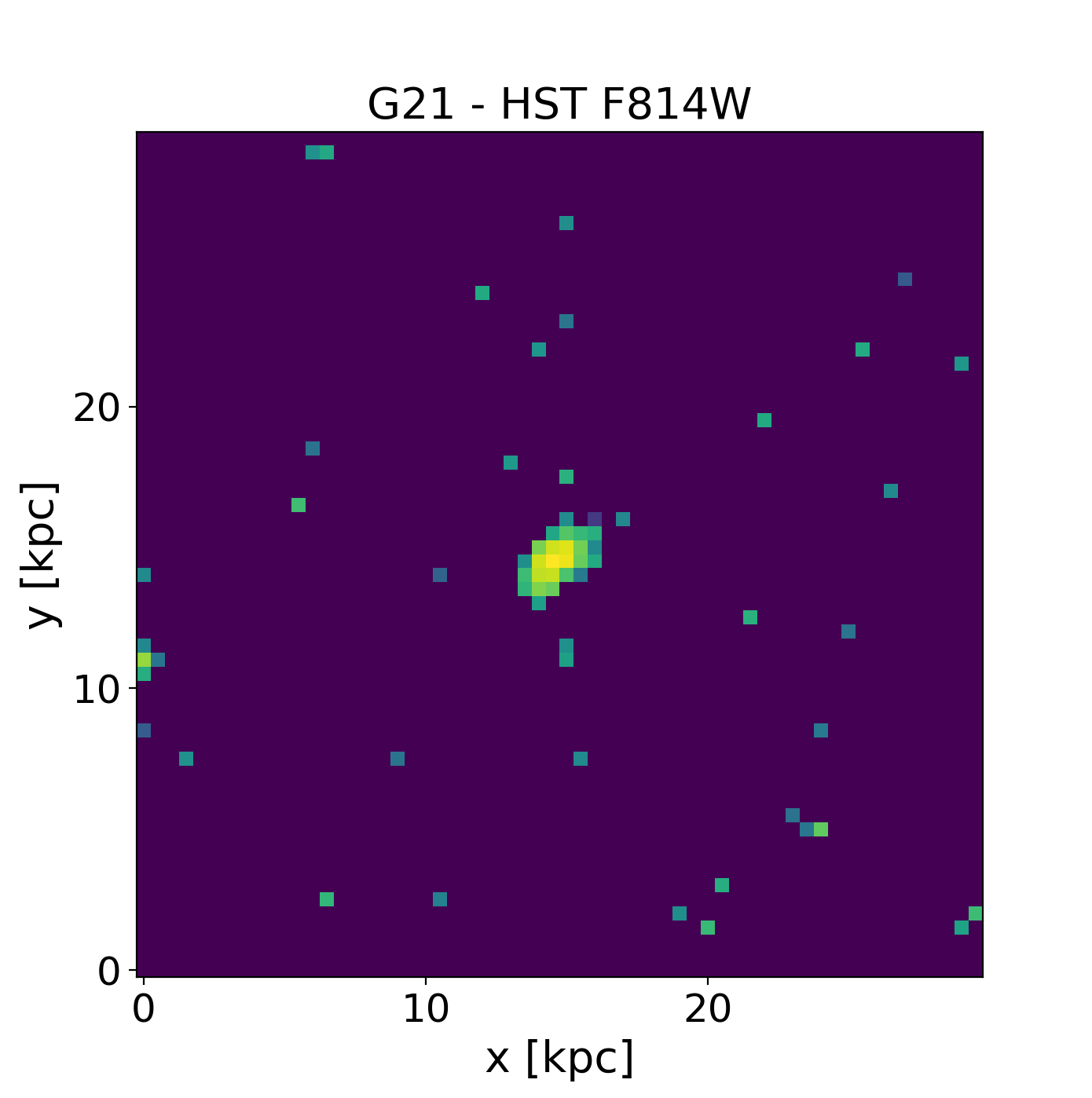}
\includegraphics[width=5.3cm, angle=0]{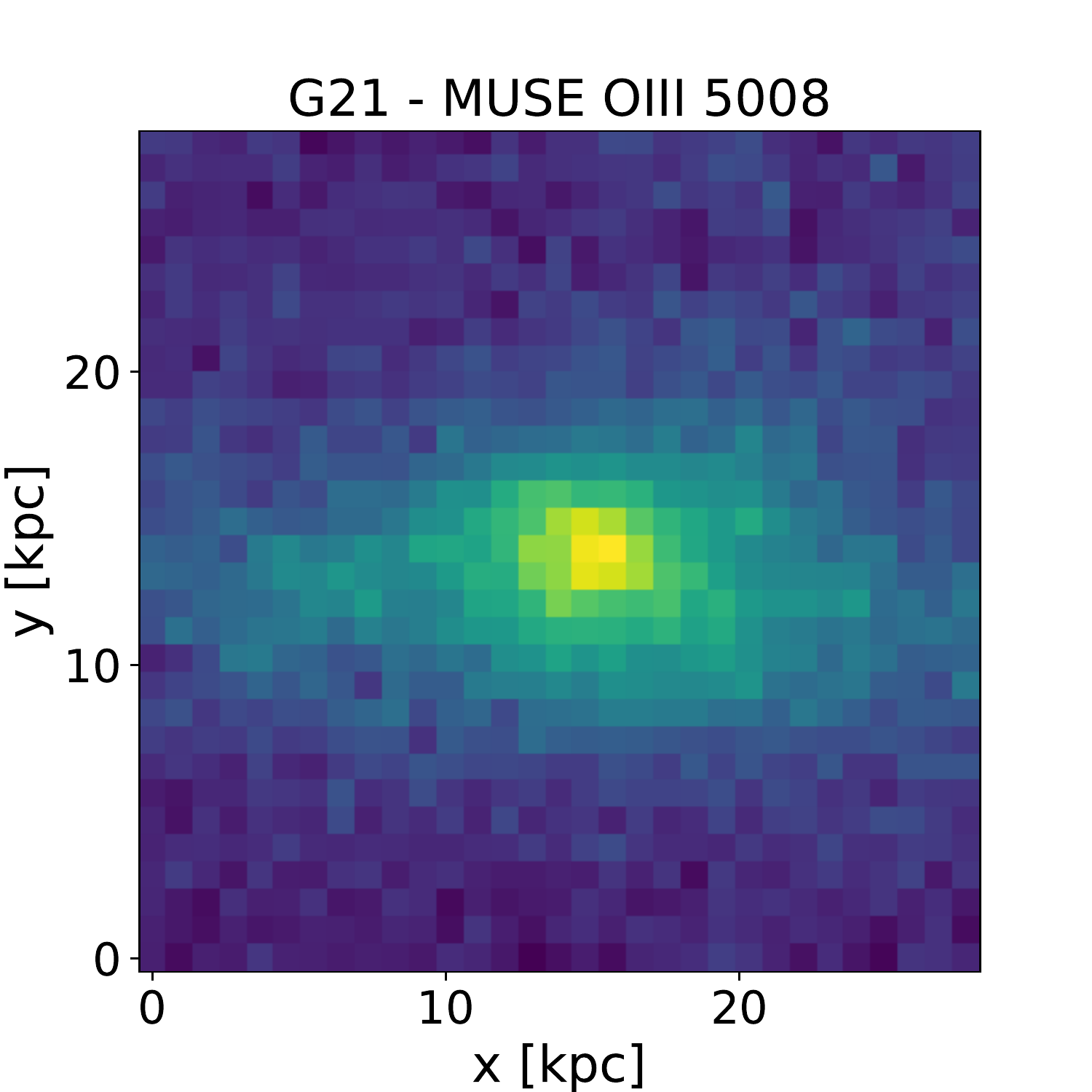}
\hspace{5.8cm}

\addtocounter{figure}{-1}
\caption{{\it continued}}
\label{f:MUSE_Kine}
\end{center}
\end{figure*}

\begin{figure*}
\begin{center}
\includegraphics[width=5.cm, angle=0]{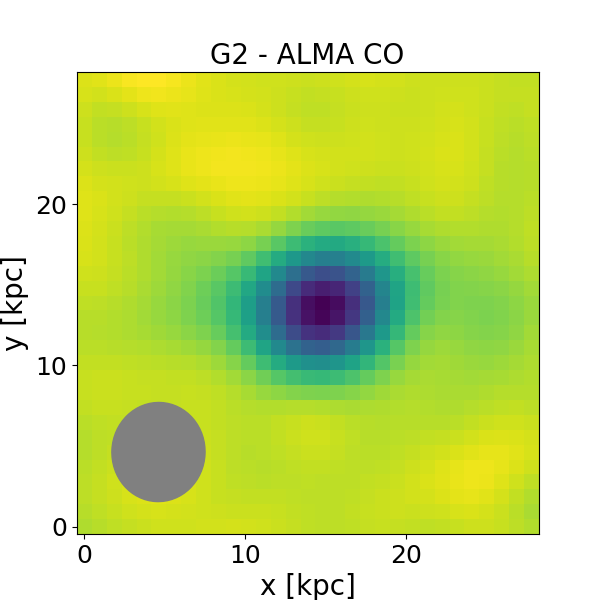}
\includegraphics[width=5.8cm, angle=0]{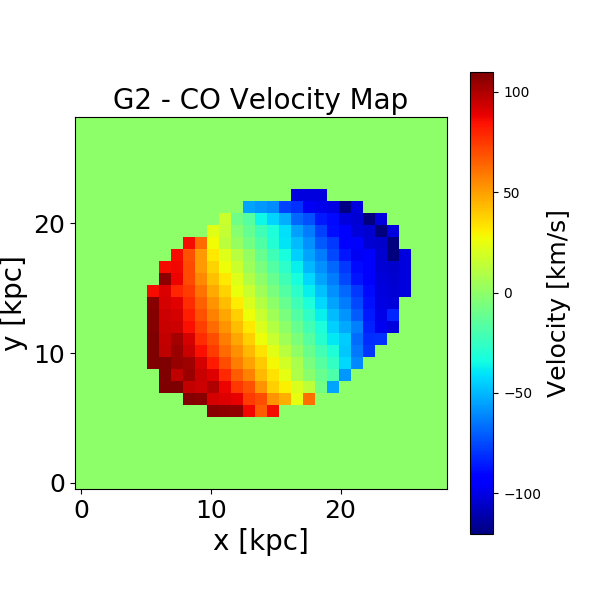}
\includegraphics[width=6.2cm, angle=0]{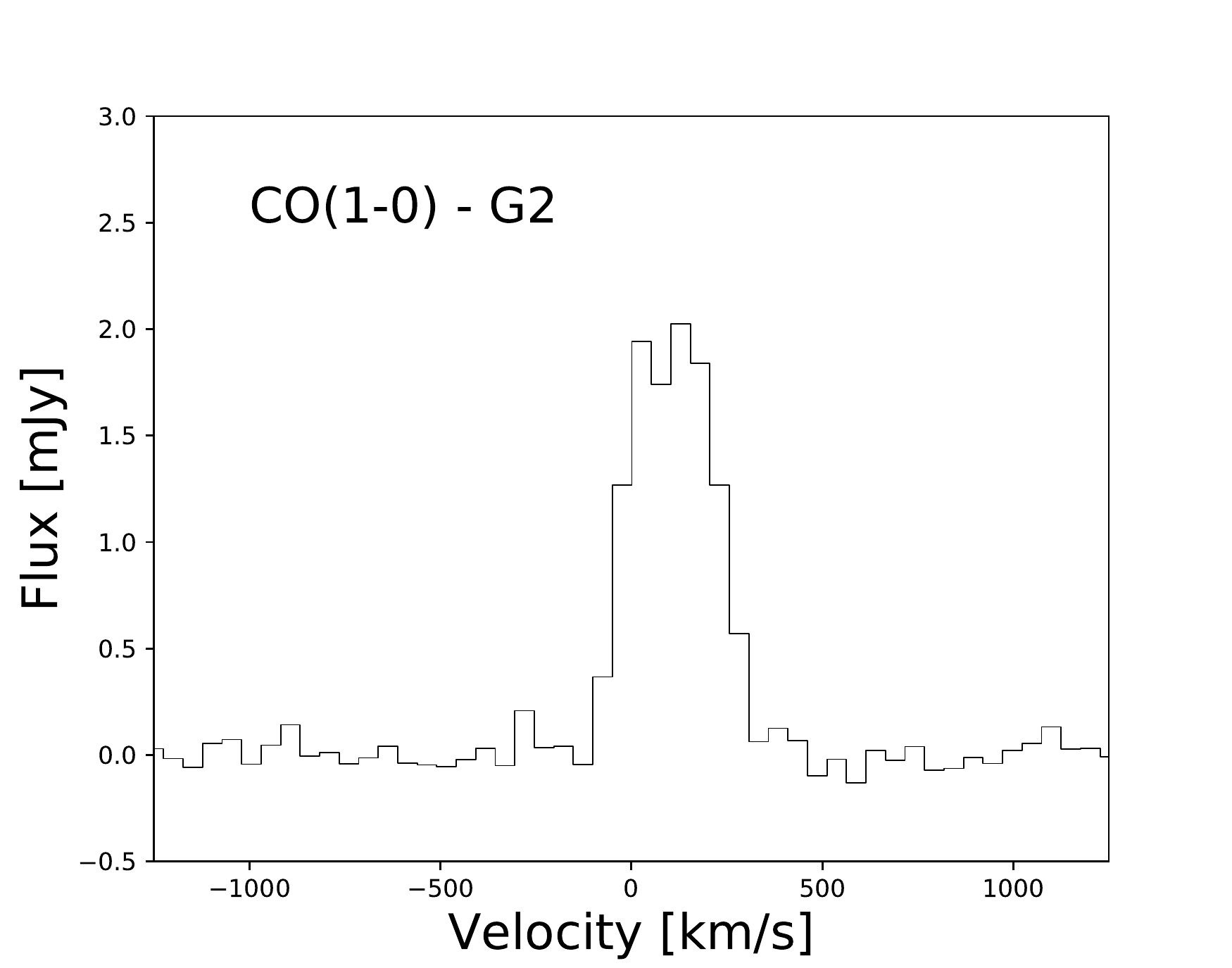}

\includegraphics[width=5.cm, angle=0]{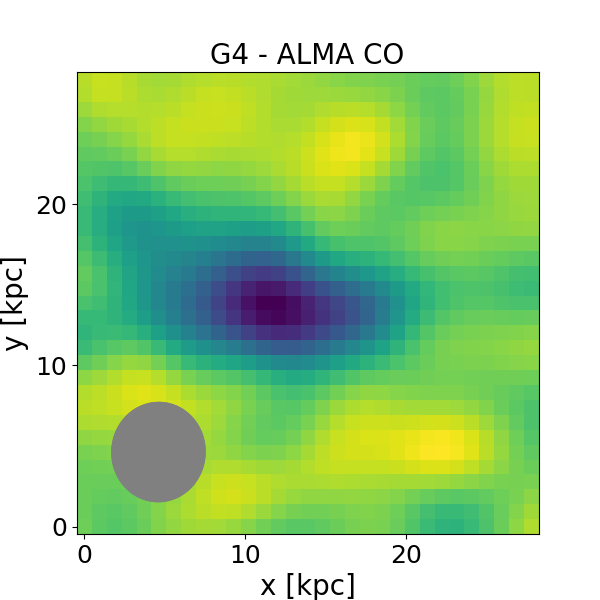}
\includegraphics[width=5.8cm, angle=0]{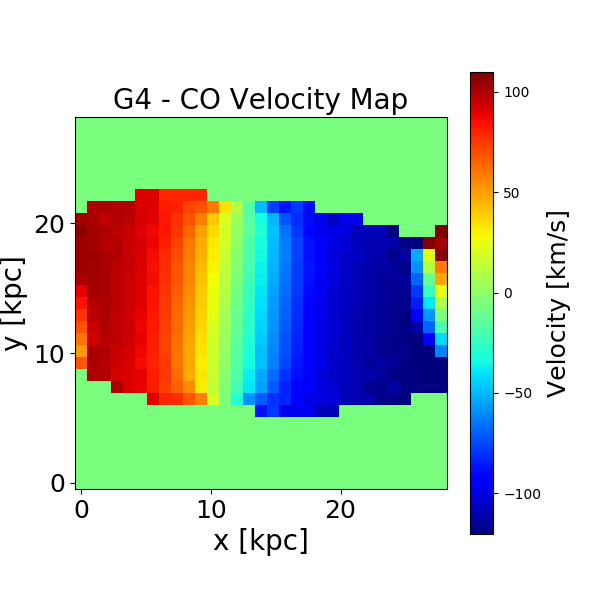}
\includegraphics[width=6.2cm, angle=0]{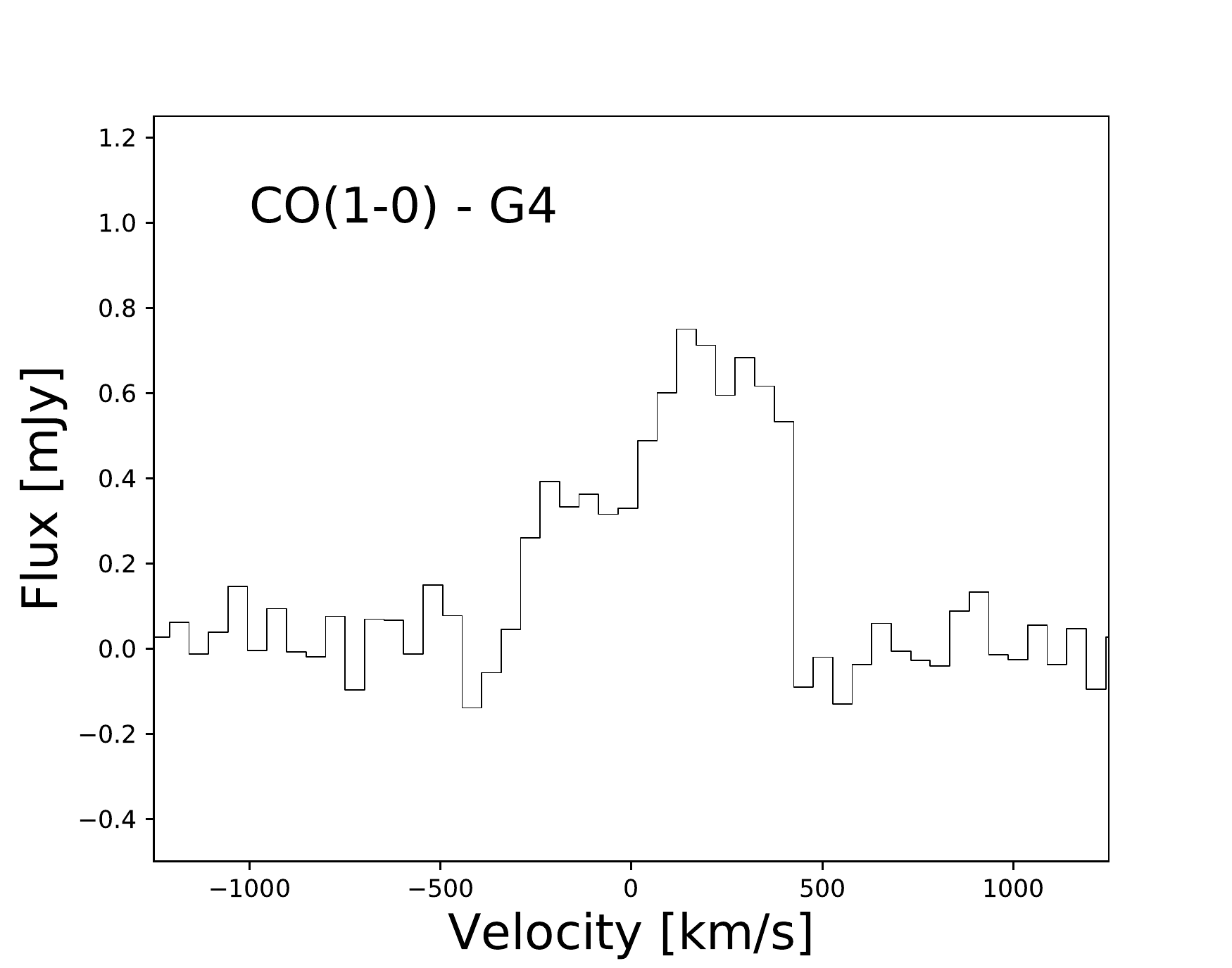}

\includegraphics[width=5.cm, angle=0]{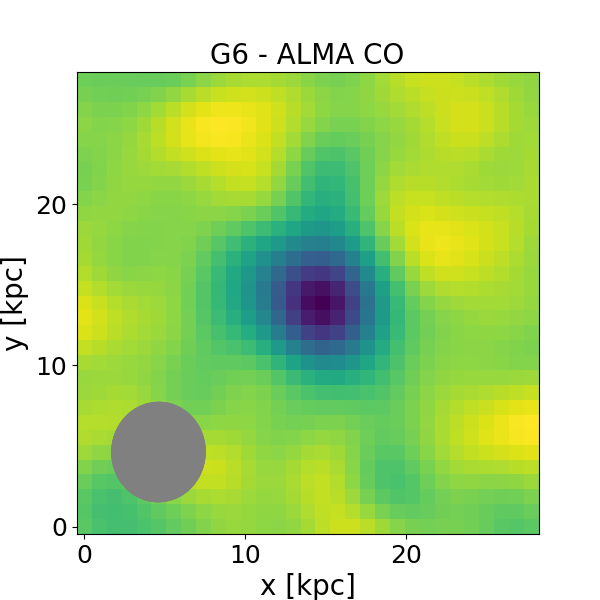}
\hspace{5.8cm}
\includegraphics[width=6.2cm, angle=0]{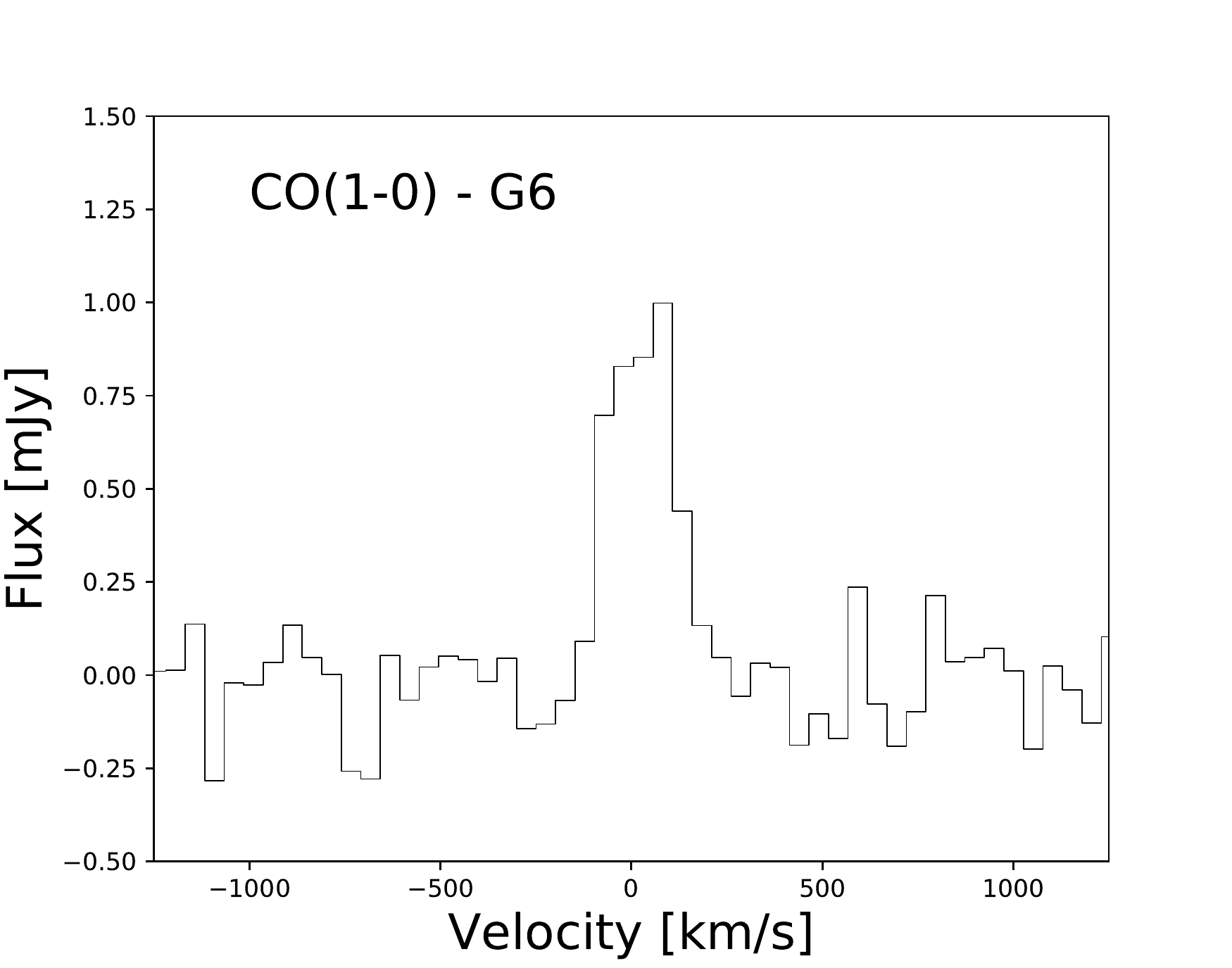}

\caption{{\bf Molecular gas content of individual galaxies detected in ALMA at z$_{\rm abs}$=0.313.} The CO(1-0) emission line map (left), modeled velocity maps convolved with the instrumental PSF (center) and extracted 1D CO(1-0) spectra (right) are shown. The grey ellipse on the left column indicates the ALMA beam size. The results of the morpho-kinematic analysis of G2 and G4 are qualitatively similar to the MUSE ones.}  
\label{f:ALMA_Kine}
\end{center}
\end{figure*}

\label{lastpage}
\end{document}